\documentclass[a4paper,fleqn,usenatbib,useAMS]{mnras}
\usepackage[T1]{fontenc}
\usepackage{ae,aecompl}
\usepackage{longtable}
\usepackage{rotating}
\usepackage{graphicx}	% Including figure files
\usepackage{amsmath}	% Advanced maths commands
\usepackage{amssymb}	% Extra maths symbols
\usepackage{multirow,bigdelim,array,pdflscape,booktabs,psfig}
\newcolumntype{C}[1]{>{\centering\arraybackslash}m{#1}}

\usepackage{float}
\usepackage{array}
\usepackage{makecell}
\usepackage{color,natbib}

\def\aj{AJ}%
\def\araa{ARA\&A}%
\def\apj{ApJ}%
\def\apjl{ApJ}%
\def\apjs{ApJS}%
\def\apss{Ap\&SS}%          % Astrophysics and Space Science
\def\aap{A\&A}%          % Astronomy and Astrophysics
\def\aapr{A\&A~Rev.}%          % Astronomy and Astrophysics Reviews
\def\aaps{A\&AS}%          % Astronomy and Astrophysics, Supplement
\def\mnras{MNRAS}%
\def\pasp{PASP}%
\usepackage[normalem]{ulem}

\title[LeMMINGs. II. Second data release.]{LeMMINGs. II. The {\it e-}MERLIN
  legacy survey of nearby galaxies. The deepest radio view of the  Palomar sample on parsec scale.}

\author[R. D. Baldi et al.]{R.~D.~Baldi$^{1,2,3,4}$\thanks{E-mail:ranieri.baldi@inaf.it},
D.~R.~A. Williams$^{5,6}$,
I.~M. McHardy$^{2}$,
R.~J. Beswick$^{5}$,
E. Brinks$^{7}$,
\newauthor B.~T. Dullo$^{8}$,
 J.~H. Knapen$^{9,10}$,
M.~K. Argo$^{5,11}$,
S. Aalto$^{12}$
A. Alberdi$^{13}$,
W.~A. Baan$^{14}$,
\newauthor G.~J. Bendo$^{5,15}$, 
S. Corbel$^{16,17}$,
D.~M. Fenech$^{18}$,
J.~S. Gallagher$^{19}$,
D.~A. Green$^{20}$,
\newauthor 
R.~C. Kennicutt$^{21,22}$,
H.-R. Kl\"{o}ckner$^{23}$,
E. K\"{o}rding$^{24}$, 
T.~J. Maccarone$^{25}$,
\newauthor T.~W.~B. Muxlow$^{5}$,
C.~G. Mundell$^{26}$, 
F. Panessa$^{4}$, 
A.~B. Peck$^{27}$,
M.~A. P\'erez-Torres$^{13}$,
\newauthor C. Romero-Ca\~nizales$^{28}$,
P. Saikia$^{29}$, 
F. Shankar$^{2}$,
R.~E. Spencer$^{5}$, 
I.~R. Stevens$^{30}$,
\newauthor E. Varenius$^{12,5}$,
M.~J. Ward$^{31}$,
J. Yates$^{18}$
\\
}

\pubyear{2020}

\begin{document}
\label{firstpage}
\pagerange{\pageref{firstpage}--\pageref{lastpage}}
\maketitle

\begin{abstract}
 %250 words limit, now 253 
  We present the second data release of high-resolution ($\leq0.2$
  arcsec) 1.5-GHz radio images of 177 nearby galaxies from the Palomar sample, observed with the {\it e-}MERLIN array, as part of the LeMMINGs (Legacy {\it e-}MERLIN Multi-band Imaging of Nearby Galaxy Sample) survey. Together with the 103 targets of the first LeMMINGs data release, this represents a complete sample  of 280 local active (LINER and Seyfert) and inactive galaxies (H{\sc ii} galaxies and Absorption Line Galaxies, ALG). This large program is the deepest radio survey of the local Universe, $\gtrsim$10$^{17.6}$ W Hz$^{-1}$, regardless of the host and nuclear type: we detect radio emission $\gtrsim$0.25 mJy beam$^{-1}$ for 125/280 galaxies (44.6 per cent) with sizes of typically $\lesssim$100 pc. Of those 125, 106 targets show a core which coincides within 1.2 arcsec with the optical nucleus. Although we observed mostly cores, around one third of the detected galaxies features jetted morphologies. The detected radio core luminosities of the sample range between $\sim$10$^{34}$ and 10$^{40}$ erg s$^{-1}$. LINERs and Seyferts are the most luminous sources, whereas H{\sc ii} galaxies are the least. LINERs show FR~I-like core-brightened radio structures while Seyferts reveal the highest fraction of symmetric morphologies. The majority of H{\sc ii} galaxies have single radio core or complex extended  structures, which probably conceal a nuclear starburst and/or a weak active nucleus (seven of them show clear jets). ALGs, which are typically found in evolved ellipticals, although the least numerous, exhibit on average the most luminous radio structures, similar to LINERs.

\end{abstract}


%%%%%%%%%%%%%%%%%%%%%%%%%%%%%%%%%%%%%%%%%%%%%%%%%%

%%%%%%%%%%%%%%%%% BODY OF PAPER %%%%%%%%%%%%%%%%%%

%-------------------------------------------------------------------

\section{Introduction}

Observational studies support the idea of co-evolution of supermassive black holes (SMBHs) and their host galaxies (e.g. \citealt{heckman14}). For example, the empirical scaling relationships between black hole mass (M$_{\rm BH}$) and both stellar velocity dispersion and host bulge luminosity (e.g. \citealt{ferrarese00,gebhardt00,tremaine02}) are indications of a coupled growth of the SMBH and its host galaxy. %The knowledge of the linked properties of the accretion onto SMBHs and of their hosts together with the M$_{\rm BH}$ function
These relationships provide some of the most basic constraints on models of SMBH and galaxy formation and evolution (e.g. \citealt{menci04}).  However, these observational constraints are poorly known (e.g. \citealt{graham13,shankar13,shankar16,shankar19} and references therein), particularly at low M$_{\rm BH}$, for which dynamical mass measurements become increasingly more challenging \citep{peterson14}. This uncertainty at low masses prevents us  from properly calibrating the scaling relations and the
prescriptions for SMBH-galaxy growth used in semi-analytical and numerical models \citep{shankar12,barausse17}.

The agreement between the accreted mass function as extracted from continuity equation arguments for M$_{\rm BH}$ and nuclear luminosity distribution, and the local BH mass function derived from local scaling relations, strongly suggest that all local massive galaxies have undergone at least one major episode of Active Galactic Nuclei (AGN) in their past (e.g., \citealt{soltan82}). It also supports the view that the vast majority of local galaxies host a central SMBH (e.g. \citealt{aller02,marconi04,shankar04}). The detection of AGN activity at the centre of galaxies is considered sufficient evidence to confirm the existence of a SMBH. The Eddington ratio\footnote{The Eddington ratio is defined as the ratio
  between the bolometric luminosity of the AGN (precisely, the disc
  luminosity, or a proxy)  and the Eddington luminosity.}  distribution of local galaxies has been measured to be extremely broad, mostly sub-Eddington, extending down to very low Eddington rates ($\sim$10$^{-6}$, e.g., \citealt{kauffmann09,schulze10}). A large portion of low mass BHs hosted in low mass galaxies are thus expected, and indeed observed, to be among the faintest luminosity AGN, with the lowest accretion rates and Eddington ratios, and extremely weak nuclear outputs (e.g. \citealt{ho99,ho08,panessa07}). These low-luminosity AGN (LLAGN) are traditionally defined as having H$\alpha$ luminosities $\leq$ 10$^{40}$ erg s$^{-1}$ \citep{ho97a}. They numerically dominate the local Universe \citep{nagar02,filho06,saikia18} and include the two main classes of active galaxies, LINERs and Seyferts, distinguished by their optical emission lines \citep{heckman80}.  While Seyfert galaxies show clear multi-band signatures of BH activity (i.e. broad emission lines, hard X-ray spectra, \citealt{maoz07,ho08}), the physical origin of the central engine in LINERs has been debated: stellar or non-stellar nature (SMBH, shocks, or hot stars; see \citealt{ho93} for a review)? The common interpretation is that a radiatively-efficient accretion disc appears to reconcile with the high-energy properties of Seyferts, while LINERs, which are typically fainter, are possibly powered by a radiatively-inefficient accretion disc \citep{kewley06,maoz07,ho08,heckman14}.

However, our view of the nuclear activity in the local Universe is partial and biased towards massive, bright and unobscured galaxies. In turn, detailed and complete studies of the low-brightness and low-mass active galaxy populations in the local Universe are still sparse. Optically weak or inactive galaxies can hide quiescent/low-accreting SMBHs, which are missed in the current local BH census. In fact, even star forming galaxies, although energetically dominated by dusty H{\sc ii} regions, and extremely enshrouded objects with high column densities can hide a compact object at their centers, which could reveal weak signatures of activity in optical, infrared and X-ray bands (e.g. \citealt{reines13,reines16,chen17,marleau17,girichidis20}). In addition, the BH activity is an episodic event: galaxies can go through periods of nuclear inactivity within their duty cycle, where their optical output is basically turned off and the SMBH becomes quiescent (e.g \citealt{woltjer59,marconi04,morganti17a}).

The best method to overcome this bias towards bright and massive galaxies is through radio observations which by virtue of not being obscured by intervening material allow the very centres of galaxies to be viewed in a less biased way (despite opacity problems in the radio band, such as synchrotron self-absorption and free-free absorption). Radio observations consent to investigate a wide range of astrophysical phenomena, from those related to the formation, evolution, and death  of stars (e.g. supernovae, SN), to  accretion onto SMBHs. In case of stellar processes, thermal and non-thermal radiation is produced by stellar ejecta (e.g. SN remnants, SNR) and photoionisation; in active SMBHs a plethora of radio-emitting mechanisms can compete (see \citealt{panessa19} for a review): jets  \citep{padovani16,blandford19}, disc winds \citep{zakamska14} or outflowing magnetically-active coronae \citep{laor08}. Radio observations provide the best single diagnostic to separate star formation (SF) and AGN components (e.g. morphology, luminosity, brightness temperature).

Long-baseline radio arrays are suitable for detecting  the low-level nuclear output of the LLAGN. These observatories, applied to nearby galaxies, provide the pc-scale resolution which is required to isolate the low-brightness nuclear emission, comparable to that of Sgr~A*, from more diffuse emission of the host galaxy. Because of its long, UK-wide baselines, and large bandwidth  {\it e-}MERLIN, is among the best radio arrays to detect compact structures, e.g., AGN cores, nuclear starburst and jets, in galaxies in the nearby Universe. A deep radio study of a complete sample of LLAGN at milli-arcsecond resolution and $\mu$Jy sensitivity with {\it e-}MERLIN array has the potential to create a census of the constituent components of galaxies at unprecedented depth and at pc-scale linear resolution.  This is the objective of the LeMMINGs (Legacy {\it e-}MERLIN Multi-band Imaging of Nearby Galaxy Sample\footnote{http://www.e-merlin.ac.uk/legacy/projects/lemmings.html}) survey \citep{beswick14}.  To reduce bias against optically active AGN as present in previous studies, we chose as our target the magnitude-limited sample of 
nearby galaxy selected by \citet{ho97a}, which is commonly known
as the 'Palomar sample'. This sample has a median distance of $\sim$20 Mpc and is statistically complete with no radio imposed
constraint or bias. This sample is by far the most widely observed, across a range of wavelength regimes ({\em Spitzer, Herschel, HST}, and with complete {\em Chandra} and VLA coverage). It includes all optical spectral classes (LINER, Seyfert, star forming and optically inactive galaxies) and morphological host types (early- and late-type galaxies), encompassing a wide range of BH masses
(from intermediate BH up to the most massive BHs of the local Universe, 10$^{4}$ -10$^{9}$ M$_{\odot}$) and accretion rates ($\sim$10$^{-6}$ -10$^{-1}$ in Eddington units, \citealt{connolly16}).

Observations at L-band (1.5 GHz) of the
first 103 galaxies of the LeMMINGs project were presented in the first data release by \citet{baldi18lem} (Paper~I hereafter). 
One of the results  reported there was the detection of pc-scale jetted structure in inactive galaxies 
down, to M$_{\rm BH}$ $\sim$10$^{6}$ M$_{\odot}$, suggesting that a (weakly) active SMBH is present at the centre of local galaxies regardless of their optical class. Here we complete the full LeMMINGs sample presenting the images of the remaining 177 galaxies at 1.5 GHz and proceed to study the survey from both data releases combined. Basic 
results on the radio properties of this population of nearby galaxies are discussed in this work.% while a deep analysis combined with optical and X-ray data will be object of following papers.

This paper is organized as follows. In Sect. 2 we present the LeMMINGs project and sample, and the updated optical classification of the sub-sample of 177 galaxies. The observations and calibration of the radio data are explained in Sect. 3. The identification of the radio cores and the general radio properties of the sub-sample are presented in Sect. 4. We discuss the results and implications of the radio emission for the entire LeMMINGs survey (280 galaxies) in Sect. 5 and draw our conclusions in Sect. 6.

\begin{figure*}
\centering
\includegraphics[width=0.6\textwidth,angle=90]{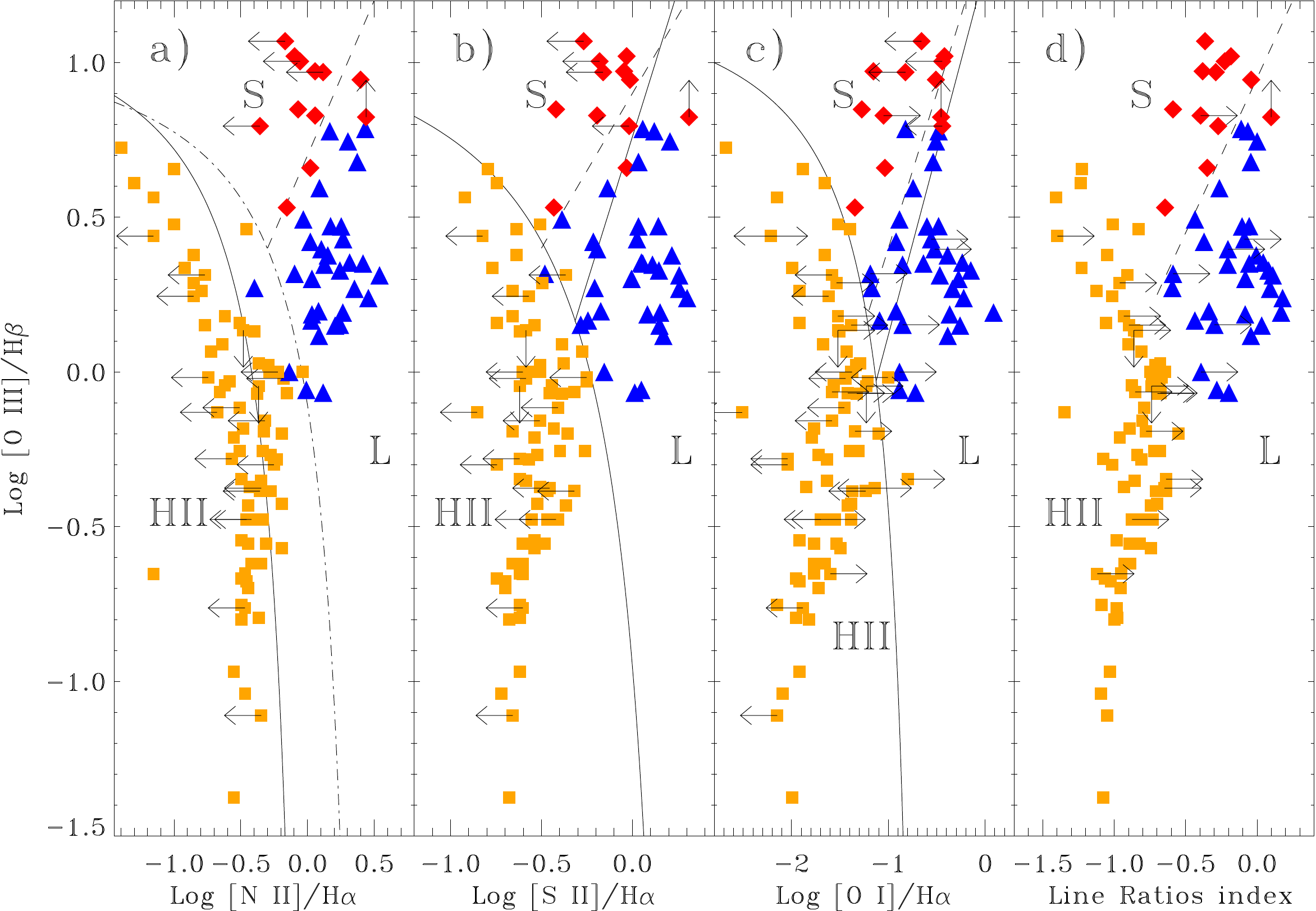}
\caption{Diagnostic diagrams based on optical spectra (named BPT, \citealt{baldwin81}): log([O~III]/H$\beta$) vs. $a$)
  log([N~II]/H$\alpha$), $b$) log([S~II]/H$\alpha$), $c$)
  vs. log([O~I]/H$\alpha$), and $d$) line ratios index (the averaged line ratios, see the definition
  in Paper~I). The data-points represent the 122 galaxies with emission line data  taken from \citet{ho97a} (the remaining 55 sources were classified based on data from the recent literature; see Table~\ref{table1}). In the first three panels, the solid lines
  separating H{\sc ii} galaxies, LINER, and Seyferts are taken
  from \citet{kewley06}. The dashed lines between Seyferts and LINERs
  in the four panels separate the two classes according to the scheme introduced by \citet{buttiglione10}. In panel $a$, the sources between the solid and the dot-dashed lines were classified as Transition galaxies by \citet{ho97a}, which we
  re-classify as LINER or H{\sc ii} galaxies based on the other
  diagrams ('$b$' and '$c$'). We mark LINERs as blue triangles, Seyferts as red diamonds, and H{\sc ii} galaxies as orange squares. Several galaxies have different classifications with respect to those from \citet{ho97a} because of the updated diagnostic diagrams (see Table~\ref{table1}).}
\label{bpt}
\end{figure*}

\section{The Palomar Sample and the LeMMINGs survey}
 
\subsection{Sample selection}
 
 The sample of this survey is a subset of the Revised Shapley-Ames
 Catalog of Bright Galaxies and the Second Reference Catalogue of
 Bright Galaxies ($\delta > 0^{\circ}$ and B$_{T} \le$ 12.5,
 \citealt{sandage81,devaucouleurs76}), which was originally observed
 by \citet{ho95} with the Hale 5m telescope at the Palomar Observatory
 \citep{filippenko85} to carry out a deep spectroscopic
 campaign. \citeauthor{ho97a} extracted the optical emission lines (H$\beta$, [O~III], [O~I], [N~II],
 H$\alpha$, [S~II] doublet) from their spectra and used the emission line ratios to classify them as H{\sc ii}, Seyfert, LINER or Transition galaxy (see Sect.~\ref{revised} for the updated classification).

The galaxies for which the active SMBH is the main photoionising source are Seyferts and LINERs, which represent the $\sim$11 and $\sim$19 per cent of the Palomar sample, respectively. The inactive galaxies are the H{\sc ii} galaxies ($\sim$42 per cent of the Palomar sample) where star forming regions populated by massive young stars mainly photoionise the surrounding gas, and the Absorption Line Galaxies (ALG, $\sim$14 per cent) which are optically inactive galaxies. The latter shows no obvious emission lines and are typically in  early-type galaxies. \cite{ho93}  introduced a further class which they named Transition galaxies ($\sim$14 per cent). These are sources with composite characteristics of both LINER and H{\sc ii} galaxies and are based on a single diagnostic diagram (of [O~III]/H$\beta$ vs.\ [N~II]/H$\alpha$; diagram '$a$' in Figure~\ref{bpt}).

The LeMMINGs survey focuses on a sub-sample of the original Palomar
catalogue, namely on those galaxies with declination $>$ 20$^{\circ}$ (280 targets), to ensure good visibility
(or {\em uv-}plane) coverage, and accessibility to the {\it e-}MERLIN
array over a wide hour-angle range. The survey was designed to carry out shallow observations at 1.5 and 5 GHz without the largest {\it e-}MERLIN antenna, the Lovell telescope. In addition, several galaxies were observed to greater depth as part of LeMMINGs (i.e., with the Lovell telescope included): M~82 \citep{muxlow10}, IC~10
\citep{westcott17}, NGC~4151 \citep{williams17}, NGC~5322
\citep{dullo18}, M~51b (NGC~5195, \citealt{rampadarath18}), and NGC~6217 \citep{williams19}.

In Paper~I a resolution of 150 mas resulted in the detection of 1.5-GHz radio emission at pc scales, reaching a sensitivity of $\sim$75 $\mu$Jy beam$^{-1}$. We detected 46 per cent (47/103) of the targets observed\footnote{Based on a more detailed analysis of NGC~147, which was the least luminous detected radio source of the sample in Paper~I, we have decided to remove this target from the group of detected galaxies (see Sect.~\ref{complete} for details).}, measuring radio dimensions typically $\lesssim$100 pc and radio luminosities in the range
$\sim$10$^{34}$-10$^{40}$ erg s$^{-1}$. Here, in analogy to Paper~I, we report our observations of the remaining 177 Palomar galaxies at 1.5 GHz, listed in Table~\ref{table1}.

\subsection{Revised optical classification}
\label{revised}

Whereas Seyferts display high ionisation lines which are indicative of photoionisation by an active nucleus, the typical emission line ratios of
LINERs can be reproduced either by AGN photoionisation, collisional excitation by shocks, photoionisation by post-AGB stars, or by combined starburst and merger-driven shock \citep{allen08,sarzi10,capetti11b,singh13,rich14}. In addition, the
Transition Galaxies are not a well defined class, being a mixture of photo-ionisation by SF and a weak AGN. Therefore a secure and net
separation between the different classes is necessary for a better and
well-refined physical interpretation of each class. For this purpose,
in analogy to what we performed in the first data release (Paper~I), we
revise the optical classifications carried out by \citet{ho97a} by using
the state-of-the-art spectroscopic diagnostic diagrams based on criteria introduced by \citet{kewley06} and \citet{buttiglione10}. The former used SDSS emission-line galaxies and BPT diagrams \citep{baldwin81} to classify mostly radio-quiet AGN galaxies, whereas the latter used optical spectra of radio galaxies from the Revised Third Cambridge Catalogue (3CR, \citealt{bennett62a}, i.e., only radio-loud AGN) obtained with the Telescopio Nazionale Galileo.  \citeauthor{ho97a} scheme marginally differ in the separation of Seyferts and LINERs from those used by \citet{kewley06} and \citet{buttiglione10}: Seyferts by \citet{ho97a} with log([O~III]/H$\beta$) $>$ 0.5 are now reclassified as LINERs. We also used the `line
ratios index' introduced by \citet{buttiglione10} as the average of three low ionisation line ratios for a more robust separation between LINERs and Seyferts. Furthermore, we opt for
removing the Transition galaxy class, by classifying the given galaxy
either as LINER or H{\sc ii} galaxy based on the other diagnostic
diagrams ('b' and 'c' diagrams in Fig.~\ref{bpt}). Finally, each target is classified as H{\sc ii}, LINER, or
Seyfert based on at least two diagnostic diagrams in case the third
criterion disagrees with the other two (see Paper~I for more details).

Of the 177 galaxies, 122 exhibit at least four detected emission lines
(i.e. having line uncertainties smaller than 50 per cent in \citealt{ho97a}),
which is enough to ensure a reliable classification in the BPT
diagrams (Fig.~\ref{bpt}). The remaining 55 sources are classified
based on recent spectra taken from the literature (see notes in
Table~\ref{table1}). The revision of their classification based on optical spectra resulted in a final sample of
89 H{\sc ii} galaxies, 60 LINERs, 14 Seyferts and 14 ALGs.

When considering  galaxy morphological type, most of the sources
are late-type galaxies (LTGs, spiral and irregular galaxies $\sim$71 per cent), with a smaller
fraction of early-type galaxies (ETGs, ellipticals and lenticulars). LINERs and ALGs are typically  in ETGs, Seyferts have  both early- and late-type  morphologies and H{\sc ii} galaxies are 
mostly late-type systems (see Sect.~\ref{complete} for more details).

\vspace{-0.4cm}
\section{Observations and Data Reduction}

Detailed information on the {\it e-}MERLIN 1.5-GHz observations can be found in Paper~I. The sub-sample presented in this work was observed from March 2017 to March 2019, divided into observing blocks of typically 10 targets, grouped based on their right ascensions to minimise the slewing of the 7 telescopes.

The observing strategy was identical to that used for the galaxies in Paper~I: it consisted in following the target and the
phase calibrator in at least six visits (cycles) to maximise {\em uv-}coverage given allocated time. A target-phase calibrator
cycle usually lasted $\sim$10 min, with $\sim$3 min on the phase calibrator
and $\sim$7 min on the target. The phase calibrators were selected from the
VLBA calibrator lists \citep{beasley02} and/or form the latest RFC catalog\footnote{available from http://astrogeo.org}, chosen for being unresolved on {\it e-}MERLIN
baseline scales. The bandpass calibrator (OQ 208) and the flux
calibrator (3C~286) were typically observed for several minutes each. In
this work, we present the data from seventeen scheduling blocks which include the 177 Palomar galaxies presented here (Table~\ref{table1}).

\subsection{Data Calibration and Imaging with CASA}

In a change from Paper~I, where the data reduction was carried out using the
\textsc{AIPS}\footnote{AIPS, the Astronomical Image Processing Software \citep{aips}, is free software available from the NRAO.} 
software package, here the data has been calibrated with \textsc{CASA} \citep{casa}, the
Common Astronomy Software Applications package. This change was prompted by the release of the {\it e-}MERLIN CASA pipeline\footnote{https://github.com/e-merlin/eMERLIN\_CASA\_pipeline/} at around the time the observations presented in this paper were conducted. In
Section~\ref{aipsvscasa} we will present the differences between the two software packages.

The \textsc{CASA} pipeline converts \texttt{`fitsidi'} format observation
files into CASA measurement sets (\texttt{`MS'}); next it performs {\em a priori}
flagging such as removing the first minutes of data when antennas are not yet all tracking a source, flagging the edges of the observing band and spectral windows, and applying any observatory flags.

The 512\,MHz wide L-band suffers substantial radio frequency interference (RFI) due to a variety of sources, such as
modems, satellites, mobile phones, and radars. Hence, to achieve the highest sensitivity possible, RFI must be removed from the outset. The CASA pipeline uses the \textsc{AOFlagger}
software \citep{AOFlagger} which calculates a limiting threshold based
on the raw data, above which any instances of RFI are automatically
flagged and subsequently removed from the dataset. To remove any
low-level RFI (the target observations are expected to be
intrinsically faint) we further inspect the data with \textsc{CASA}
task \texttt{`plotms'}, plotting the amplitude and phase of the  visibilities as a
function of time and channel/frequency to detect any amplitude spikes,  dropouts, or intervals with null phase. We have estimated that the
flagged data represents typically $\sim$15-20 per cent of the  raw data.

 The data are subsequently averaged down in frequency by a factor of four
 to reduce the data volume and improve calibration speed without losing any significant information that might affect our scientific objectives. The pipeline then allows any additional manual flags to be added by the user, before proceeding with calibrating the data as follows. First, the flux for the primary
 calibrator 3C~286 is set using a model by
 \texttt{`setjy'}. This is followed by delay and phase-only calibration with the task \texttt{`gaincal'} using a solution
 interval of $\sim$10\,s. Next is amplitude and phase calibration with a solution
 interval of 2-3 min in amplitude and phase.
 An initial bandpass response table was created to account for the
 changes of sensitivity across the band with the task
 \texttt{`bandpass'}. At this point further automatic identification and removal of data outliers in the
 time-frequency plane indicating RFIs is done with
 the task \texttt{`flagdata'} (with \texttt{`tfcrop'} mode). The flux density
 scale of 3C~286 is then bootstrapped to the secondary calibrator and
 target sources using \texttt{`fluxscale'}, taking into account that
 3C~286 is slightly resolved by the longest {\it e-}MERLIN baselines. A
 final bandpass table is then recalculated using the spectral
 information obtained from the previous step. Final phase and
 amplitude solutions were recalculated after the bandpass correction.
 The phase solutions are usually delimited within $\pm$20$^{\circ}$
 independent of baseline, while the amplitude solutions show typical
 variations within a range of 10$-$20 per cent. The phase and amplitude
 solution tables are then applied to the data.  A final step of
 flagging the data using \texttt{`flagdata'} (\texttt{`tfcrop'} mode) is run
 as part of the pipeline to remove any RFI from the target fields. The
 solutions from the amplitude and phase calibration and the bandpass
 table were applied to the data to assess the data quality with
 \texttt{`possm'}. Diagnostic plots are produced and uploaded to
 a `weblog' which allows for checking the quality of the calibration by
 showing such plots as the per-antenna flagging percentage
 at each step, calibration tables at each step and 
 images of the calibrators and targets. These plots were inspected for
 any signs of remaining RFI or poor solutions in the calibration tables, with
 the option of running the pipeline again with manually input flags
 post averaging, with the aim of removing any subsections of the data that are of poor quality. After
 inspection of the diagnostic plots and manual excision of RFI in the
 target fields, the latter were 'split' from the now calibrated measurement set to create a single-source data file which is more manageable for the data imaging stage. In conclusion, the entire procedure achieved a maximum of 20 per cent calibration error in L-band.

Imaging of the {\it e-}MERLIN data (Stokes I) was also performed within the \textsc{CASA} environment, using the task \texttt{`tclean'} on the 'split', 
calibrated datasets as described above. This task includes the possibility of using the
\texttt{`mtmfs'} deconvolver mode, which allows to reconstruct images from
visibilities using a multi-term (multi-scale) multi-frequency
approach \citep{rau11}. After Fourier transforming the visibilities into an image, the latter is then 'cleaned', essentially deconvolved quasi simultaneously at a small number of characteristic scales, taking into account the no-null frequency dependence of the emission in the different sub-bands\footnote{We used 2 Taylor coefficients in the spectral model, which then corresponds to a spectrum defined by a
straight line with a slope at the reference frequency of 1.5 GHz. The spectral image has not been considered in this work.}. We used three different scales: the smallest scale size is recommended to be 0 (point source), the second the size of the synthesized beam
and the third 3-5 times the synthesized beam. Since the nominal beam
size of the {\it e-}MERLIN L-band observations is 150 mas, we used a cell
size of 50 mas and we set the corresponding scales array to [0, 3, 10] pixels. The
images were produced with natural baseline weighting, mapping a field of 1024 $\times$ 1024 pixels ($\sim$0.85 arcmin $\times$ $\sim$0.85
arcmin, i.e. 0.73 arcmin$^{2}$ which corresponds to an area of $\sim$11.6 kpc$^2$ at the median distance of the sample).

For the targets with flux densities higher than 5 mJy, we carried out a
few rounds of self-calibration in phase and a final one in phase and
amplitude, using 1-2 min integration times and using a
3-$\sigma$ minimum threshold for valid solutions. This procedure had the effect of increasing the
signal-to-noise of the final maps and reducing the scatter in phase and gain solutions. Bright sources in the fields were mapped in parallel with the targets by using separate, small fields centred on their location, thus reducing the level of their side-lobes and their contribution to the effective noise floor for the fields on the targets.

Several images were created with different resolutions to explore the
presence of diffuse low-brightness radio emission and to possibly
detect a target in case of no detection in full
resolution. Lower-resolution maps were obtained using different values
of the \texttt{`uvtaper'} parameter in \texttt{`tclean'}. This parameter specifies the width of
the Gaussian function in the {\em uv-}plane to down-weight the contribution by the longer baselines. We chose values ranging
between 300 and 750\,k$\lambda$. The narrower the Gaussian, the less weight is given to the longer baselines and hence the lower the resolution of the resulting maps. A value of 300 k$\lambda$
corresponds to a beam size typically $3 - 4$ times larger
(i.e. $0.45 - 0.6$ arcsec) than that reached at full resolution. Furthermore, the angular resolution of the images strongly depends on the  {\em uv-}plane coverage, hence the inclusion of the data from all
seven antennas. Extreme data flagging can consequently result in degradation of the resolution.  The range of restoring beam sizes is between 0.12 and 0.50 arcsec at full
resolution.

 Figures \ref{ident-maps} and \ref{unident-maps} (the full sets of figures for identified and unidentified sources will be as online supplementary data) present the full and low-resolution (\textit{uv}-tapered) maps of the detected sources. For each detected galaxy we present one or two \textit{uv}-tapered images chosen among those obtained with the highest \texttt{`uvtaper'} parameters (typically 750-500 k$\lambda$) judged to be the best for illustrating the radio structure. The radio images have a large
 dynamic range which highlights their quality. For a small fraction of
 sources (in blocks 08 and 12), the image quality is modest, usually
 due to an antenna `drop-out', but still adequate for the purposes of
 our survey (see Table~\ref{table1}).

To analyse the source parameters in the radio maps, we  used \texttt{`imfit'}, part of the CASA \texttt{`viewer'}, which fits two-dimensional Gaussians to an intensity
distribution on a region selected interactively on the map, providing
the position, the deconvolved size, peak flux density, integrated flux
density, and position angle (PA) of the source (all listed in
Table~\ref{tabdet} and Table~\ref{tabsfr}). Whilst this procedure is valid for compact or slightly-resolved components, a simple technique to estimate the total brightness of an extended component is by
interactively marking the region around the irregular shape of the
source. Similarly, we estimate the rms noise of the maps by selecting a region
around the target free from any significant emitting source. The
 maps show a range in effective rms noise of between $\sim$30 and 250
$\mu$Jy beam$^{-1}$. The higher values indicate we are for the brighter sources limited by the dynamic range of the data. An extreme example is the brightest source NGC~1275 in our sample with an integrated flux density of 47.1\,Jy and an effective rms noise of 156 mJy beam$^{-1}$, or a dynamic range of about 300:1 (Tab.~\ref{contours}).

\subsection{\textsc{AIPS} vs  \textsc{CASA}}
\label{aipsvscasa}

\begin{figure}
\includegraphics[width=0.35\textwidth,angle=90]{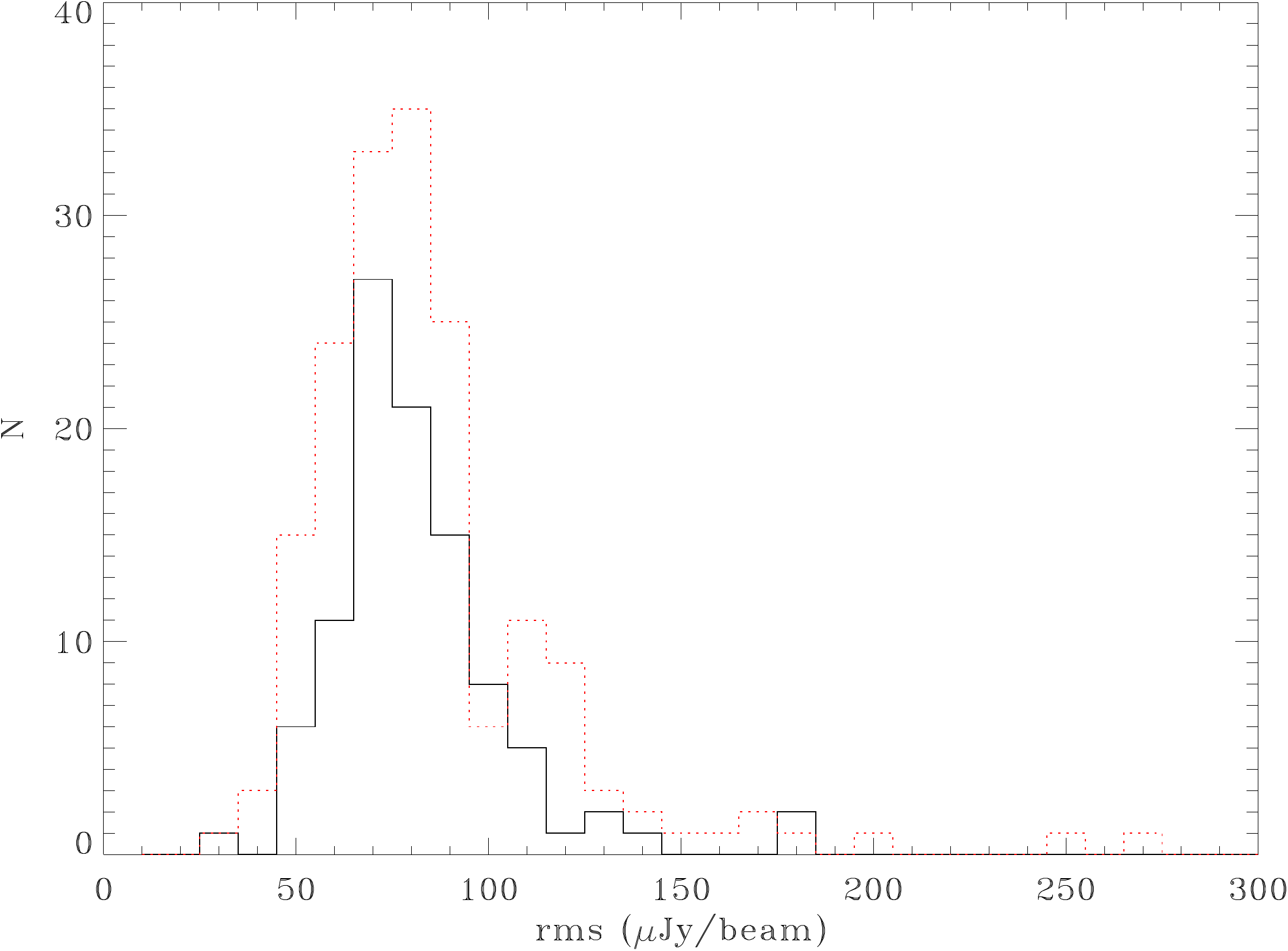}
\caption{The effective rms noise distribution of the data (103 targets) presented in
  Paper~I calibrated and imaged with \textsc{AIPS} (solid black line) and
  the data presented here (177 targets) calibrated and imaged with
  \textsc{CASA} (dashed red line). The result of a KS test is that
  the two distributions are not drawn from different parent
  populations at a confidence level greater than 95. Four targets with effective rms
  higher than 300 $\mu$Jy\,beam$^{-1}$ are from images that are dynamic range limited and not plotted.}
\label{rms}
\end{figure}

Since this Legacy Survey has been processed with two techniques based
on \textsc{AIPS} and \textsc{CASA} softwares, a comparison of the data
calibration and results between the two procedures is needed. Here we
report the main differences and crucial similarities between Paper~I and this work:

\begin{itemize}
\item Data calibrated and imaged via both \textsc{CASA} and \textsc{AIPS} (Paper~I) were averaged in frequency and/or time by the same amount (averaged to 0.5 MHz channels and 2-seconds integrations) prior calibration procedures.

\item Both \texttt{`SERPENT'}, the auto-flagging code \citep{peck13}
  written in ParselTongue \citep{kettenis06}  used in Paper~I, and \textsc{AOFlagger} are
  based on similar algorithms to assess the data quality and flag
  any instances of RFI. However, the updated version
  \textsc{AOFlagger} is more efficient than the \textsc{SERPENT} routine,
  because scripts searching for low-level RFIs have been
  optimised.

\item In Paper~I the calibration procedures began with a fit to the delay
  offsets among the distant antennas using the \textsc{AIPS}
  task \texttt{`FRING'}. Only recently a fringe fitting procedure became available in \textsc{CASA}, a Python-based global
  fringe fitter (task \texttt{`fringefit'}, \citealt{vanbemmel19}) developed specifically for long-baseline
  interferometers. The {\it e-}MERLIN \textsc{CASA} pipeline implements a simplified antenna-based delay correction.  Moreover, since {\it e-}MERLIN utilises a single central clock, unlike VLBI arrays, no additional rate correction is essentially required.

\item Each spectral window was used for amplitude and phase calibration, apart from their edge channels  which are noisier. In Paper~I we used the central 80 per cent of channels as recommended by the {\it e-}MERLIN cookbook, but the \textsc{CASA} pipeline uses the inner 90 per cent.

\item The data published in Paper~I underwent several rounds of phase-only self-calibration on
  the phase calibrators plus a final round of self-calibration in
  phase and amplitude. The {\it e-}MERLIN \textsc{CASA} pipeline performs a  self-calibration on the calibrators, but it defaults to a point source model assumption rather than iteratively creating maps as in \textsc{AIPS}. In the classical assumption of a phase calibrator structure being dominated by a compact point source, the two routines are essentially the same.

\item In the imaging technique, the \textsc{AIPS} task \texttt{`imagr'} uses the H\"ogbom clean method \citep{hogbom74}, which amounts to a
brute force deconvolution by subtracting the brightest
points in the map until it reaches a simply noisy map. Instead the deconvolution used by \texttt{`tclean'} with \texttt{`mtmfs'} mode
offers multi-frequency synthesis of the wide-band data.  \textsc{AIPS} task \texttt{`imagr'} also offers a simplified multi-scale option in the cleaning phase, but \textsc{CASA} \texttt{`tclean'} permits a better control of the parameters of the image deconvolution.

\item In order to extract the source parameters from the detected
  components, \textsc{AIPS} task \texttt{`jmfit'} and \textsc{CASA}
  task \texttt{`imfit'} work in a similar way by fitting a 2-D Gaussian
  across the source.
 \end{itemize}

\begin{figure}
\includegraphics[width=0.5\textwidth]{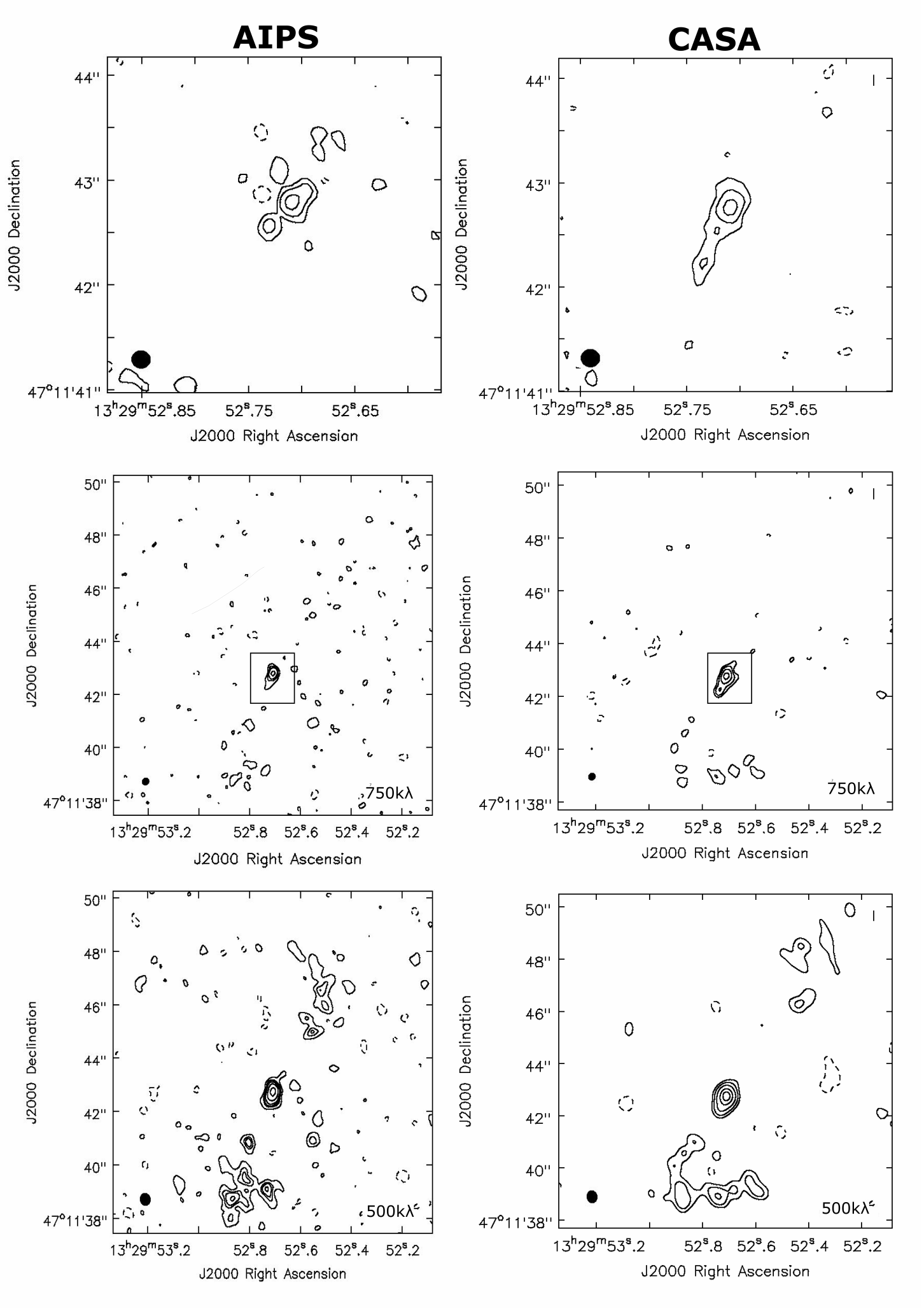}
\vspace{-0.5cm}
\caption{Radio maps (with natural weighting) of NGC~5194 for a comparison between the
  calibration and imaging performed with \textsc{AIPS} (left panels, taken from Paper~I) and with \textsc{CASA} (right panels). The upper panels are at full resolution, the medium panels at 750k$\lambda$ (the boxes represent the regions depicted in the upper panels at full resolution) and the lower ones at 500k$\lambda$ with matched restoring beams. The rms noise of the \textsc{CASA} maps are 68 $\mu$Jy beam$^{-1}$, while those of the \textsc{AIPS} maps are 78 $\mu$Jy beam$^{-1}$. The contour levels are 3$\times$rms$\times$N, where N = [-1,1,2,4], [-1,1,2,3,6], and [-1,1,2,2.5,3.3,5,8] respectively for the three sets of panels from the top to the bottom.}
\label{ngc5194}
\end{figure}

In the following we carry out a comparison between the two methods.  First, we compare the rms noise obtained for the radio images of the 103 sources from Paper~I and those calibrated in this work. Figure~\ref{rms} shows the two rms noise distributions. The median values of two distributions are 
similar, 81 and 84 $\mu$Jy beam$^{-1}$, for Paper~I and this
work, respectively. We also evaluate a two-sample Kolmogorov-Smirnov (KS) test,
to compare the cumulative distributions of the two
datasets. This test confirms there is no significant difference between the rms noise distribution of the two samples.

In addition, as a further test, we calibrated with \textsc{CASA} a randomly selected observing block of good quality, LEM~10, already published in Paper~I. We run the \textsc{CASA} pipeline on the dataset by tuning the calibration steps and further flagging the RFIs and
noisy visibilities. The calibrated dataset returns a lower rms
noise level of $\sim$68 $\mu$Jy beam$^{-1}$ than the $\sim$78 $\mu$Jy beam$^{-1}$ achieved with the
\textsc{AIPS} routine. The targets which were not detected with \textsc{AIPS} remain as such with \textsc{CASA} (NGC~4914, NGC~5055, NGC~5112, and NGC~5297). The restoring beams obtained with the two procedures are consistent with each other. The detected targets (NGC~5005, NGC~5194, NGC~5195, NGC~5377, and NGC~5448) display
similar but not identical radio structures identified in the \textsc{AIPS} maps (see next paragraph and Figure~\ref{ngc5194} for a detailed comparison). At full resolution the peak flux densities of the unresolved components (with matched synthesised beams) are typically $\sim$87 per cent of the estimates obtained with \textsc{AIPS} task
\texttt{`imagr'}. The difference is insignificant in the lower resolution maps. The integrated fluxes of the extended structures are consistent with the measurements obtained with \textsc{AIPS}. In terms of resolved structures, the wide-band imaging with \textsc{CASA} resulted in emission features that were better defined, reducing the confusion between noisy artifacts and genuine regions of emission. The differences revealed between the two procedures are probably the result of two
aspects: i) the \textsc{AOFlagger} software provides a better flagging
of data affected by RFI, returning a lower rms noise level; ii) the
multi-scale multi-Frequency synthesis of the \textsc{CASA} task \texttt{`tclean'} probes the spatial and spectral characteristics present in the data in a more accurate mode than the \textsc{CASA} task \texttt{`imagr'} . Furthermore, the different self-calibration routes of the two procedures could also lead to a little influence on the dynamic range of the target itself, if the object is highly detected (high signal-to-noise ratio) and can be self-calibrated. Conversely, the inability of self-calibrating because of a weak or non-detected target could largely affect the noise levels of the map and probably contribute to the different noise levels observed between the maps produced with \textsc{CASA}  and \textsc{AIPS}. 

 Figure~\ref{ngc5194} shows an example of two sets of radio images for the galaxy NGC~5194 obtained from the same raw data calibrated and imaged with the two software packages. As discussed before, the neater structures (i.e, see the edge of the southern radio lobe) and the lower rms noise observed in the \textsc{CASA} images of this target probably result from the combination of different steps of the two procedures. This example provides the level of reliability of the low-brightness structures observed in the maps of our sample and suggests caution in interpreting weak source structures, obtained either with or without self-calibration.

We can conclude that the two data calibration and imagining techniques
based on \textsc{AIPS} and \textsc{CASA} produce consistent results in
terms of flux densities and detected emission regions, with minor differences which are still within the absolute flux calibration error of $<$20 per cent.

\section{Results}
\subsection{Radio maps and source parameters}

As one of the key goals of the LeMMINGs Survey is to study of the 1.5-GHz emission ascribed to the central SMBH, we examine the innermost region of the galaxy in the full-resolution and
$uv$-tapered images near the optical/infrared identification of the
supposed nucleus. Practically, we search for significant radio emission, i.e. we detect a source component if its
flux density is above 3$\sigma$ of the local noise in analogy to Paper~I\footnote{We are aware that 3$\sigma$ limit might include spurious detections for particularly weak sources.}. In the cases of
diffuse low-level emission and no detection, we extract a 3$\sigma$ upper-limit to the core
flux density at full
resolution. Moreover note that the larger beams of the $uv$-tapered maps could cause the appearance of additional components not present at full resolution due to increased signal-to-noise ratio.

The median rms noise of the final naturally-weighted full-resolution images
is 84$\mu$Jy beam$^{-1}$, with a median ratio between the peak flux
density and the rms noise of five. We detected radio emission for 78 out of
the 177 targets with flux densities $\gtrsim$0.25 mJy. For such
sources, we also derive the flux densities of their counterparts in
the $uv$-tapered maps. The source
parameters (e.g. peak/integrated flux, deconvolved major/minor axes, position) for the detected sources are
listed in Table~\ref{tabdet} and \ref{tabsfr}. The contour levels and the restoring beam parameters  are listed in Table~\ref{contours}. For the remaining 99 objects, no significant radio emission ($>$ 3$\times$rms
level) was detected in the imaged fields, neither at full- nor at
low-resolution.

The morphology of the detected radio structures is varied,
ranging from pc-scale unresolved cores to jetted to extended and complex
shapes. The lower resolution images generally reveal a more extended
morphology than the full-resolution images.  The radio structures vary in size
from 150\,mas at the smallest scale for unresolved sources to up to
17 arcsec for sources like NGC~5548 ($\sim6$ kpc). With a mean
size for the radio structures in the survey being 0.5 arcsec, i.e.,  a physical size $\sim$3 - 550 pc (median $\sim$100 pc), most of the sources are slightly resolved or unresolved. All of the radio sizes for the resolved
sources are listed in Table~\ref{tabdet} and \ref{tabsfr}.

In agreement with Paper~I, we identify the radio core in each source as the unresolved central component, which might be associated with the active SMBH or nuclear star-forming region. We used the specific radio morphologies of each source combined with the optical centre of the galaxy obtained from the NASA Extragalactic Database (NED\footnote{https://ned.ipac.caltech.edu/}) to gauge the distance of the radio core from the optical centre. For sources with a symmetric structure, the central unresolved component is assigned to be the radio core, but for asymmetric sources the brightest component is used. The distance
between the optical galaxy centre and the closest possible radio
component is the main criterion for a core identification (see more
details in Paper~I).

There have been several previous VLA observations of subsets of the Palomar sample which, with the final goal of studying the nuclear activity, were successful in detecting radio cores at the centre of the galaxies, at resolution of $\sim$1--2 arcsec \citep[e.g.][]{nagar02,filho00,nagar05}. The VLA core position generally matches with the optical centre of the galaxy within the VLA beam width. This spatial coincidence sets the upper limits of the radius from the optical centre where an active SMBH could be located. 

The previous VLA observations did not resolve the {\it e-}MERLIN cores, but on the contrary {\it e-}MERLIN resolves out part of the extended radio emission visible in VLA maps. This is also because of the longer baselines of the {\it e-}MERLIN array which results in the {\it e-}MERLIN angular resolution being five times better than that of most previous VLA surveys. This higher resolution in turn complicates for identifying the correct position of the radio-emitting SMBH within the VLA beam width around the optical galaxy centre without a more precise localisation, like from VLBI. Therefore, optical and {\it e-}MERLIN astrometries play a crucial role in pinpointing the active SMBH. The {\it e-}MERLIN astrometry is set by the International Celestial Reference Frame to an accuracy a few 10\,mas. We note that the positional uncertainties are therefore now dominated by the positional uncertainty in the optical observations. In fact, absolute optical positions provided by NED are limited by seeing and the light profile across the nuclear region observed by the optical telescopes catalogued in NED. The NED target positions typically refer to the position accuracy of the Two Micron All Sky Survey, i.e. $\sim$0.3 -- 0.5 arcsec\footnote{{\scriptsize https://old.ipac.caltech.edu/2mass/releases/second/doc/sec6\_7f.html}}. \normalfont  Considering the relative optical-radio astrometry, systematic errors and plausible degradation of the {\it e-}MERLIN resolution/astrometry due to low phase reference quality and/or a low signal-to-noise ratio of some sources, we set 1.5 arcsec as a conservative, maximum offset from the optical nucleus to search for a radio core in the {\it e-}MERLIN maps in analogy to Paper~I.

For 66 out of the 78 detected sources in the sample presented in this
work, evident radio cores are identified within 1.2 arcsec from
the optical centre of the galaxy: we refer to them as `identified'
sources (see Sect.~\ref{core-ident}). For 12 of the 78 detected sources we cannot clearly identify the radio source with the nucleus of the galaxy either because the optical-radio separation is $>$2 arcsec or because there are multiple radio components within 1 arcsec of the optical centre. In addition to those, three galaxies which have a detected identified core (NGC~2832, NGC~3077, NGC~4111) reveal additional significantly bright structures in the field, associated with galaxy companions or other sources of ambiguous nature (named 'identified+unidentified'). These 15 (12 galaxies + 3 sources in the field of three identified galaxies) sources have been called 'unidentified' hereafter (see Sect.~\ref{unidentified}). The core-identification tags for the sample are listed in Table~\ref{table1} and are described in details in the next sub-sections.

\subsection{Identified sources}
\label{core-ident}

The full and low-resolution maps of the 66 detected and identified
galaxies are presented in Figure~\ref{ident-maps}, along with the
tables including source characteristics (Table~\ref{tabdet}), radio
contours and restoring beams (Table~\ref{contours}).

To ensure that the `identified' sources are genuine, we calculated the
probability of detecting a radio source above the 0.25-mJy
detection limit of the LeMMINGs survey within a given area of sky. We
use the source count distribution obtained from the 1.4\,GHz {\it e-}MERLIN
legacy programme SuperCLASS \citep{battye20} over an area of $\sim$1
square degree centred on the Abell~981 super-cluster to provide an upper limit on the number of background confusing sources. We find that when observing 177 galaxies, statistically at most one unrelated radio source falls within a circular radius of $\sim$2.6 arcsec of the optical centre. Hence, given this result, a radio sources detected
 within a 1.2 arcsec circular aperture can with a high degree of confidence likely be identified with the central optical nucleus, e.g., with an SMBH.

Radio cores were detected at full resolution for all 66 identified sources with the exception of one, NGC~4369, which is undetected at full resolution but reveals a radio core coincident with the optical galaxy centre in the lower resolution radio images. Most of the sample have
peak core flux densities $\sim$1 mJy beam$^{-1}$ . The brightest source is NGC~1275 which reaches a 10.5 Jy beam$^{-1}$ peak flux density. Most of the central components
can be considered unresolved or compact as the deconvolved source sizes are much
smaller than the beam size. For those sources the peak flux densities of
the radio core components are usually consistent with the integrated
flux densities to within a factor of $\sim$2. Those which have
significantly larger integrated flux densities than their peak flux
densities include sources that are extended or contain multiple components.

For 17 identified sources, clear extended radio structures are
observed, which have been preferentially
interpreted as originated from a compact jet. There are several reasons why compact jets might be preferentially detected: the high spatial resolution (150 mas) and the lack of short-spacings of the {\it e-}MERLIN array, and the use of snapshot imaging of the LeMMINGs program which produces sparse {\em uv-}coverage, cause loss of sensitivity to diffuse low-brightness emission, such as expected by a galaxy disc \citep{brown61,kennicutt83} (see Section~\ref{anysfg} for
discussion). We estimate that, based on the properties of our observations (array configuration and snapshot imaging), {\it e-}MERLIN appears to resolve out up to  75 per cent of the radio structure detected with VLA with 1-arcsec resolution: based on previous VLBI and VLA studies (e.g. \citealt{hummel82,falcke00,nagar05,panessa13}), LINERs are less affected by this issue than Seyferts and H{\sc ii} galaxies, which are generally associated with more extended low-brightness radio emission than LINERs. However, this interpretation does not preclude the possibility that the radio emission arises from  compact radio emission from circumnuclear SF (e.g. \citealt{linden20}) or circumnuclear disc (e.g  \citealt{carilli98}).

The 66 identified sources were divided into five distinct classes,
based on their radio morphology in both the full- and low-resolution
maps (Table~\ref{table1}), in analogy to Paper~I. The five morphologies
are discussed below:
\begin{itemize}
\item{{\it core/core--jet}, marked as A (49 galaxies): these sources show bright unresolved or slightly resolved cores and often show
  a protrusion \citep{conway97}. A few radio components could be aligned in the
  same direction of the possible jet. Some examples include NGC~1161
  and NGC~1275.}
\item{{\it one-sided jet}, marked as B (3 galaxies): the one-sided jets show a clear asymmetric extended jet structure
  with multiple one-directional components emerging at different resolutions, possibly
  due to relativistic beaming of the jet. Some examples include
  NGC~5322.}
\item{{\it triple-source}, marked as C (8 galaxies):
  Triple-sources have three aligned components. These components are interpreted as the radio core and
  adjacent jets/lobes. These sources may appear as twin-symmetric
  jets in the lower resolution images. Some examples include NGC~4036
  and NGC~4589.}
\item{{\it double-lobed}, marked as D (0 galaxies): these sources have two large radio lobes over extended scales in
  either the full or low resolution images (see NGC~5005 from Paper~I). A possible overlap with
  C-type morphologies in case of triple source with two weak
  unresolved radio lobes (see NGC~3348).}
\item{{\it jet+complex}, marked as E (6 galaxies): these sources show a complicated radio morphology with several
  components dispersed around a core. They could hide a possible jet
  interacting with the interstellar medium (ISM) or be an extended star-forming region. Some examples
  include NGC~1186 and NGC~2964.}
\end{itemize}

To further discriminate the radio sample, the radio sources which show
`jet-like' morphologies (e.g. one-sided, two-sided, triple,
double-lobed sources) are hereafter named `jetted' and those without
a clear jet, `non-jetted'. Note that the radio classification can be
equivocal because of the low-brightness of the radio structures.

\subsection{Unidentified sources: background sources, M\,82 and Arp\,299}
\label{unidentified}

Fifteen sources (12 galaxies and 3 radio sources appeared in three core-identified galaxies) are considered as `unidentified'. The online supplementary data include the full and low-resolution maps for these
`unidentified' sources (for radio contours and restoring beam see
Table~\ref{contours}) and their radio source parameters are listed in
Table~\ref{tabsfr}.

Seven objects show off-nuclear radio sources further than 4 arcsec from the optical centre and as far away as $\sim$38 arcsec, but still falling within the extent of the optical galaxy. The morphology varies from multiple compact components, double/triple
sources, to a single unresolved component on a typical scale of a few arcseconds with low flux densities ($<$1 mJy beam$^{-1}$). The nature of this off-nuclear
emission is ambiguous, whether related to star-forming regions or
background AGN. Using the same approach as in
Section~\ref{core-ident}, it is possible to estimate the likelihood of
radio sources falling within 4 arcsec and 38 arcsec radii (the offsets
measured above) and hence consider whether they are likely related to
the nuclear core. The expected number of unrelated sources detected within those radii are 3.8 and 65.2 respectively, suggesting that the off-nuclear sources are potentially background objects. However, we cannot rule out the possibility that those off-nuclear sources nonetheless belong to the galaxy concerned.

Five galaxies show several components in the nuclear region
($<$4 arcsec) and it is therefore difficult to pinpoint unequivocally the
core (NGC~2750, NGC~3034, NGC~3690, NGC4631 and NGC~5012).  The low resolution maps of NGC~2750, NGC4631 and NGC~5012 reveal a few components near the optical
centre. The  cases of NGC~3034 and NGC~3690 deserve more attention and are presented below.

NGC~3034 (M~82) is a dusty star forming galaxy which lacks clear evidence of an active nucleus so far. We detect a mix of SNe, SNRs, and H{\sc ii} regions, which have
been previously identified by (e-)MERLIN (e.g. \citealt{muxlow94,beswick06,fenech08,fenech10,muxlow10,gendre13a,varenius15} and references therein).  Amongst the stellar remnants we find   SN~2008iz which is the brightest source in the LeMMINGs images of M~82 and has a peak luminosity of $\sim$60 mJy at 1.5 GHz (epoch 19 April 2017). This bright radio source has been interpreted as synchrotron emission due to an expanding SN shock which encounters clumpy dense medium \citep{brunthaler10,kimani16}. One of the interesting sources detected in the {\it e-}MERLIN map of M~82 is also 41.95+575, which has a double-lobed structure at VLBI
resolution and has been decreasing by $\sim$8 per cent per year for
the last 5 decades. Hence, 41.95+575 may be a remnant of a
Gamma Ray Burst instead of a conventional SNR \citep{muxlow05}. Several faint sources ($<$ a few sub-mJy) detected in previous radio observations are missing from our map due to low sensitivity and resolution, such as X-ray Binaries (XRB) and, in particular, a transient discovered by \citet{muxlow10} at RA 09h55m52.5083s, Dec. +69$^{\circ}$40$\arcsec$45$\farcs$410 (J2000).

NGC~3690, an interacting system also known as Arp~299 (or Mrk~171), is an interesting case, composed of two galaxies: the eastern, brightest member
NGC~3690A, and the western member NGC~3690B. We use the names
Arp~299-A and Arp~299-B to refer to NGC~3690A and NGC~3690B, respectively,
in agreement with the nomenclature used in Fig.~1 from \citealt{romero11}.
In our {\it e-}MERLIN maps we detected three sources, which correspond to i)
Arp~299-A,
ii) Arp~299-B, and iii) Arp~299-C which could be a satellite galaxy
taking part in the
merger event or a vigorous off-nuclear star-forming region related to
the merging system
\citep{tarchi11}. These sources have been confirmed and largely studied
by previous
continuum and spectroscopic radio, infrared and X-ray observations (see
\citealt{alonso00,perez09,tarchi11,romero11,bondi12,romero14,kankare14,anastasopoulou16} and references therein).  Specifically, Arp299-A was resolved into
several components with VLBI: an extremely prolific SN factory
\citep{neff04,perez09,ulvestad09}, and a LLAGN \citep{perez10}. In
addition, Arp~299-B has been resolved into two main  components: B1,
which includes another SN factory, although less extreme than the one in
Arp-299A \citep{ulvestad09,romero11,sliwa12}, an AGN ( detected in hard X-rays, \citealt{ptak15}) and a bright transient source (Arp299-B AT1;
\citealt{mattila18});  and B2 \citep{neff04,alonso00}, a weaker
component with no reported compact radio sources in it. Subsequent VLBI
and infrared observations of the transient AT1 in the nucleus of Arp~299-B,
showed an increase in luminosity over several years, which was later
identified with a jetted tidal disruption event (TDE;
\citealt{mattila18} and references
therein). We have not resolved the nucleus of Arp~299-B into its compact
components,
but we clearly detect it with a flux density of 14.1 mJy (on 18 April
2017), so it is likely that we have detected the late time radio
emission of AT1 at 1.5 GHz, together with the quiescent emission of its
host Arp~299-B. Two of the several well-known SNe present in this system, SN~2010O and
SN~2010P
\citep{romero11,romero14,kankare14} are not detected in our map with a
flux density upper limit of $<$0.6 mJy beam$^{-1}$. The current and past
observations of NGC~3690 appear to confirm that both starburst and AGN
can co-exist in this merging system, as pointed out by, e.g.,
\citet{perez10} and \citet{romero11}.

In conclusion, it is worth mentioning that the source confusion level is high in NGC~3690 as well as in M~82 (and probably in other sources of our sample) even at the {\it e-}MERLIN resolution, where both stellar and SMBH processes are probably embedded in dusty regions and difficult to disentangle within a few tens of parsec at the centre of a (obscured) galaxy.

Three galaxies (NGC~2832, NGC~3077 and NGC~4111) which have been
core-identified, show additional radio sources in the field. For example the radio map of NGC~2832 reveals the radio core component of its companion NGC~2831. The off-nuclear radio source in NGC~3077 is a potential SNR \citep{rosa05,leonidaki10}.

 \subsection{Undetected galaxies}

Most of the LeMMINGs sample (99/177) have not been detected. The vast majority of our undetected sources were also not detected by previous VLA campaigns \citep{nagar02,nagar05}. However, there are some galaxies for which {\it e-}MERLIN have not detected any core, which instead was detected by the VLA (i.e. NGC~3780) and viceversa (i.e. NGC~6340) in different radio observation frequencies (1 -- 15 GHz) and epochs.

It is important to note that undetected as well as `unidentified' galaxies, may still conceal an AGN. Indeed, radio-emitting active SMBHs may be below the detection limit of the LeMMINGs survey or not identified in
the more complex structures listed above (see for example the cases of M~82 and Arp~299). Radio core variability on a time scale of a few years, episodic accretion onto the BH and nuclear inactivity expected within a typical duty cycle of  $\lesssim$10$^{8}$ years could also account for their current none detection (e.g. \citealt{mundell09,morganti17a,alexander20}). Future {\it e-}MERLIN 5\,GHz
observations of the LeMMINGs targets with higher resolution and sensitivity will be able to possibly pinpoint the core with higher accuracy than L-band data.

\begin{figure}
	\includegraphics[width=0.5\textwidth]{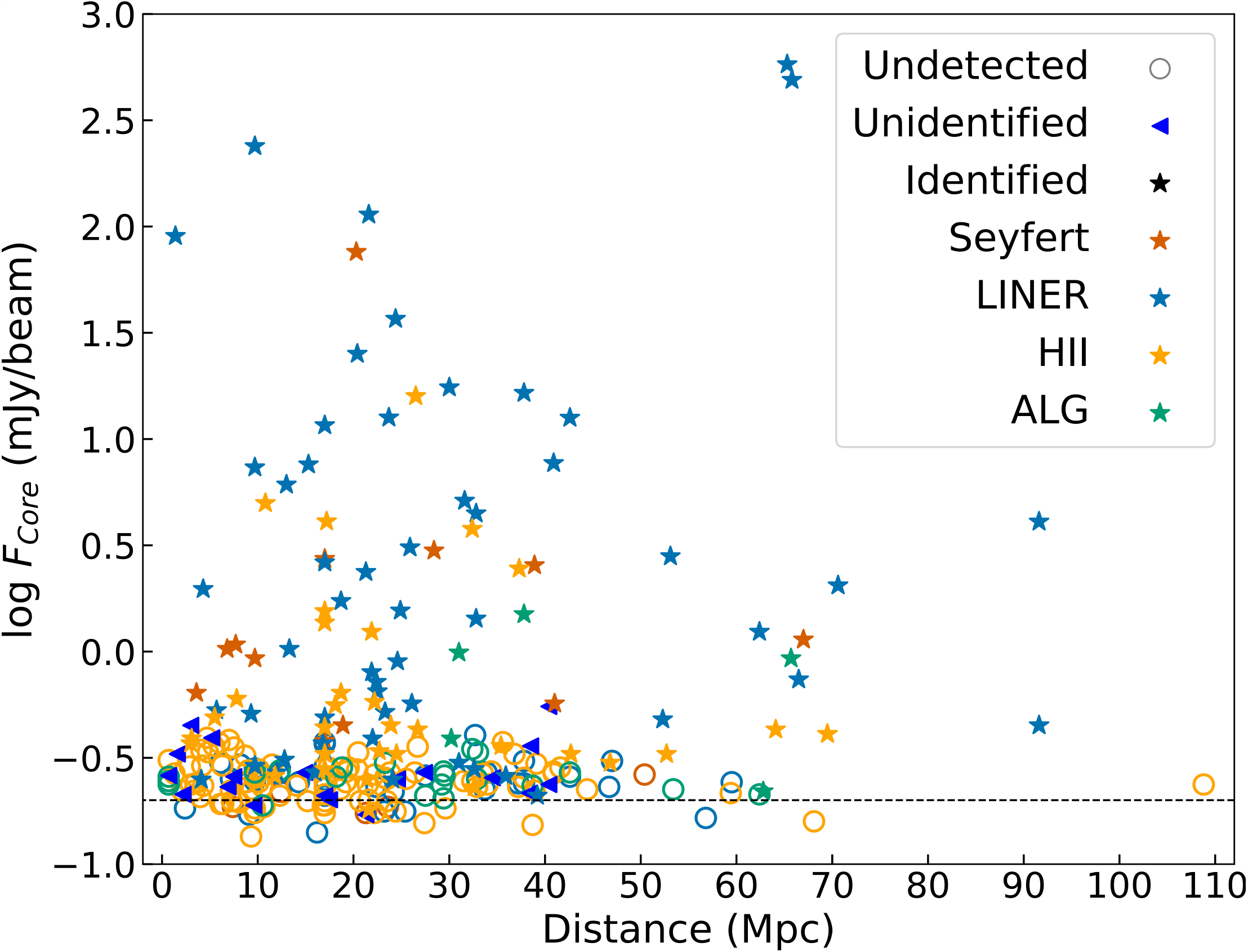}
    \includegraphics[width=0.5\textwidth]{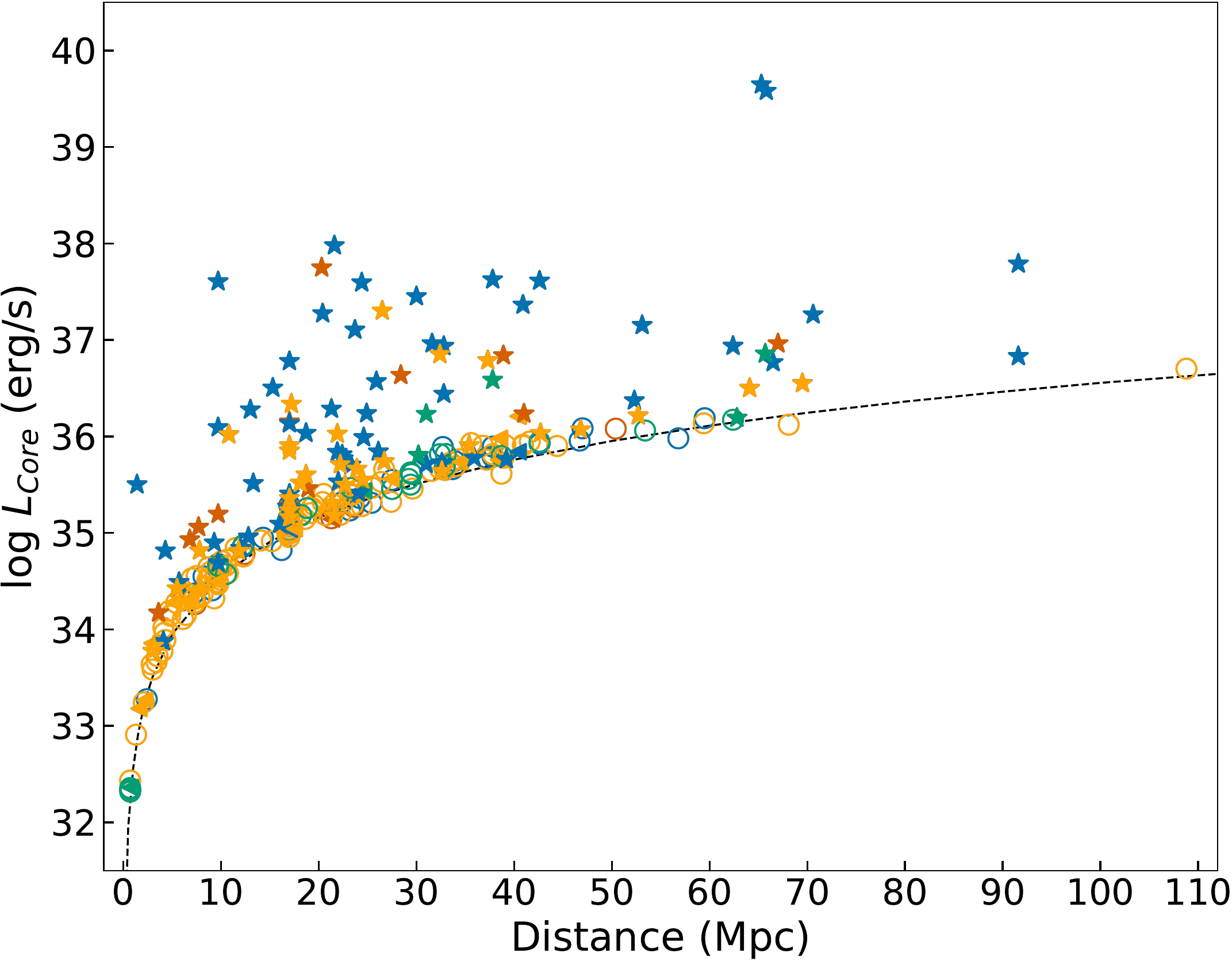}
    \caption[]{Radio core flux density (F$_{\rm core}$
      in mJy beam$^{-1}$, upper panel) and its luminosity (L$_{\rm core}$, integrated over the {\it e-}MERLIN L-band, 1.244 -- 1.756 GHz, in
      erg s$^{-1}$, lower panel) as a function of 
      distance (Mpc) for the full sample.  The dashed lines
      represents the 3$\sigma$ flux density limit of this
      survey (0.25 mJy beam$^{-1}$) and the corresponding luminosity, respectively in the upper and lower panels. The symbols in the legend indicate the status of their detection (open circles for undetected and filled stars for detected) and their optical class (red Seyferts, light blue LINERs, yellow H{\sc ii} galaxies and green ALGs). For the unidentified (blue left-pointing triangles) and
      undetected radio sources,
      the values on the y-axis should be understood as upper limits. The typical error bar on flux densities and luminosities is 20 per cent of the values. To convert the radio luminosities from erg s$^{-1}$ to W Hz$^{-1}$ at 1.5 GHz, an amount +16.18 should be subtracted from the logarithm of the luminosities presented in the graph.}
    \label{distance}
\end{figure}

\section{The complete LeMMINGs Legacy survey}
\label{complete}

In the following we present the radio flux and luminosity distributions, radio morphology and brightness temperatures for the full sample of LeMMINGs galaxies combining the first data release (Paper~I) with the observations presented in this work. The detailed scientific results for the full LeMMINGs sample and for each optical class will be reported in forthcoming papers.

In the entire Legacy survey {\it e-}MERLIN detected radio emission
in 125 targets (typically $<$ a few mJy) at scales of a few tens of
parsecs in a statistically-complete sample of 280 nearby galaxies
($\sim$44.6 per cent)\footnote{NGC~147 was classified as `identified' in Paper~I,
  but a more careful analysis of its HST position indicates that the
  radio detection lies at ($>$ 2 arcsec) from the optical nucleus. Therefore its updated classification
  is `unidentified'.}. The 3$\sigma$ flux density limit of this
survey is $\sim$0.25 mJy beam$^{-1}$ and the range of detected peak flux densities F$_{\rm core}$ covers three order of magnitudes from a fraction of a mJy to a few Jy beam$^{-1}$. The galaxies are detected out to $\sim$100\,Mpc
(Fig.~\ref{distance}, upper panel), but the fraction of radio detections increases with the
distance to the targets, with only $\sim$20 per cent of galaxies nearer than $\sim$20\,Mpc (the median distance of the sample) being detected but $\sim$40 per cent within 50\,Mpc. The core luminosity (F$_{\rm core}$ integrated over the {\it e-}MERLIN L-band, 1.244 -- 1.756 GHz) L$_{\rm core}$ vs.\ distance plot of the LeMMINGs program is presented in Fig.~\ref{distance} (lower
panel). The identified sources lie above the black dotted curve which corresponds to the detection threshold as shown in the top panel of Fig.~\ref{distance}, while the undetected and unidentified sources straddle the curve.

The core luminosities\footnote{The radio luminosities have been presented in units of erg s$^{-1}$. To convert the radio luminosities from erg s$^{-1}$ to monochromatic luminosities (W Hz$^{-1}$) at 1.5 GHz, an amount +16.18 should be subtracted from the logarithm of the luminosities.} range between $\sim$10$^{33.8}$ to 10$^{40}$
erg s$^{-1}$ ($\sim$10$^{17.6}$ - 10$^{23.8}$ W Hz$^{-1}$), or down to
10$^{32}$ erg s$^{-1}$ when including our upper limits for the non-detected cores. A Kaplan-Meier (censored) mean value \citep{kaplan58} of 1.5$\times$10$^{34}$ erg s$^{-1}$ is found. The radio luminosities of the 177 galaxies presented in this work (Table~\ref{table1}) are similar to those of the 103 galaxies from Paper~I. Compared to previous surveys of the Palomar surveys,
the LeMMINGs legacy presents the deepest survey, by extending to lower
luminosities by a factor of at least 10 with respect to previous ones
\citep{nagar02,filho06,panessa13}. It reaches a depth similar to the 15-GHz Palomar survey by \citet{saikia18} but our work is on a larger sample of active and inactive galaxies and at lower frequencies. Our legacy program is sensitive to sources about a factor of 100 times more luminous than Sgr~A* ($\sim$1 Jy, $\sim$10$^{15.5}$ W
Hz$^{-1}$, \citealt{krichbaum98}),  and represents the deepest radio
survey of the local Universe at 1.5 GHz.

\setcounter{table}{4}
\begin{table*}
	\centering
	\caption[Spectral--radio morphological classification breakdown of the LeMMINGs sample]{Spectral--radio morphological classification breakdown of the LeMMINGs sample.}
	\label{fraction}
	\begin{tabular}{cl|cccc|c} 
		\hline
            &                        & \multicolumn{4}{c|}{optical class} \\
\hline
 & radio class                     &   LINER & ALG & Seyfert  &  H{\sc ii}  &  Tot  \\
		\hline

 \parbox[t]{2mm}{\multirow{6}{*}{\rotatebox[origin=c]{90}{ core identified}}} & {\tiny core/core--jet (A)}       &    37 (29)  &  3 (0)  &  6 (6) &  18 (14)  &  64 (49) \\
 & {\tiny one-sided jet (B)}        &    2 (2)    &  0 (0)  &  1 (1) &  2 (0)    &  5 (3)  \\
 & {\tiny triple (C)}               &    13 (4)   &  2 (1)  &  3 (2) &  4 (1)    &  22 (8) \\
 & {\tiny doubled-lobed (D)}        &    3 (0)    &  0 (0)  &  1 (0) &  0 (0)    &  4 (0) \\
 & {\tiny jet+complex (E)}          &    1 (0)    &  0 (0)  &  1 (0) &  9 (6)    &  11 (6) \\
 
  \cmidrule(rl){2-7}
 & Tot {\tiny core-identified}      &   56 (35)  &  5 (1)  &  12 (9) & 33 (21) & 106 (66)   \\
\hline	
 & unidentified                      &    2 (1) &  2 (1)  &  1 (0)  &   14 (10)    &  19 (12)  \\
\hline
 & Tot detected                  &    58(36)   &  7(2)    & 13(9) & 47(31) & 125(78) \\
 & undetected                        &  36 (24)  &  21 (12) & 5 (5)    &  93 (58) & 155 (99) \\      
	\hline
	\hline
 & Tot                               &  94 (60) &  28 (14) & 18 (14) & 140 (89) & 280 (177) \\	
 
\hline
\end{tabular}
\begin{flushleft}
  Notes. The sample is divided into morphological radio (core/core-jet, one-sided jet, triple, double-lobed source, and complex source) and spectroscopic optical classes (LINER, ALG, Seyfert, H{\sc ii} galaxies) based on
  their radio detection, core-identification or non-detection. The
  numbers are related to the total LeMMINGs sample (280 objects),
  whilst the numbers in parenthesis are related to the sub-sample of
  177 galaxies reported here.
\end{flushleft}
\end{table*}

Table~\ref{fraction} summarises the number of core-identified galaxies, core-unidentified sources, and undetected sources, listed by optical
classes. Seyferts have the highest detection rate (13/18, $\sim$72.2 per cent) in the sample, a
more robust number than that measured in the first data release
because of the poor coverage of this class (only 4 Seyferts in
Paper~I). This fraction is similar to the radio detection rate measured in type-1 Seyferts (72 per cent) from previous VLA studies \citep{ho08}. LINERs have the largest number of detections, 58/94,
$\sim$61.7 per cent, a rate that is comparable to what was achieved in
Paper~I and similar to what was obtained in previous VLA studies for type-1 LINERs, 63 per cent, \citep{ho08}. However we note that our radio detection rates relative to type-II AGN are higher than what measured in previous VLA studies \citep{ho08}. H{\sc ii} galaxies have a smaller detection rate, 47/140
($\sim$33.6 per cent) and for ALG the detection rate is even lower, at 7/28 (25 per cent). Overall 106/280, or $\sim$37.9 per cent of the detected sources are  core-identified; 19 detected sources (two ALG,
one Seyfert, two LINERs and fourteen H{\sc ii} galaxies) are
unidentified. The final identified radio core fractions are therefore: LINERs 56/94 ($\sim$59.6 per cent), Seyferts 12/18 ($\sim$66.7 per cent), ALGs 5/28 ($\sim$17.9 per cent) and H{\sc ii} galaxies 33/140 ($\sim$23.6 per cent).

The LeMMINGs sample covers all radio morphological types across
all of the optical classes (see Table~\ref{fraction}). Furthermore,
the radio classes which suggest the presence of a radio jet (B, C and
D), enclose all the optical classes. LINERs exhibit a variety of
morphologies, but are most commonly observed as core/core--jet and
triple structures, similar to the ALGs. Seyferts cover the whole variety of
morphologies, but with the highest fraction of extended jetted structures among the classes. In contrast, the H{\sc ii} galaxies show primarily
compact cores or extended complex structures.

Although H{\sc ii} galaxies are classified as star forming based on their location on the BPT diagrams, the presence of
a jet is not precluded in these sources. In fact, there are seven LeMMINGs H{\sc ii} galaxies that show clear 'jet-like' morphologies, we refer to them as {\it jetted} H{\sc ii} galaxies\footnote{The {\it jetted}
  H{\sc ii} galaxies are NGC~972, NGC~3665, UGC~3828, NGC~7798,
  UGC~4028, NGC~2782, and NGC~3504.}. One of these sources is NGC~3665
which exhibits a Fanaroff-Riley type-I (FR~I) radio morphology extended over $\sim$3 kpc at
the VLA scale \citep{parma86}. NGC~3504 appears extended in VLBI maps
with a core of 3 mJy beam$^{-1}$ \citep{deller14}. NGC~7798 shows only
a bright core ($\sim$6 mJy) at 8.5 GHz with the VLA
\citep{schmitt06}. NGC~2782 shows an extended radio source which matches
a previous {\it e-}MERLIN observation with an elongated core (a peak flux
density of 1.4 mJy beam$^{-1}$, lower than our detected value 2.5 mJy
beam$^{-1}$), resolved in a twin-jet morphology by EVN observations
(peak flux density of 0.4 mJy beam$^{-1}$, \citealt{krips07}). The
radio emission for the remaining {\it non-jetted} H{\sc ii} galaxies is more probably related to SF instead of a jet (single cores or complex
morphologies).

The brightness temperature (T$_{B}$) is the temperature needed for a black-body (thermal) radiator to produce the same specific intensity as the observed point source. In astrophysical phenomena, below 10$^5$ K, the radio emission can be explained by a large contribution from free-free emission \citep{condon91} but above 10$^6$ K synchrotron emission from relativistic particles
(e.g. from jets or AGN) is required to explain such high brightness temperatures \citep{condon92}. Using half the beam width at the {\it e-}MERLIN resolution at 1.5 GHz as the deconvolved size of the cores (see Table~\ref{table1}), a flux density of $\gtrsim$5 mJy beam$^{-1}$ corresponds to a T$_{B}$ $\gtrsim$10$^{6}$ K (Paper~I). Twenty-one LeMMINGs galaxies meet this requirement and are
associated with all types of radio morphologies observed and are for
the majority LINERs.  The brightness temperatures broadly reflect the
flux density distribution. However, T$_{B}$ lower than
10$^6$K do not preclude an SMBH origin of the radio emission, as
 weak LLAGN emit low brightness. VLBI observations are required to give a robust measurement of cores' brightness temperatures.

\begin{figure}
%\vspace{-1cm}
	\includegraphics[width=0.43\textwidth,height=0.56\textwidth]{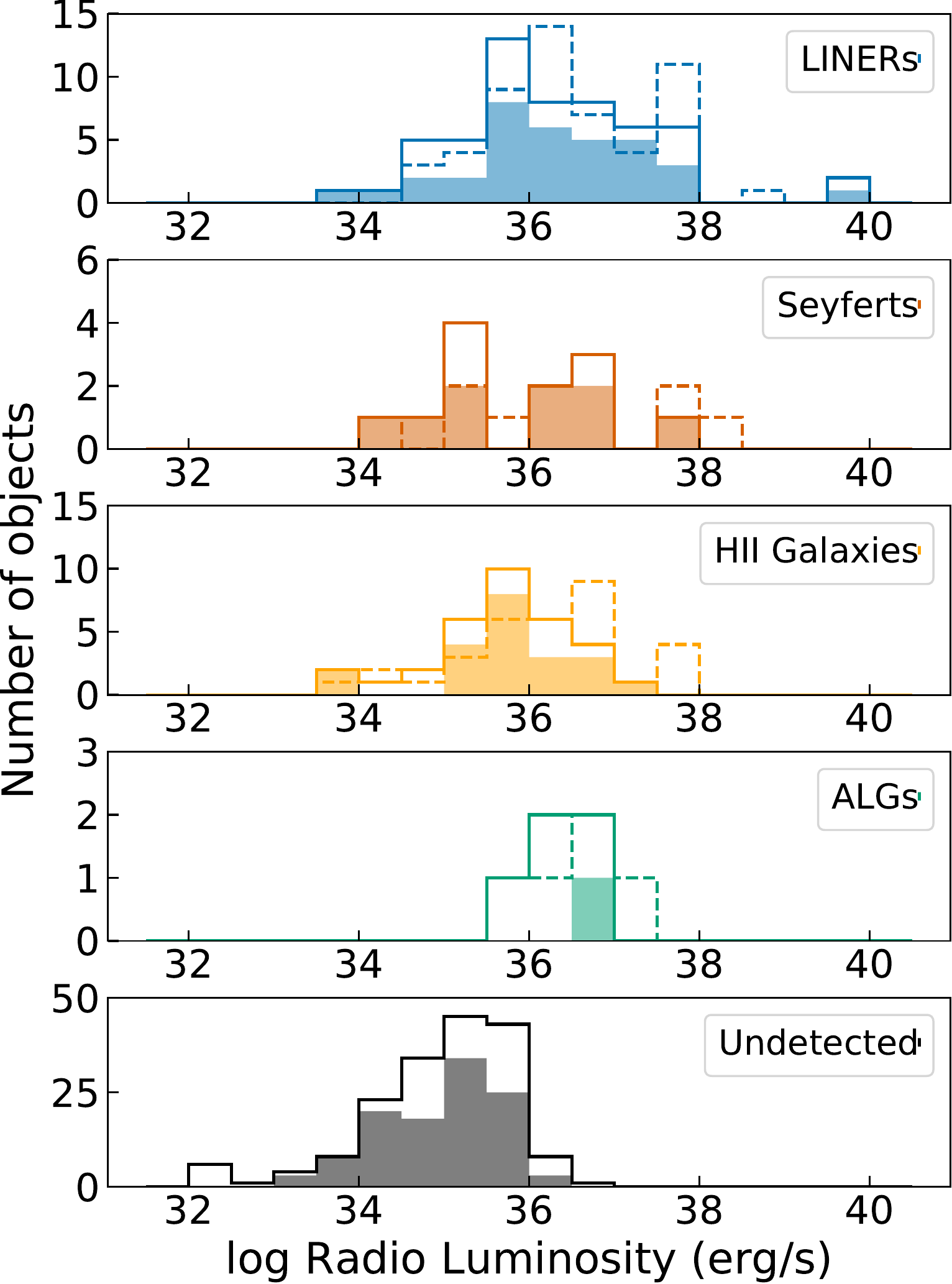}
	%\vspace{-1.2cm}	
	\includegraphics[width=0.43\textwidth,height=0.56\textwidth]{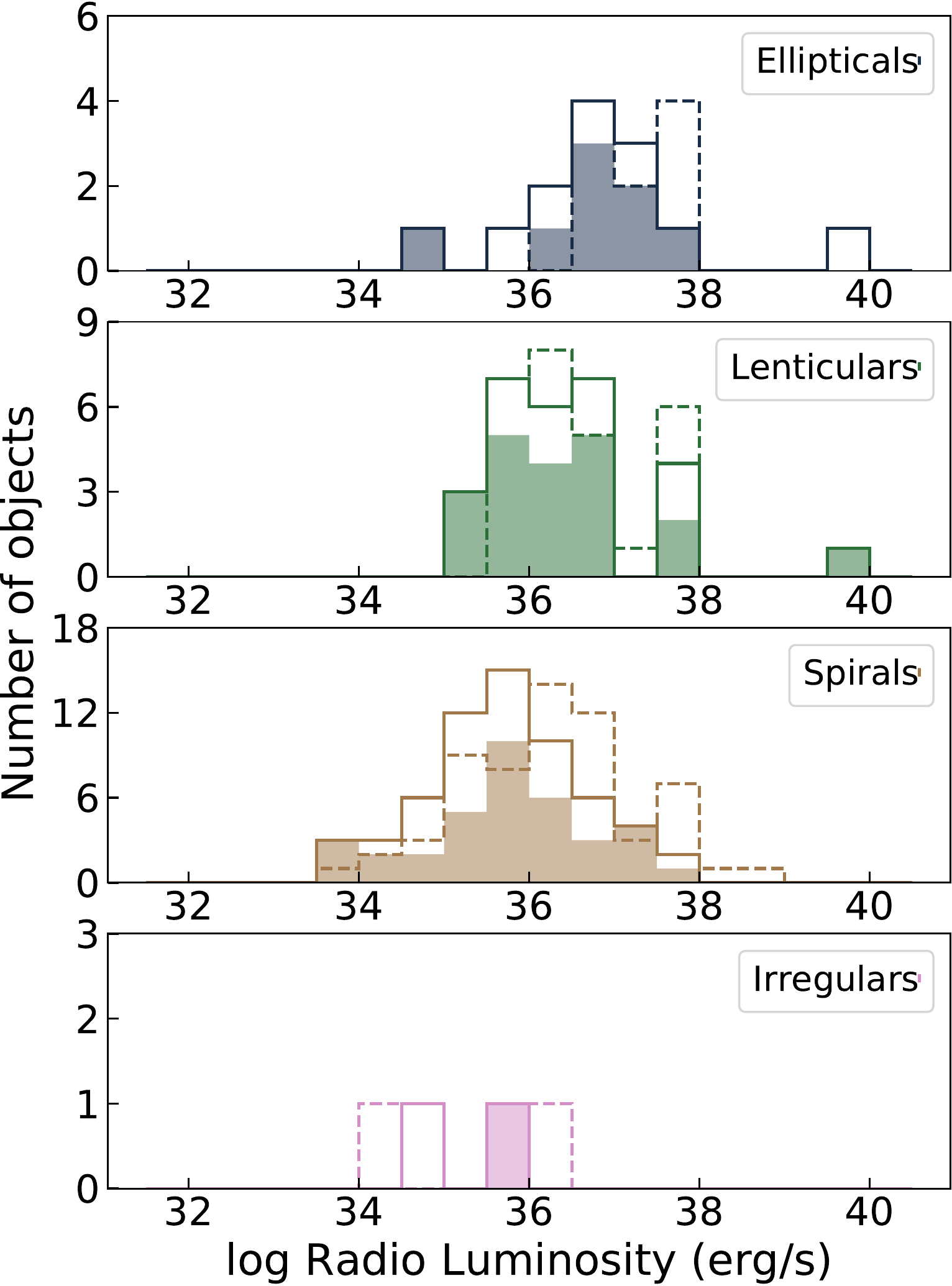}
%	\vspace{-0.8cm}	
        \caption[]{Radio
          luminosity distribution (erg s$^{-1}$) divided per optical class (upper figure) and
          host morphological type (lower figure). The radio core luminosity
          and the total radio luminosity distributions are outlined by the
          solid-line and the dashed-line histograms, respectively. The bottom panel of the upper figure depicts the $3\,\sigma$ upper-limit radio luminosity distribution for the undetected and unidentified sources. The filled histogram represents the 177 sources presented in this work.  To convert the radio luminosities from erg s$^{-1}$ to W Hz$^{-1}$ at 1.5 GHz, an amount +16.18 should be subtracted from the  logarithm of the luminosities presented in the graph.}
    \label{histo}
\end{figure}

The luminosity distributions for the LeMMINGs sample are presented in
Figure~\ref{histo}: both the peak radio core luminosities and the
total luminosities (L$_{\rm Tot}$) integrated over the radio-emitting region, split by optical class (upper panel) and host type (lower panel). The detected
cores have a mean (uncensored) luminosity of 10$^{36.10 \pm 0.11}$ erg s$^{-1}$. In
general, LINERs are among the most
luminous sources of the sample, with a censored mean value of
1.8$\times$10$^{35 \pm 0.24}$ erg s$^{-1}$. Seyferts show the largest
mean core luminosity (2.6$\times$10$^{35 \pm 0.26}$ erg s$^{-1}$). In
contrast, the ALG and H{\sc ii} galaxies have the lowest censored mean core
powers, at 5.0$\times$10$^{34 \pm 0.30}$ and 3.1$\times$10$^{34 \pm
  0.17}$ erg s$^{-1}$, respectively. Yet, when detected ALGs show the highest core luminosities, $>$10$^{36}$ erg s$^{-1}$.  The undetected/unidentified
galaxies have upper-limit radio luminosities between 10$^{32}$ to
10$^{37}$ erg s$^{-1}$, with a median radio luminosity of
9.06$\times$10$^{34 \pm 0.07}$ erg s$^{-1}$. The different median luminosities of the optical classes are not due to different sensitivities or distances because the sources are randomly distributed in the space volume of the survey and observed in similar conditions.

 In analogy to the core luminosities, the total luminosities estimated from the low-resolution radio images for the 177 galaxies described in this work (Table~\ref{table1}) are comparable with the values obtained from the first sub-sample from Paper~I (Fig.~\ref{histo}):  the median value is 4$\times$10$^{36}$ erg s$^{-1}$. The total luminosities equal the core luminosities in case of an unresolved core or can be larger up to a factor 100 in case of bright extended emission (see Table~\ref{table1}).  LINERs and Seyferts are once again the most powerful radio sources and the H{\sc ii} galaxies are the weakest. The core dominance, defined as the ratio of the radio core
 power to the total flux density changes  with each class, but has a large variance: LINERs and ALGs are the most core dominated
 ($\sim$75 per cent), followed by Seyferts with moderate core dominance
 ($\sim$40 per cent), whereas the H{\sc ii} galaxies have the smallest core
 dominance ($\sim$35 per cent) (Figure~\ref{histo}, upper panel). On average the radio core 
 contributes half of the total radio emission.

\setcounter{table}{5}
\begin{table}
	\centering
	\caption[Spectral--host classification breakdown of the LeMMINGs sample]{Spectral--host-radio classification breakdown of the LeMMINGs sample.}
	\label{hostfraction}
	\begin{tabular}{l|cccc|c} 
		\hline
            &                        & \multicolumn{4}{c}{optical class} \\
\hline
 host type                     &   LINER & ALG & Seyfert  &  H{\sc ii}  &  Tot  \\
		\hline

  {elliptical} &    14 (11)  &  13 (3)  &  0 (0) &  0 (0)  &  27 (14) \\
 {lenticular}        &    29 (21)    &  14 (2)  &  6 (3)    &  6 (2)  & 55 (28) \\
   \cmidrule(rl){2-6}
  Tot ETG                              &   43 (32)  &  27 (5)  &  6 (3) &  6 (2) & 82 (42)   \\
\hline	
 {spirals}  &    50 (23)   &  1 (0)  &  12 (9) &  122 (29)    &  185 (61) \\
  {irregular}          &    1 (1)    &  0 (0)  &  0 (0) &  12 (2)    &  13 (3) \\
  \cmidrule(rl){2-6}
  Tot  LTG                 &   51 (24)  &  1 (0)  &  12 (9) & 134 (31) &  198 (64)   \\
\hline	
	\hline
 Tot                               &  94 (56) &  28 (5) & 18 (12) & 140 (33) & 280 (106) \\	
 
\hline
\end{tabular}
\begin{flushleft}
  Notes. The sample is divided into galaxy types (elliptical and lenticular [ETG], spiral and irregular [LTG]) and spectroscopic optical classes (LINER, ALG, Seyfert, H{\sc ii} galaxies) based on their radio detection. The numbers in parenthesis are related to the sub-sample of
  106 core-identified galaxies.
\end{flushleft}
\end{table}

In terms of the host morphological classes (see Table~\ref{hostfraction}), although spiral galaxies are the
most abundant host type in the LeMMINGs sample (185/280, $\sim$66 per cent), approximately 33 per cent of spiral galaxies are core-detected and are associated with all types of radio
morphology. Detected spirals host Seyfert, LINERs and H{\sc ii} galaxies in decreasing order of radio fraction. Of the 13 irregular galaxies in our sample, three are core-detected in our radio survey and host LINER or H{\sc ii} galaxies. Conversely, ellipticals and lenticulars are the most detected radio sources ($\sim$51 per cent) and are usually associated with the
jetted radio morphologies. LINERs in ETGs have the highest core detection fraction. In summary, approximately one third of LTGs (spiral and irregular) and half of the ETGs (elliptical and lenticular) are detected in our survey. In terms of radio luminosities, Figure~\ref{histo}, lower panel,
depicts the radio luminosity distribution per host type. Ellipticals
and lenticulars show the highest radio luminosities
($\sim$10$^{35-40}$ erg s$^{-1}$). Spiral galaxies have intermediate radio
luminosities of $\sim$10$^{36}$ erg s$^{-1}$.  The irregular galaxies
 have the lowest core and total luminosities.

\section{Discussion}
\label{overview}

\subsection{CASA versus AIPS}

In this work we present the {\it e-}MERLIN observations of 177 Palomar
galaxies from the LeMMINGs survey. The data were calibrated with
\textsc{CASA} which is different from the 103 sources presented
in Paper~I which were analysed with \textsc{AIPS}. The rms noise distribution of the resulting maps presented here is not statistically different from
the values obtained with \textsc{AIPS}. However, for a single observation
block calibrated with both software packages, the rms noise from \textsc{CASA} is
lower than that from \textsc{AIPS} whereas the integrated flux
densities obtained from the maps are consistent with each other to within a 20 per cent calibration error. Moreover, we note marginal differences in terms of low-brightness structures between the maps created with \textsc{AIPS} and \textsc{CASA} procedures, with the latter exhibiting neater radio structures than the former. This implies that caution is required when 
interpreting faint radio sources.

A potential difference in rms noise between the
two calibration techniques could skew the 3$\sigma$ detection limit so as to
have more galaxies detected in the \textsc{CASA} sample. However, this
can not be reconciled with the slightly higher detection fraction of galaxies mapped with \textsc{AIPS} (45.6 per cent) than in this work where we used \textsc{CASA}
(44.6 per cent). However, if one considers these detection statistics as a
binomial distribution, the  detection fractions of the two sub-samples presented in Paper~I and here are consistent with one another within $\lesssim$1 per cent. In conclusion, the two calibration methods produce generally consistent flux densities and radio structures, which minimise a possible bias in the results.

\subsection{General characteristics of the survey}

The LeMMINGs survey stands for the deepest radio study of the local
Universe represented by the Palomar galaxy sample, reaching an average rms noise of $\sim$0.8 mJy beam$^{-1}$. More importantly, the LeMMINGs
sample probes pc-scale activity in all types of galaxies, irrespective of the
nuclear properties set by their optical emission line ratios. This
characteristic makes the LeMMINGs survey unbiased towards the presence
of an active SMBH and different from previous programs which partially
focused only on Seyferts and LINERs.

Of the complete sample of 280 Palomar galaxies, 125
sources ($\sim$44.6 per cent) were detected. This corresponds to a detection rate similar to previous VLA/VLBA
campaigns of the Palomar active galaxies (see \citealt{ho08} for a
review). For 106/280 ($\sim$37.9 per cent) of the sample, radio emission has been detected within the central 1.2 arcsec from the galaxies' optical centre, possibly due to a radio-emitting active nucleus.

Direct comparison with previous radio studies is non-trivial since it was decided to reclassify all galaxies based on the updated BPT
diagrams. Nonetheless, more than half of the LINERs and Seyferts, i.e. of the line-emitting active galaxies, in our sample show compact radio core or jetted structures. Previous VLA campaigns targeting  active galaxies
detected radio emission in half of them
(e.g. \citealt{nagar00}), with a more prominent detection fraction for
Seyferts and type-I AGN in general. One quarter of the detected sources
which are not identified as powered by active SMBH by their
emission-line ratios show evident jetted structures (one-side, twin
jets, triple and double sources), excluding those with complex
morphology. This corroborates the idea that at least one fourth ($\sim$28.6 per cent) of the Palomar galaxies emit pc-scale radio
emission possibly related to a LLAGN. If we include those sources with a intricate radio morphology which could hide jet emission, and also the core-jet H{\sc ii} galaxies, the possible fraction of radio-emitting
LLAGN rises to one third ($\sim$32.5 per cent).

The most common radio morphology observed in the LeMMINGs galaxies with centrally detected radio emissions is the 
core/core--jet morphology (64/106). This is unsurprising as the
high-resolution of {\it e-}MERLIN can resolve out some of the diffuse
emission associated with jets and star-forming regions. The nature of these unresolved radio components, whether
an unresolved radio jet base or a star-forming nuclear region on scales of
$<$100 pc, can be explored by using further diagnostics such as
[O~III] and X-ray luminosities when clear jet-like
structures lack. We will discuss the radio core origin per optical class in upcoming papers.

Half of the early-type galaxies (elliptical and lenticular) in the LeMMINGs sample
host a radio source coincident with the optical centre, that is
determined to be an active SMBH. Regarding late-type galaxies
(spirals and irregulars) $\sim$32 per cent  has a radio-emitting central active SMBH. Quantitatively, the detection fraction decreases with later Hubble type, as only $\sim$17 per cent of Sc--Sd galaxies are
detected. Hence, the LeMMINGs survey agrees with the basic results of previous radio studies of the nearby galaxies
\citep{ho01a,ulvestad01a,nagar05,ho08}, which detected flat-spectrum
radio cores predominantly at the centre of massive (typically early-type) galaxies
\citep{sadler89,wrobel91b,capetti09,miller09,nyland16}.

\subsection{Radio evidence of nuclear star formation}
\label{anysfg}

In our Legacy survey only a small number of galaxies
show evidence of intense bursts of SF, based on the radio properties. The clearest examples are the well-known galaxies/systems M~82 and Arp~299. Only
three sources (NGC~4013, NGC~4102, and NGC~5273) clearly exhibit
diffuse low-brightness radio emission, consistent with a stellar disc
or ring, similar to what is expected from diffuse extended star forming galaxies \citep{muxlow10,murphy18,herrero17}. Single radio components were detected at the centre of most of H{\sc ii}
galaxies. We note that complex radio morphologies, unidentified off-nuclear sources
and non-jetted galaxies might conceal nuclear star-forming regions, though. Furthermore,  ambiguous morphologies (see e.g. NGC~2964, NGC~2273, NGC~2639) might be caused by the interaction of a jet with a dense ISM, which in turn could trigger SF \citep{silk05,gaibler12}.

The very low fraction of clear star forming regions observed in our
survey is most probably due to the sparse $u-v$ coverage of the observations because of the long baselines of {\it e-}MERLIN and of the snapshot imaging technique of our program. For most of the sample, which is 
further than $\sim$4 Mpc, the
spatial frequencies covered by {\it e-}MERLIN are appropriate to detect compact bright young SN/SNR and H{\sc ii} regions ($<$ 400--500 yr, \citealt{westcott17}) and are not suited for detecting diffuse,
low-brightness radio emission (T$_{B}$ $<$10$^{5}$ K), typical of old
SNR. At distances $\lesssim$4 Mpc, VLA data are required to study long-lived diffuse H{\sc ii} complexes, which instead would be resolved out with {\it e-}MERLIN. We have estimated that our {\it e-}MERLIN observations can lose up to 75 per cent of the radio structures detected with VLA with 1-arcsec resolution. In fact, adding shorter spacing (VLA) data to {\it e-}MERLIN visibilities will thus increase the ability to detect
more diffuse lower surface brightness emission from SF products. Dedicated deep {\it e-}MERLIN radio observations combined with shorter-baseline datasets of compact star forming regions, similar to that performed on M~82
\citep{wills99,muxlow10} has the potential to diagnose the nature of the radio emission in H{\sc ii} galaxies. This will be the goal of one of our future works.

\vspace{-0.4cm}
\section{Summary and Conclusions}

This  paper present  the  second data release from the {\it e-}MERLIN legacy survey, LeMMINGs, aimed at studying a sample of nearby 280 (active and quiescent) galaxies. Here we show the observations of 177 sources from the Palomar sample \citep{ho97a} at 1.5 GHz. By combining this release with the first one \citep{baldi18lem}, the complete survey represents the deepest, least unbiased view of the galactic nuclei of the local Universe ($<$110 Mpc) in the radio band, reaching a sensitivity of $\sim$0.80 mJy beam$^{-1}$ and an angular resolution of $\sim$150 mas, which corresponds to a physical scale of $\lesssim$100 pc. This program revealed a large population of local radio-emitting LLAGN and nuclear starbursts ($\lesssim$10$^{17.6}$ W Hz$^{-1}$).

After updating the optical spectroscopic classifications of the 280 galaxies of the survey, the entire sample consists of 94 LINERs, 18 Seyferts, 140 H{\sc ii} galaxies and 26 ALG.

Our radio survey has detected significant radio emission with flux densities $\gtrsim$0.25 mJy in the innermost region (0.73 arcmin$^{2}$) of 125 galaxies  (44.6 per cent): 58/94 LINERs, 16/18 Seyferts, 47/140 H{\sc ii} 
galaxies and 7/28  ALGs. For 106 of the 125 detected sources we identified the core within the radio structure, spatially associated with the optical galaxy centre. We resolved parsec-scale radio structures with a broad variety of morphologies: core/core–jet, one-sided jet, triple sources, twin jets, double-lobed, and complex shapes with sizes of 3 -- 6600 pc. The compact cores (64/106) are the most common morphology. There are 31 sources with clear jets, roughly half (18/31) of which are LINERs. This jet fraction could be higher because the complex morphologies (11/106) could possibly hide diffuse jets interacting with the ISM, similar to what is seen in nearby LLAGN and star forming galaxies (e.g. \citealt{mould00,croft06,gaibler12}).

The detected radio cores have been interpreted as a sign of nuclear activity and their luminosities range between $\sim$10$^{34}$ and 10$^{40}$ erg s$^{-1}$. The lower end of this interval explicits the depth of this survey, greater than that reached by previous radio surveys, $\sim$10$^{35}$ erg s$^{-1}$ \citep{nagar05,filho06}. 
The total radio luminosities determined by integrating the extended radio structures 
 are on average double the core radio luminosities, although they can be up to a factor of 100 times the core luminosity  for jetted sources and those with complex morphologies.

Concerning the host type, approximately half of the early-type galaxies and one third of the late-type galaxies are detected in our survey. The jetted sources are typically related to elliptical, lenticulars or bulged-dominated spirals.

Based only on the radio properties (brightness temperatures, luminosities, morphologies) and spectroscopic classification, the origin of the radio emission from the LeMMINGs galaxies is probably ascribed to active SMBHs in one third of the sample, precisely in the generic population of LINER and Seyferts.  Conversely, SF is the most plausible physical process of radio-emission production in H{\sc ii} galaxies. For ALG the nature of the radio emission is more controversial, but the lack of clear SF favours an AGN origin. Nonetheless, adding multi-band data to the radio analysis will better address the question on the nature of the radio emission in each single galaxy and will be subject of upcoming papers.

{\it LINERs} reveal narrow structures of rapidly declining brightness at increasing distance from the nucleus, i.e. core-brightened morphology, similar to small FR~I radio galaxies. They have the highest brightness temperatures (some $>$10$^{6}$ K) and are among the most luminous galaxies, suggesting a synchrotron emission from a (mildly?) relativistic jet. They tend to live in ellipticals and lenticulars, another analogy with classical radio-loud AGN \citep{heckman14}.

{\it Seyferts} exhibit the highest fraction of detections and double-lobed radio outflows, echoing the 'edge-brightened' morphology observed in nearby radio-quiet Seyferts (e.g. \citealt{kukula95,wrobel00}). Along with LINERs, they are among the most luminous sources of the sample and are found in both galaxy types but more frequently in late types. Similar to the conclusion in Paper~I regarding Seyferts, their symmetric (two-sided) radio morphology and their association with spirals recall the `spin paradigm' \citep{wilson95,sikora07,tchekhovskoy10,dotti13} which suggests that SMBH in spiral/disc galaxies may host (on average) lower-spinning SMBHs than those in giant elliptical galaxies. This argument has been interpreted as one of the possible conditions which prevents from launching faster jets in late-type galaxies than in early-type galaxies, although largely under debate.

We typically detected the cores of {\it H{\sc ii} galaxies}, with brightness temperature $<$10$^{6}$ K and with sub-kpc sizes, probably
representing nuclear starburst as similar to local star-forming galaxies (e.g. \citealt{herrero17}). Although this class encompasses the least luminous objects, a small sub-group of seven H{\sc ii} galaxies is associated with core-brightened jetted structures similar to jets seen in LINERs. This association suggests the presence of an active SMBH, optically outshined by the nuclear SF, but able to
support the launch of a jet. These star forming galaxies with active nuclei are still consistent with the picture of LLAGN \citep{nagar01,ulvestad01b}. In addition, H{\sc ii} galaxies have the highest fraction of complex morphologies and multiple components (see M~82 and Arp~299), plausibly related to diffuse SF and SN factories.

Only 7 out of 28 {\it ALGs} have been detected and only 2 reveal clear jets. They are typically associated with massive ellipticals and when detected, they are the most luminous sources. Their radio and host properties are similar to those of the LINER population \citep{baldi10b}. The absence of a clear emission-line nucleus and their low radio activity chime with a picture of a population of evolved galaxies with dormant SMBHs which occasionally trigger AGN activity \citep{morganti17a}.

The nuclear components revealed by our {\it e-}MERLIN survey suggest that the detected pc-scale radio cores, which unlike in previous radio surveys could now be resolved, represent the brightest and main parts of the entire galaxy. In one third of the nuclei the emission can plausibly be ascribed to a central, active SMBH in a low-accretion stage and/or to a disc emitting at low-radiative efficiency, as expected in LLAGN able to launch pc-scale jets \citep{ho08,mezcua14}. However, it is clear that sub-mJy radio cores can conceal both strong SF and an active, low-brightness SMBH \citep{padovani16}, even in a flaring or dimming stage of accretion, such as in the case of a TDE, as observed in our target NGC~3690 (Arp~299, \citealt{mattila18}). Therefore, by eventually disentangling SF and SMBH activity and assessing the origin of the radio emission at the centre of our galaxies, the next LeMMINGs papers will make use of the optical and X-ray data along with our radio observations to address the following astrophysical open issues: the disc-jet connection in LLAGN \citep{merloni03}, the contribution from SF (stellar processes and XRBs) in the GHz band and possible core variability due to transient phenomena \citep{mundell09,alexander20}.

\section*{Acknowledgements}

The authors thank the anonymous referee for his/her helpful comments to improve the manuscript. AA and MAPT acknowledge support from the  Spanish MCIU through grant
PGC2018-098915-B-C21 and from the State Agency for Research of the
Spanish MCIU through the “Center of Excellence Severo Ochoa” award
for the Instituto de Astrofísica de Andalucía (SEV-2017-0709).  B.T.D acknowledges support from a Spanish postdoctoral fellowship `Ayudas 1265 para la atracci\'on del talento investigador. Modalidad 2: j\'ovenes investigadores.' funded by Comunidad de Madrid under grant number 2016-T2/TIC-2039. B.T.D also acknowledges support from  grant `Ayudas para la realizaci\'on de proyectos de I+D para jóvenes doctores 2019.' funded by Comunidad de Madrid and Universidad Complutense  de Madrid under grant number PR65/19-22417. J.H.K. acknowledges financial support from the European Union's Horizon
2020 research and innovation programme under Marie Sk\l odowska-Curie
grant agreement No 721463 to the SUNDIAL ITN network, from the State
Research Agency (AEI-MCINN) of the Spanish Ministry of Science and
Innovation under the grant "The structure and evolution of galaxies and their central regions" with reference
PID2019-105602GB-I00/10.13039/501100011033, and from IAC project
P/300724, financed by the Ministry of Science and Innovation, through
the State Budget and by the Canary Islands Department of Economy,
Knowledge and Employment, through the Regional Budget of the Autonomous Community.JSG thanks the University of Wisconsin-Madison and its Foundation for support of this research through his Rupple Bascom Professorship. FS acknowledges partial support from a Leverhulme Trust Research fellowship. CGM acknowledges support from the University of Bath and Jim and Hiroko Sherwin. {\it e-}MERLIN is a National Facility operated by the University of Manchester at Jodrell Bank Observatory on behalf of STFC, part of UK Research and Innovation.

\section*{DATA AVAILABILITY}

The data on which this paper is based are publicly available from the {\it e-}MERLIN archive. Calibrated image products are available upon reasonable request to the corresponding author. These, along with other LeMMINGs survey products, will be publicly hosted in association with upcoming publications.

%%%%%%%%%%%%%%%%%%%%%%%%%%%%%%%%%%%%%%%%%%%%%%%%%%



\bibliographystyle{mn2e}
\bibliography{my} % if your bibtex file is called example.bib

%%%%%%%%%%%%%%%%%%%%%%%%%%%%%%%%%%%%%%%%%%%%%%%%%%

\section*{Affiliations}
$^{1}$ Istituto di Radioastronomia - INAF, Via P. Gobetti 101, I-40129 Bologna, Italy\\
$^{2}$ School of Physics and Astronomy, University of Southampton, Southampton, SO17 1BJ, UK\\
$^{3}$ Dipartimento di Fisica, Universit\'a degli Studi d Torino, via Pietro Giuria 1, 10125 Torino, Italy\\
$^{4}$ INAF - Istituto di Astrofisica e Planetologia Spaziali, via Fosso del Cavaliere 100, I-00133 Roma, Italy\\
$^{5}$ Jodrell Bank Centre for Astrophysics, School of Physics and Astronomy, The University of Manchester, Manchester, M13 9PL, UK\\
$^{6}$ Department of Physics, University of Oxford, Denys Wilkinson Building, Keble Road, Oxford, OX1 3RH, UK \\
$^{7}$ Centre for Astrophysics Research, University of Hertfordshire, College Lane, Hatfield, AL10 9AB, UK\\
$^{8}$ Departamento de Astrofisica y Ciencias de la Atmosfera, Universidad Complutense de Madrid, E-28040 Madrid, Spain\\
$^{9}$ Instituto de Astrofisica de Canarias, Via Lactea S/N, E-38205, La Laguna, Tenerife, Spain\\
$^{10}$ Departamento de Astrofisica, Universidad de La Laguna, E-38206, La Laguna, Tenerife, Spain\\
$^{11}$ Jeremiah Horrocks Institute, University of Central Lancashire, Preston PR1 2HE, UK\\ 
$^{12}$ Department of Space, Earth and Environment, Chalmers University of Technology, Onsala Space Observatory, 43992 Onsala, Sweden\\
$^{13}$ Instituto de Astrofisica de Andaluc\'ia (IAA, CSIC), Glorieta de la Astronom\'ia s/n, 18008-Granada, Spain\\
$^{14}$ Netherlands Institute for Radio Astronomy, ASTRON, Dwingeloo, The Netherlands\\
$^{15}$ UK ALMA Regional Centre Node, Jodrell Bank Centre for Astrophysics\\
$^{16}$ AIM/CEA Paris-Saclay, Universit\'e de Paris, CNRS, F-91191 Gif-sur-Yvette, France\\
$^{17}$ Station de Radioastronomie de Nan\c{c}ay, Observatoire de Paris, PSL Research University, CNRS, Univ. Orl\'{e}ans, 18330 Nan\c{c}ay, France\\
$^{18}$ Department of Physics \& Astronomy, University College London, Gower Street, London WC1E 6BT, UK\\
$^{19}$ Department of Astronomy, University of Wisconsin-Madison, Madison, Wisconsin, USA\\
$^{20}$ Astrophysics Group, Cavendish Laboratory, 19 J.~J.~Thomson Avenue, Cambridge CB3 0HE, UK\\
$^{21}$ Steward Observatory, University of Arizona, Tucson, AZ  85721-0065, USA\\
$^{22}$ George P. and Cynthia W. Mitchell Institute for Fundamental Physics \& Astronomy, Texas A\&M University, College Station, TX USA\\
$^{23}$ Max-Planck-Institut f\''{u}r Radioastronomie, Auf dem H\''{u}gel 69, 53121 Bonn, Germany \\
$^{24}$ Department of Astrophysics/IMAPP, Radboud University, P.O. Box 9010, 6500 GL Nijmegen, The Netherlands\\
$^{25}$ Department of Physics, Box 41051, Science Building, Texas Tech University, Lubbock, TX 79409-1051, US\\
$^{26}$ Department of Physics, University of Bath, Claverton Down, Bath, BA2 7AY, UK\\
$^{27}$ International Gemini Observatory/NSF's NOIRLab, 670 N. A'ohoku Pl, Hilo, HI 96720 USA\\
$^{28}$ Institute of Astronomy and Astrophysics, Academia Sinica, 11F of Astronomy-Mathematics Building,AS/NTU No. 1, Sec. 4, Roosevelt Rd, Taipei 10617, Taiwan, R.O.C\\
$^{29}$ Center for Astro, Particle and Planetary Physics, New York University Abu Dhabi, PO Box 129188, Abu Dhabi, UAE\\
$^{30}$ School of Physics and Astronomy, University of Birmingham, Edgbaston, Birmingham B15 2TT, UK\\
$^{31}$ Centre for Extragalactic Astronomy, Department of Physics, Durham University, Durham DH1 3LE.

%%%%%%%%%%%%%%%%%%%%%%%%%%%%%%%%%%%%%%%%%%%%%%%%%%

%%%%%%%%%%%%%%%%% APPENDICES %%%%%%%%%%%%%%%%%%%%%

%%%%%%%%%%%%%%%%% APPENDICES %%%%%%%%%%%%%%%%%%%%%

\appendix

\section{Radio data}
\label{app}

\begin{figure*}
\centering
{\includegraphics[width=0.93\textwidth]{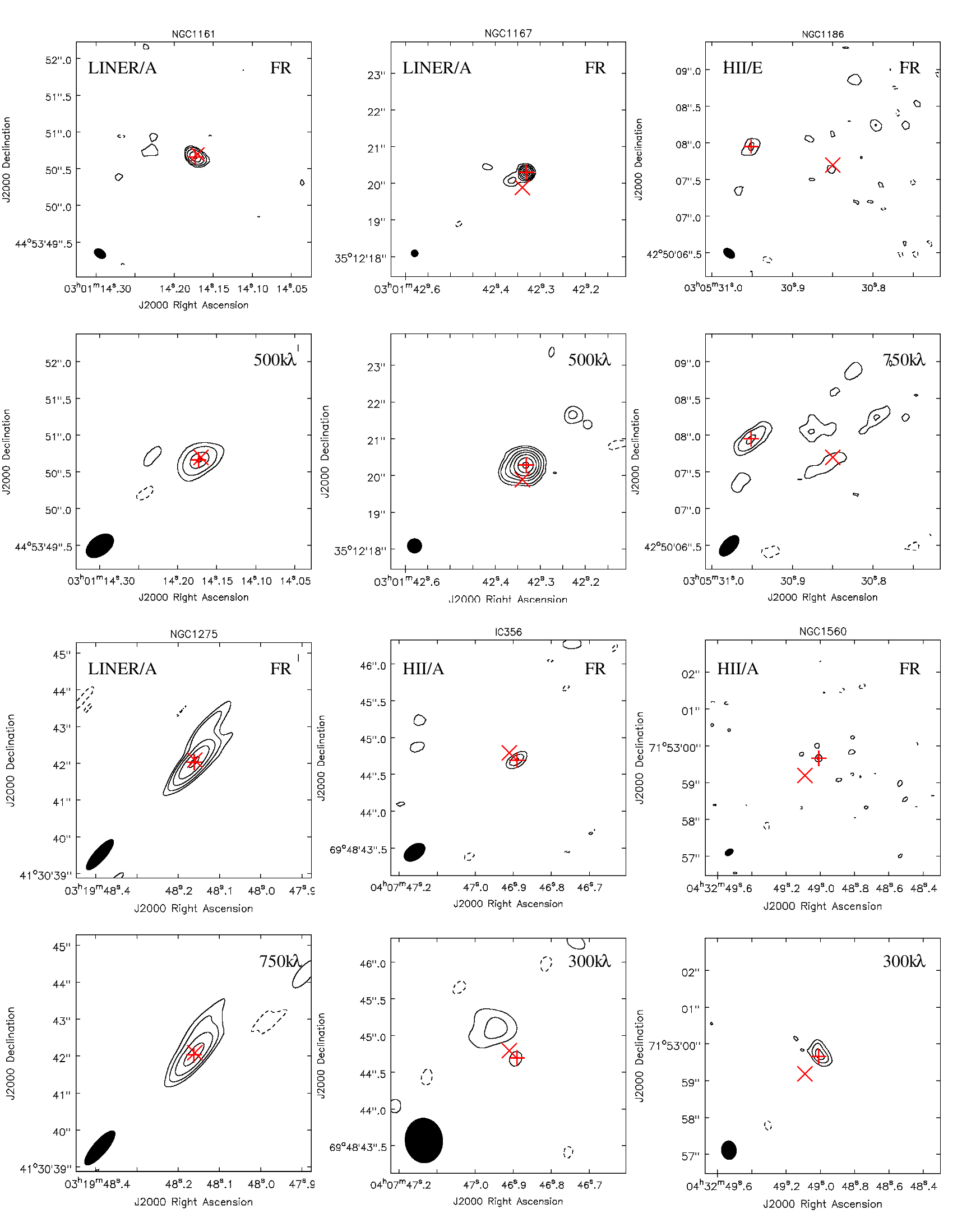}}
\caption[]{{\it e-}MERLIN 1.5-GHz radio maps of the detected and core-identified galaxies. For each galaxy two panels are shown. The upper panel depicts the full-resolution map, while
the lower panel shows the low-resolution map obtained with a $uv$-tapered scale written in the panel (in k$\lambda$), all on a same physical scale. For four galaxies (NGC~3516, NGC~4036, NGC~5322, and NGC~5548) a third radio map is presented corresponding to a lower resolution map (see the scale and map parameters in Tab~\ref{contours}). The restoring beam is presented as a filled ellipse at one of the corners of each of the maps. The contour levels of the maps are presented in Table~\ref{contours}. The $\times$ marks indicate the optical galaxy centre taken from NED, while the $+$ symbol marks the radio core position, if identified. In the upper panels, the optical (LINER, Seyfert, H{\sc ii}, and ALG) and radio (A, B, C, D, E, see Section 4.2 for description) classifications of the sources are reported. The full sets of Figures are available online.}
\label{ident-maps}
\end{figure*}

In the Appendix \ref{app} we present the radio images of the 78
detected Palomar galaxies (out of 177) studied here:  Fig.~\ref{ident-maps} for the 66 galaxies where we identified the core and Fig.~\ref{unident-maps} for the 15 galaxies without a core identification. Table~\ref{tabdet} and Table~\ref{tabsfr} list the source parameters of the radio components detected in the images for the identified and unidentified sources, respectively. Table~\ref{contours} provides the radio contours and the properties of the restoring beams of the radio maps of all the detected galaxies.

\addtocounter{figure}{-1}
\begin{figure*}
\centering
{\includegraphics[width=0.93\textwidth]{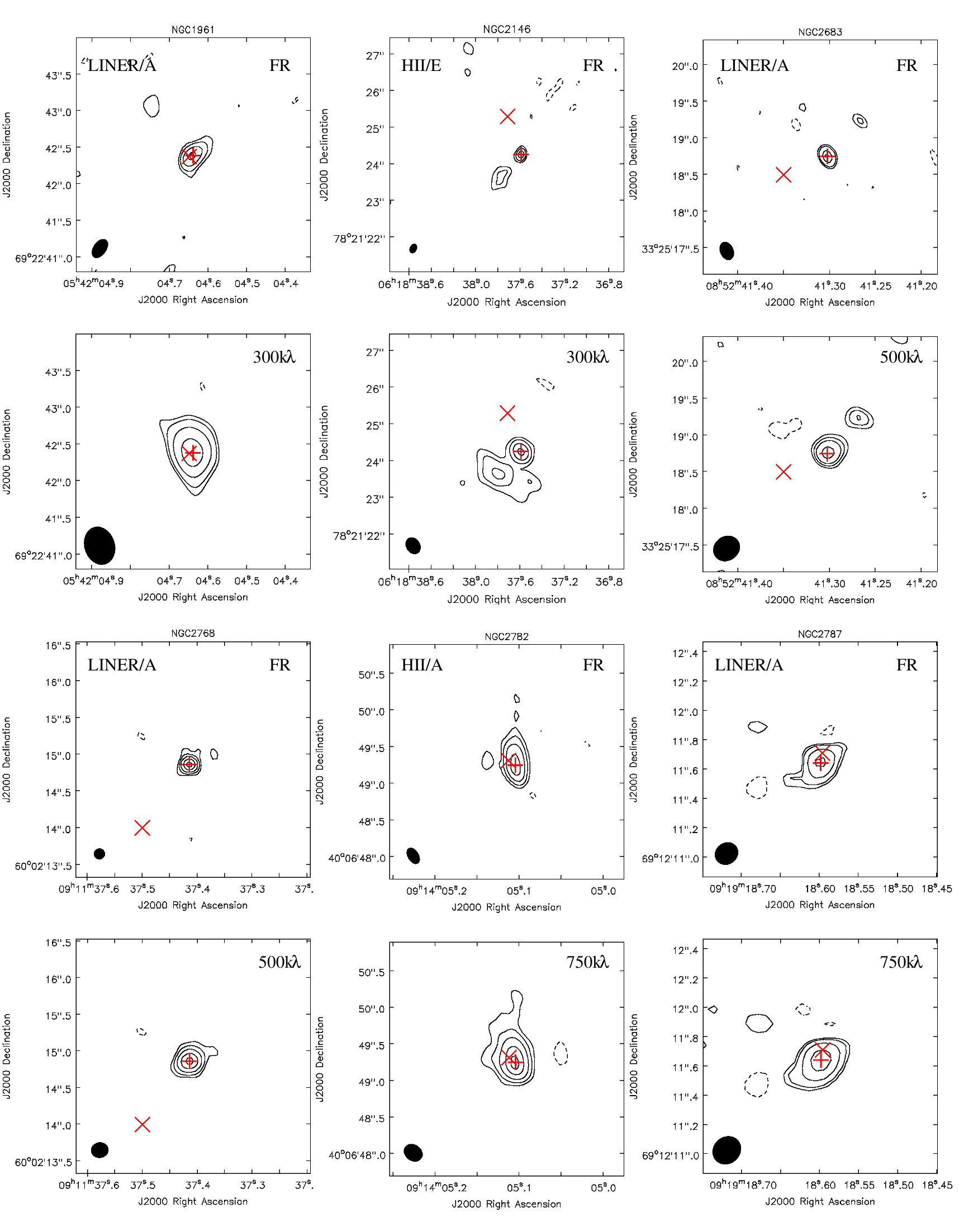}}
\caption{e-MERLIN 1.5-GHz maps of the detected and core-identified galaxies. See first page of this figure for details.}
\end{figure*}

\addtocounter{figure}{-1}
\begin{figure*}
\centering
{\includegraphics[width=0.93\textwidth]{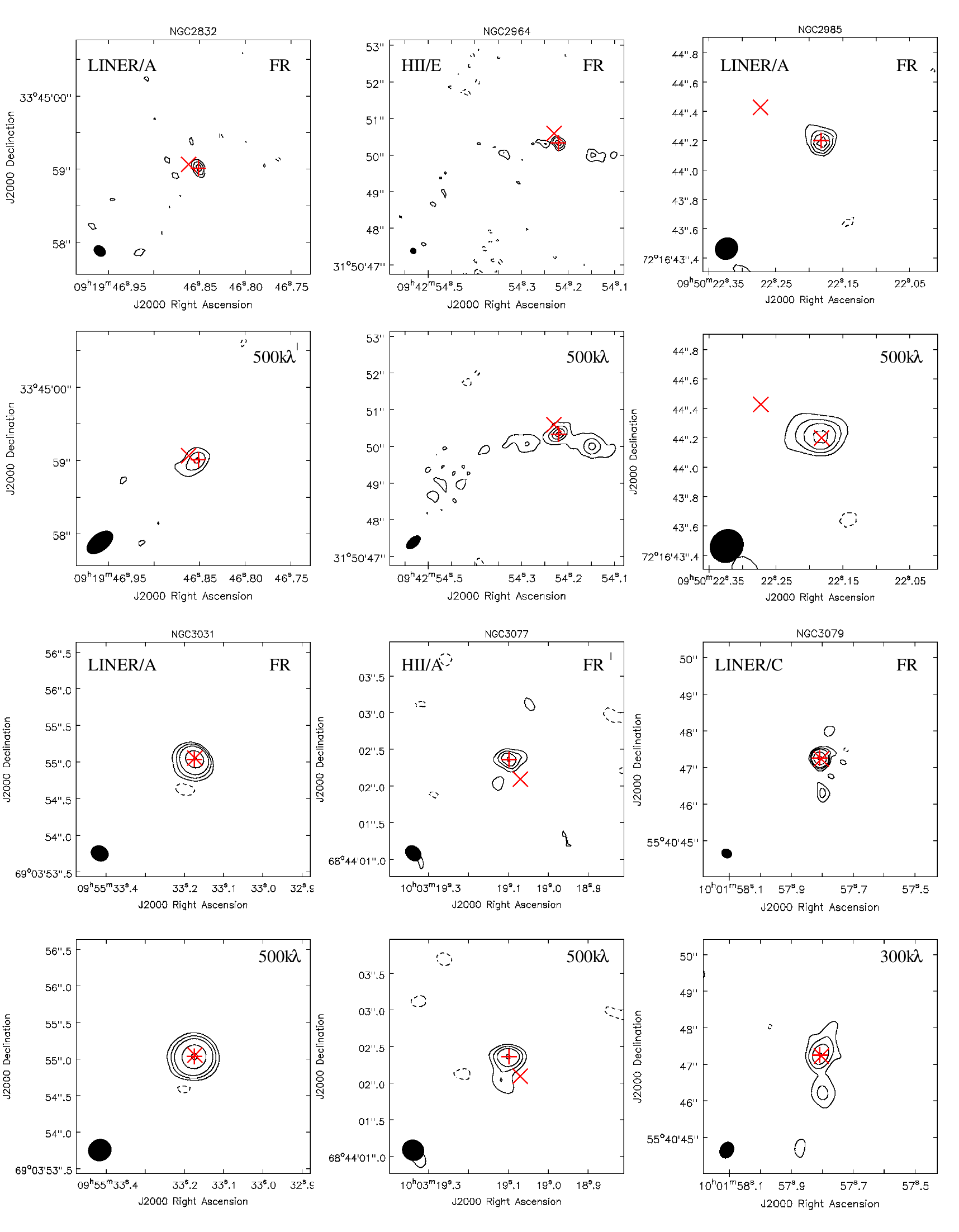}}
\caption{e-MERLIN 1.5-GHz maps of the detected and core-identified galaxies. See first page of this figure for details.}
\end{figure*}

\addtocounter{figure}{-1}
\begin{figure*}
\centering
{\includegraphics[width=0.93\textwidth]{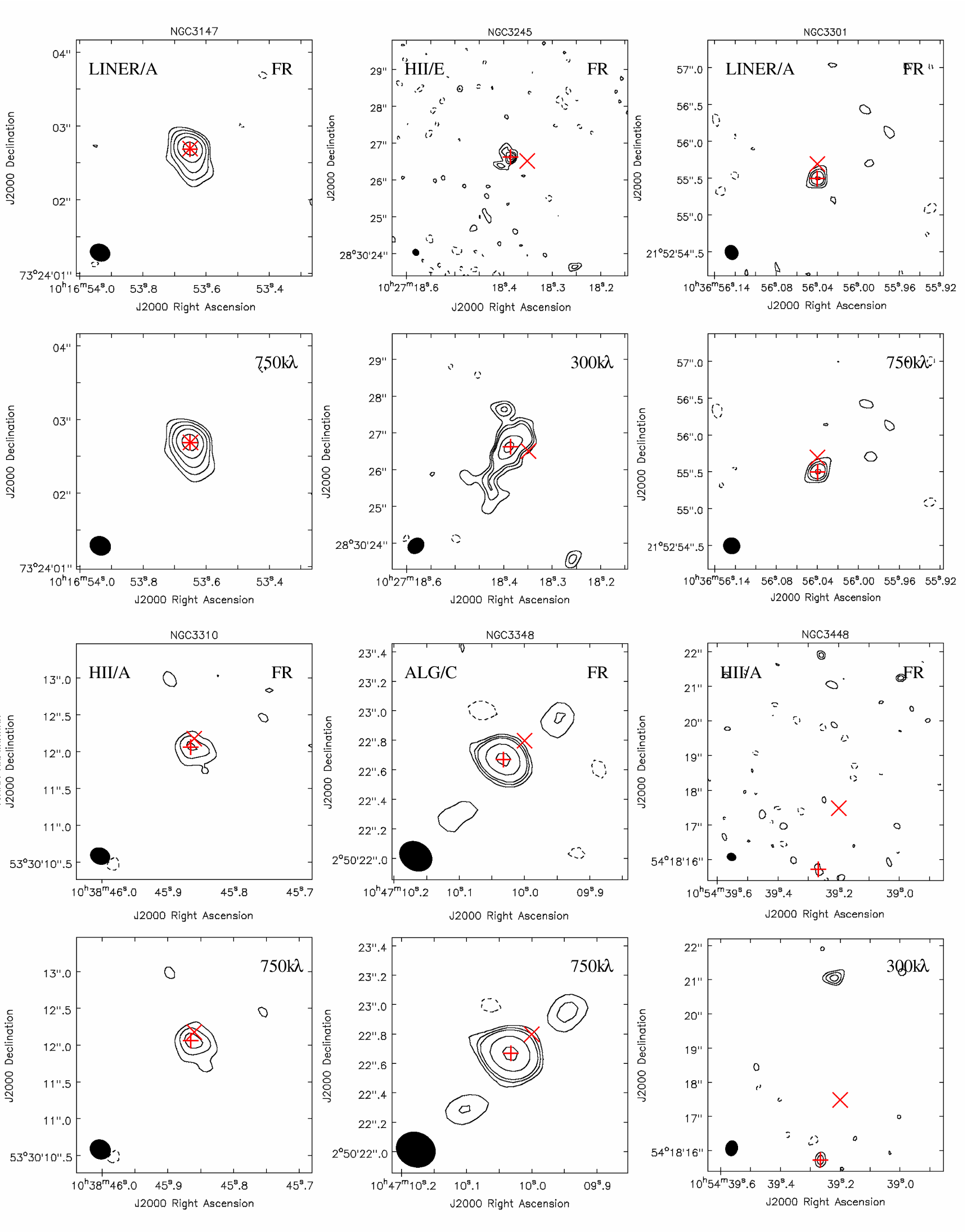}}
\caption{e-MERLIN 1.5-GHz maps of the detected and core-identified galaxies. See first page of this figure for details.}
\end{figure*}

\addtocounter{figure}{-1}
\begin{figure*}
\centering
{\includegraphics[width=0.93\textwidth]{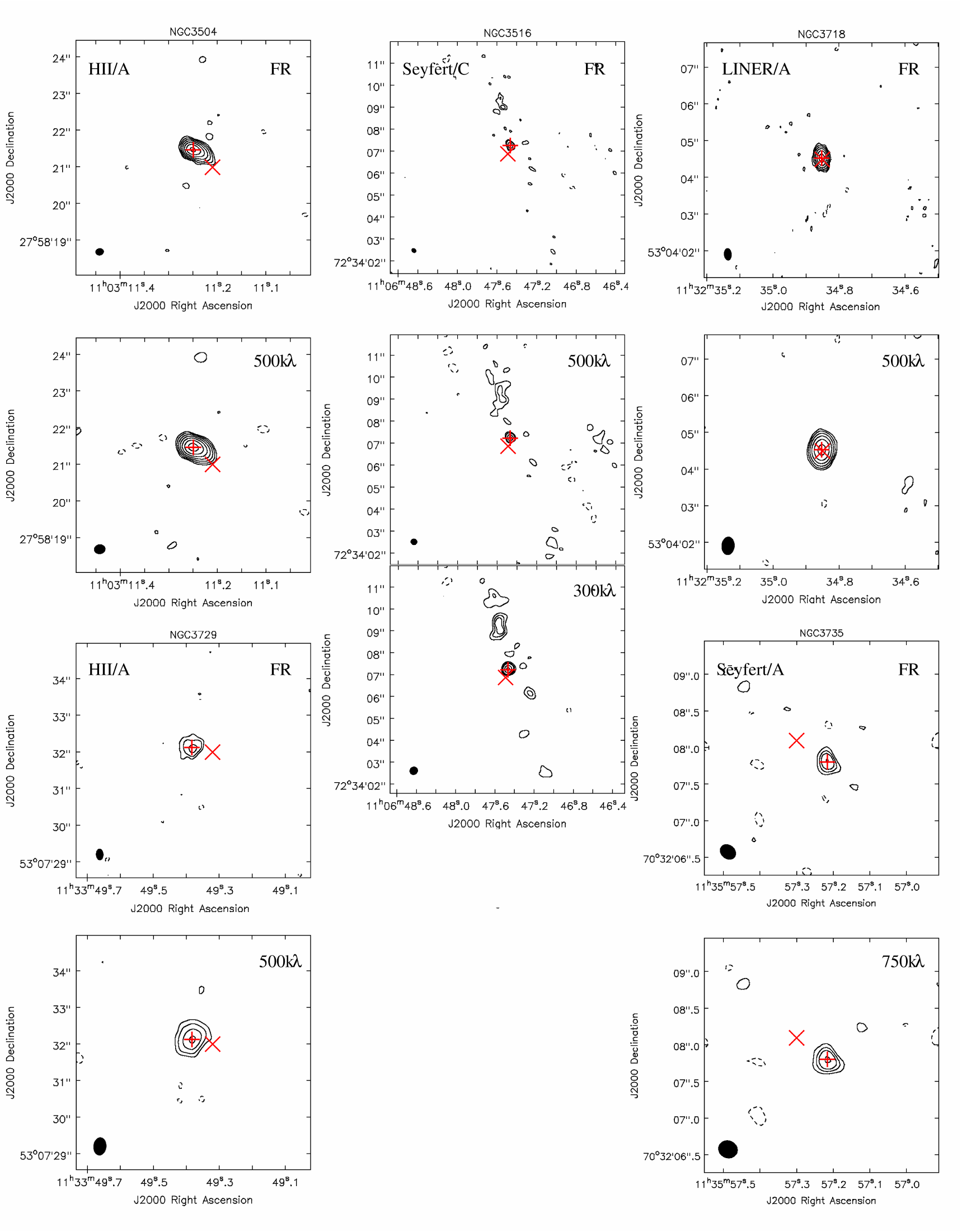}}
\caption{e-MERLIN 1.5-GHz maps of the detected and identified galaxies. See first page of this figure for details.}
\end{figure*}

\addtocounter{figure}{-1}
\begin{figure*}
\centering
{\includegraphics[width=0.93\textwidth]{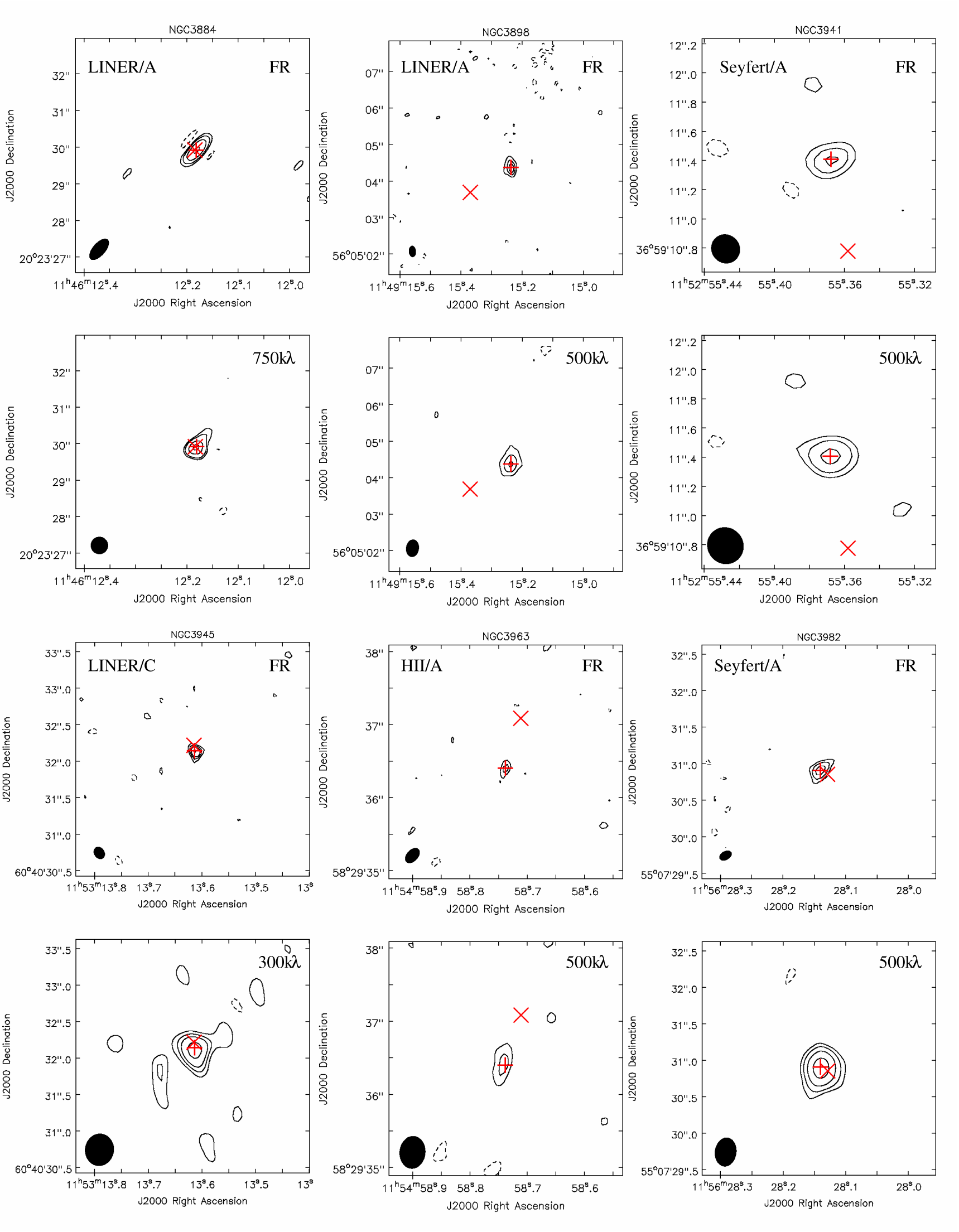}}
\caption{e-MERLIN 1.5-GHz maps of the detected and core-identified galaxies. See first page of this figure for details.}
\end{figure*}

\addtocounter{figure}{-1}
\begin{figure*}
\centering
{\includegraphics[width=0.93\textwidth]{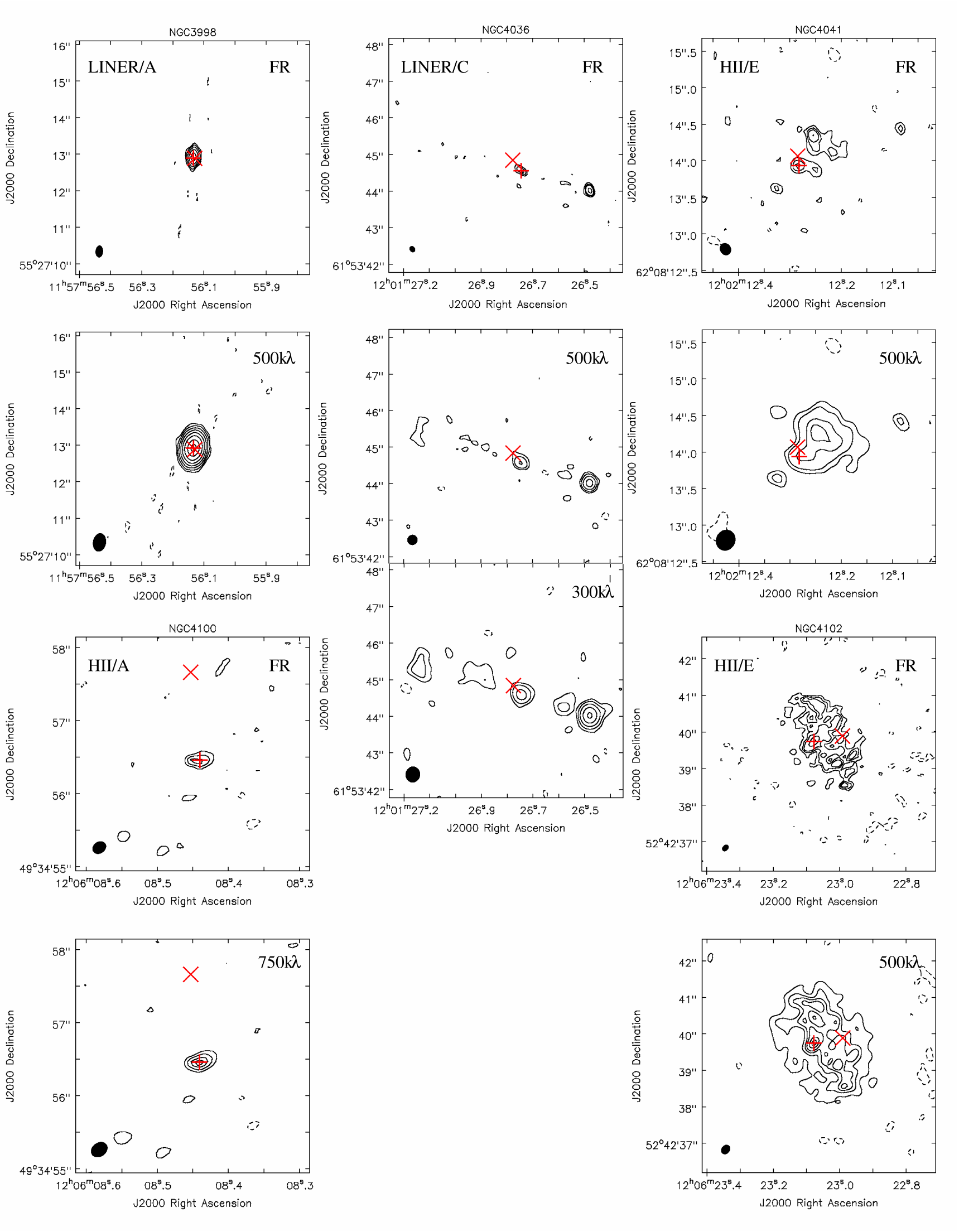}}
\caption{e-MERLIN 1.5-GHz maps of the detected and core-identified galaxies. See first page of this figure for details.}
\end{figure*}

\addtocounter{figure}{-1}
\begin{figure*}
\centering
{\includegraphics[width=0.93\textwidth]{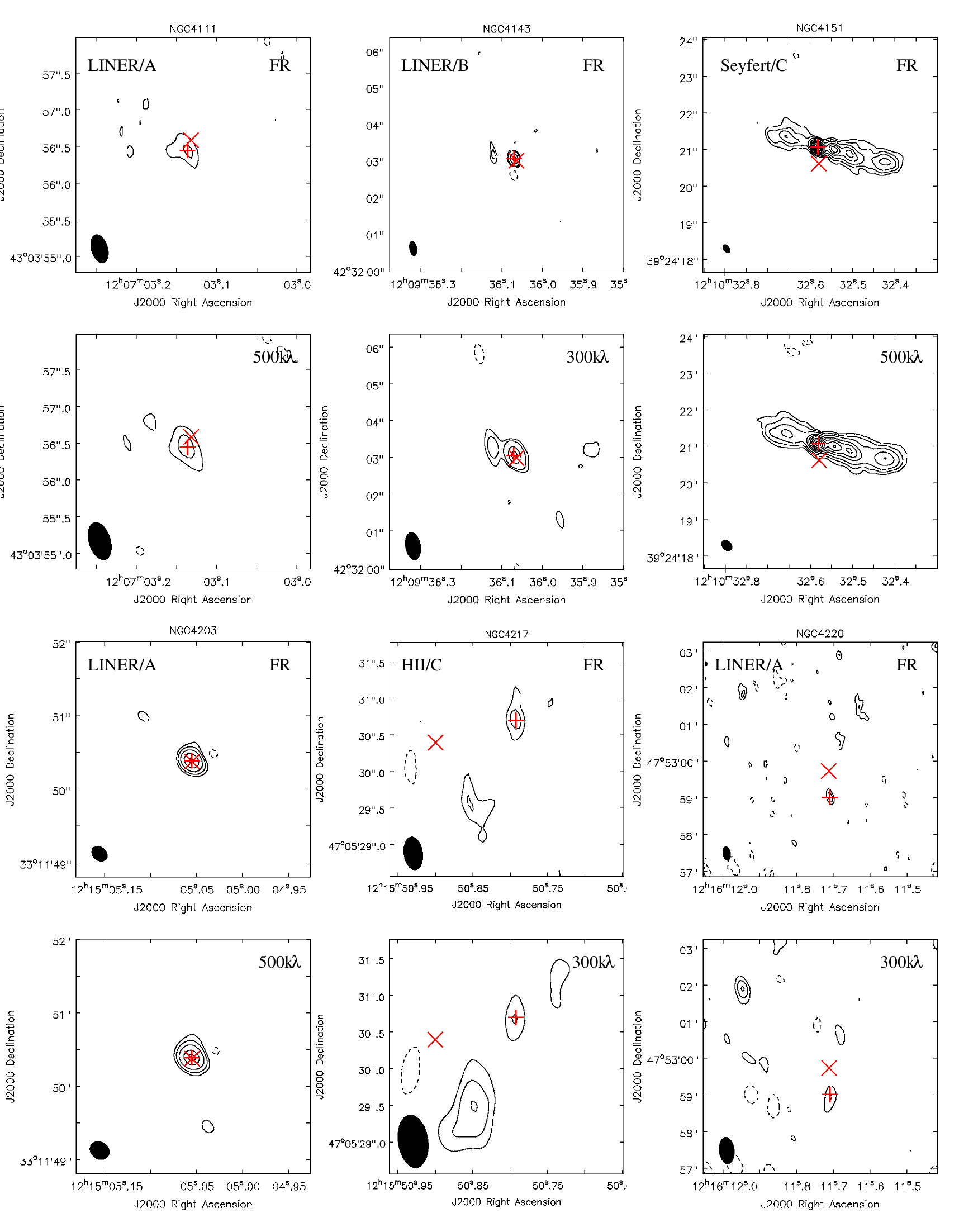}}
\caption{e-MERLIN 1.5-GHz maps of the detected and core-identified galaxies. See first page of this figure for details.}
\end{figure*}

\addtocounter{figure}{-1}
\begin{figure*}
\centering
{\includegraphics[width=0.93\textwidth]{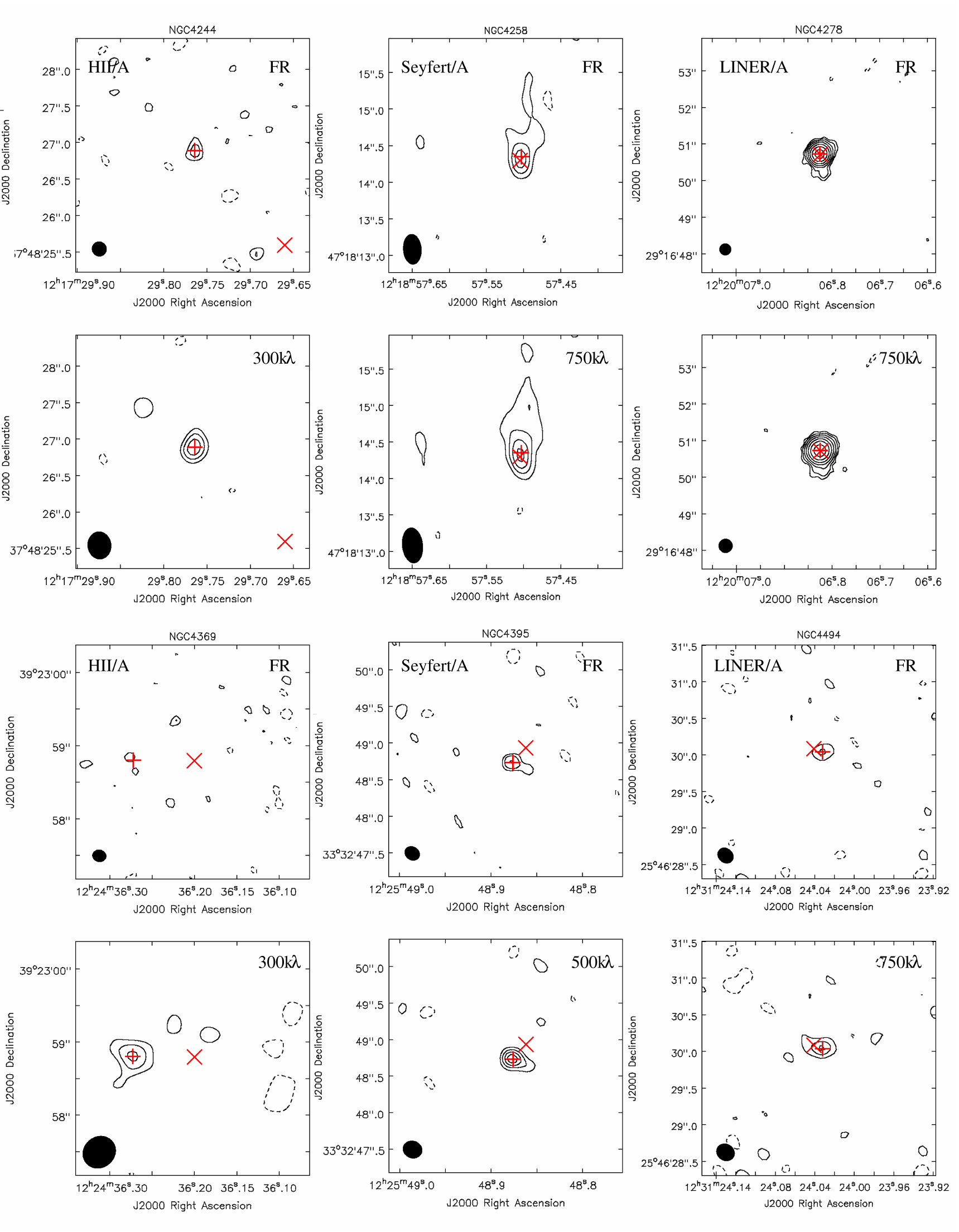}}
\caption{e-MERLIN 1.5-GHz maps of the detected and core-identified galaxies. See first page of this figure for details.}
\end{figure*}

\addtocounter{figure}{-1}
\begin{figure*}
\centering
{\includegraphics[width=0.93\textwidth]{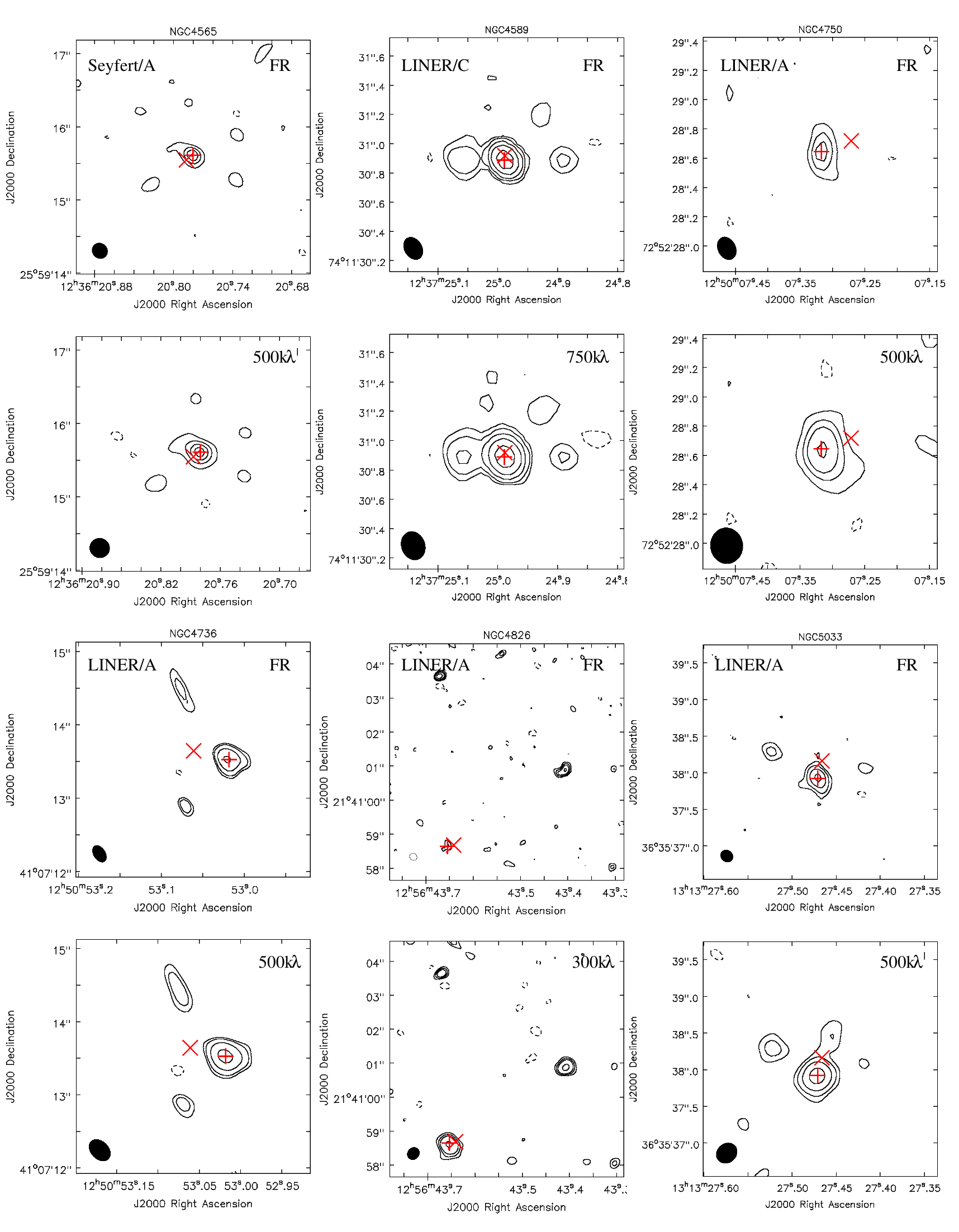}}
\caption{e-MERLIN 1.5-GHz maps of the detected and core-identified galaxies. See first page of this figure for details.}
\end{figure*}

\addtocounter{figure}{-1}
\begin{figure*}
\centering
{\includegraphics[width=0.93\textwidth]{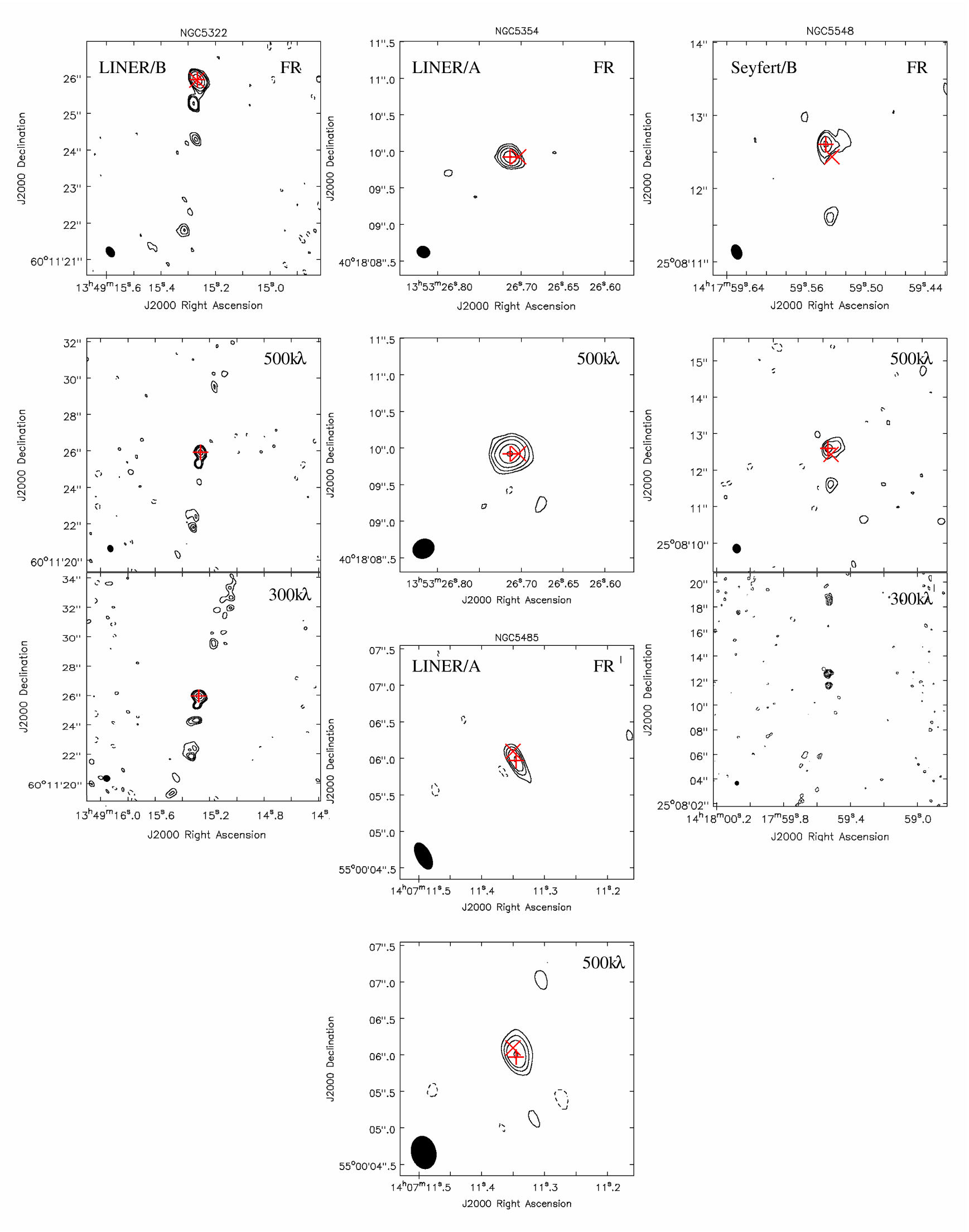}}
\caption{e-MERLIN 1.5-GHz maps of the detected and core-identified galaxies. See first page of this figure for details.}
\end{figure*}

\addtocounter{figure}{-1}
\begin{figure*}
\centering
{\includegraphics[width=0.93\textwidth]{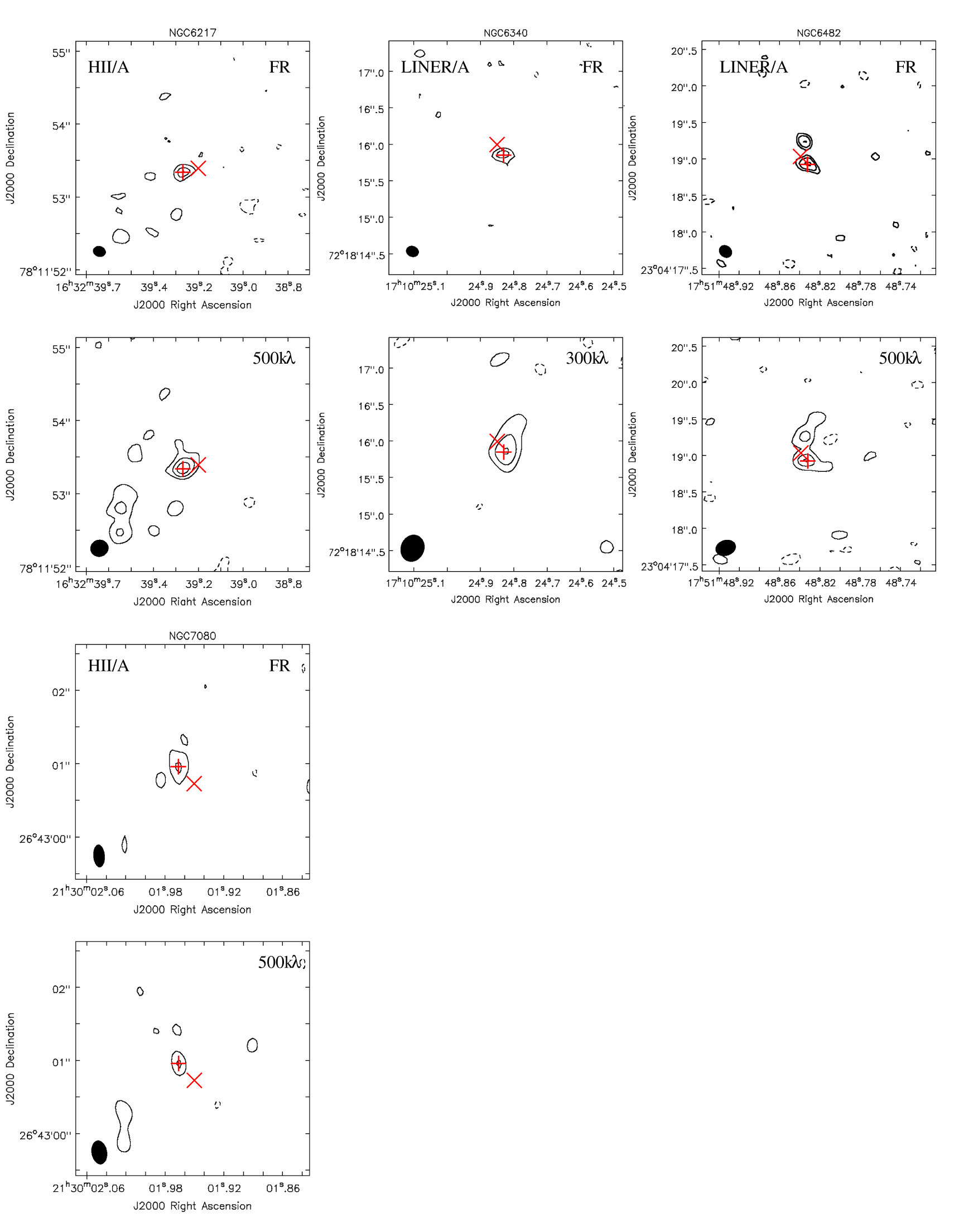}}
\caption{e-MERLIN 1.5-GHz maps of the detected and core-identified galaxies. See first page of this figure for details.}
\end{figure*}

\begin{figure*}
\centering
{\includegraphics[width=0.93\textwidth]{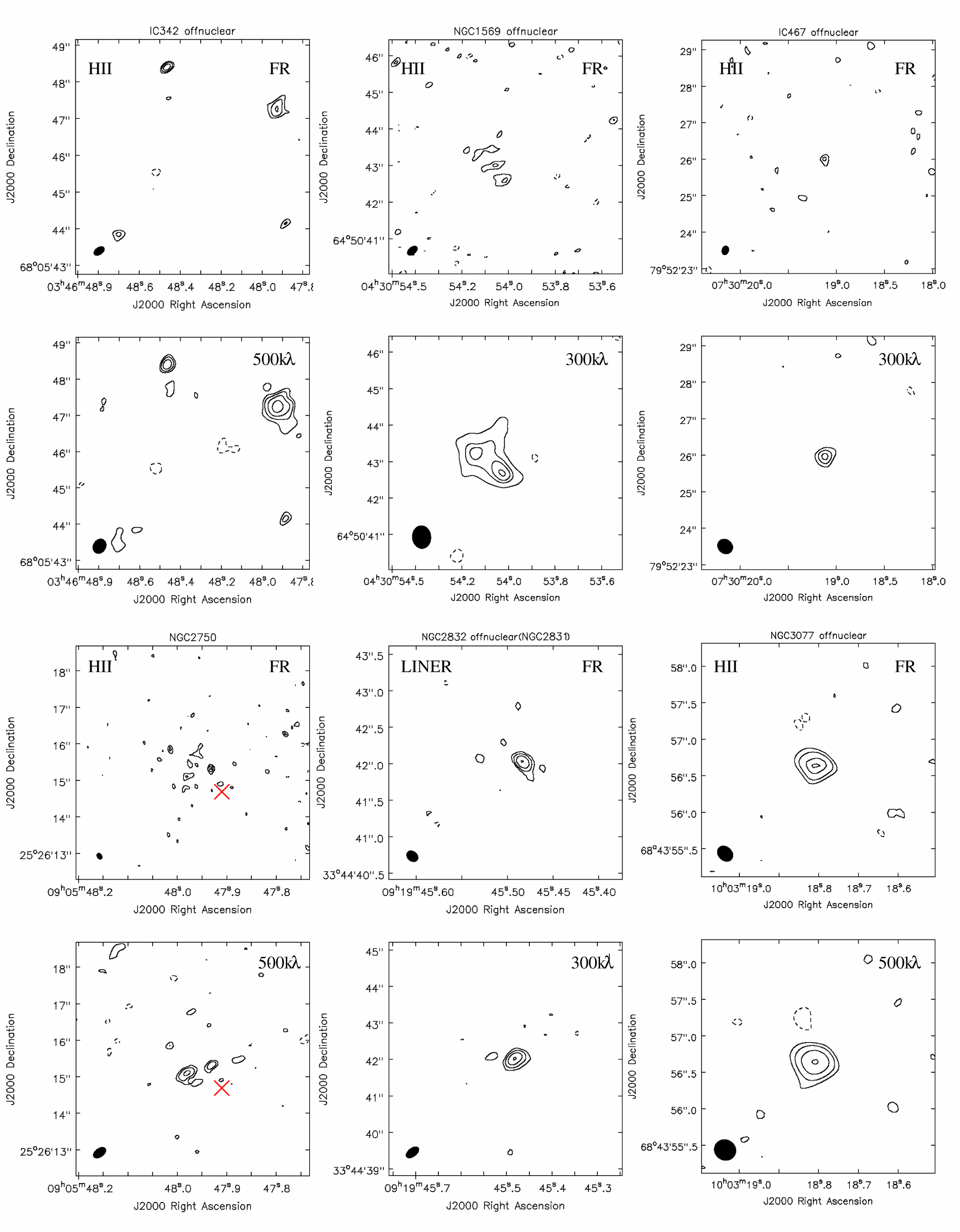}}
\caption{{\it e}-MERLIN 1.5-GHz radio images of the galaxies, which are detected but no radio core is identified. For each galaxy two panels are shown. The upper panel shows the full-resolution map, while the lower panel shows the low-resolution map obtained with a uv-tapered scale written in the panel (in k$\lambda$). The maps for the each source are generally on the same physical scale. The restoring beam is presented as a filled ellipse at one of the corners of each of the maps.  The contour levels of the maps are presented in Table~\ref{contours}. The $\times$ marks indicate the optical galaxy centre taken from NED, while the $+$ symbol marks the radio core position, if identified. In the upper panels, the optical (LINER, Seyfert, H{\sc ii}, and ALG) and radio (A, B, C, D, E, see Section 4.2 for description) classifications of the sources are reported.}
\label{unident-maps}
\end{figure*}

\addtocounter{figure}{-1}
\begin{figure*}
\centering
{\includegraphics[width=0.93\textwidth]{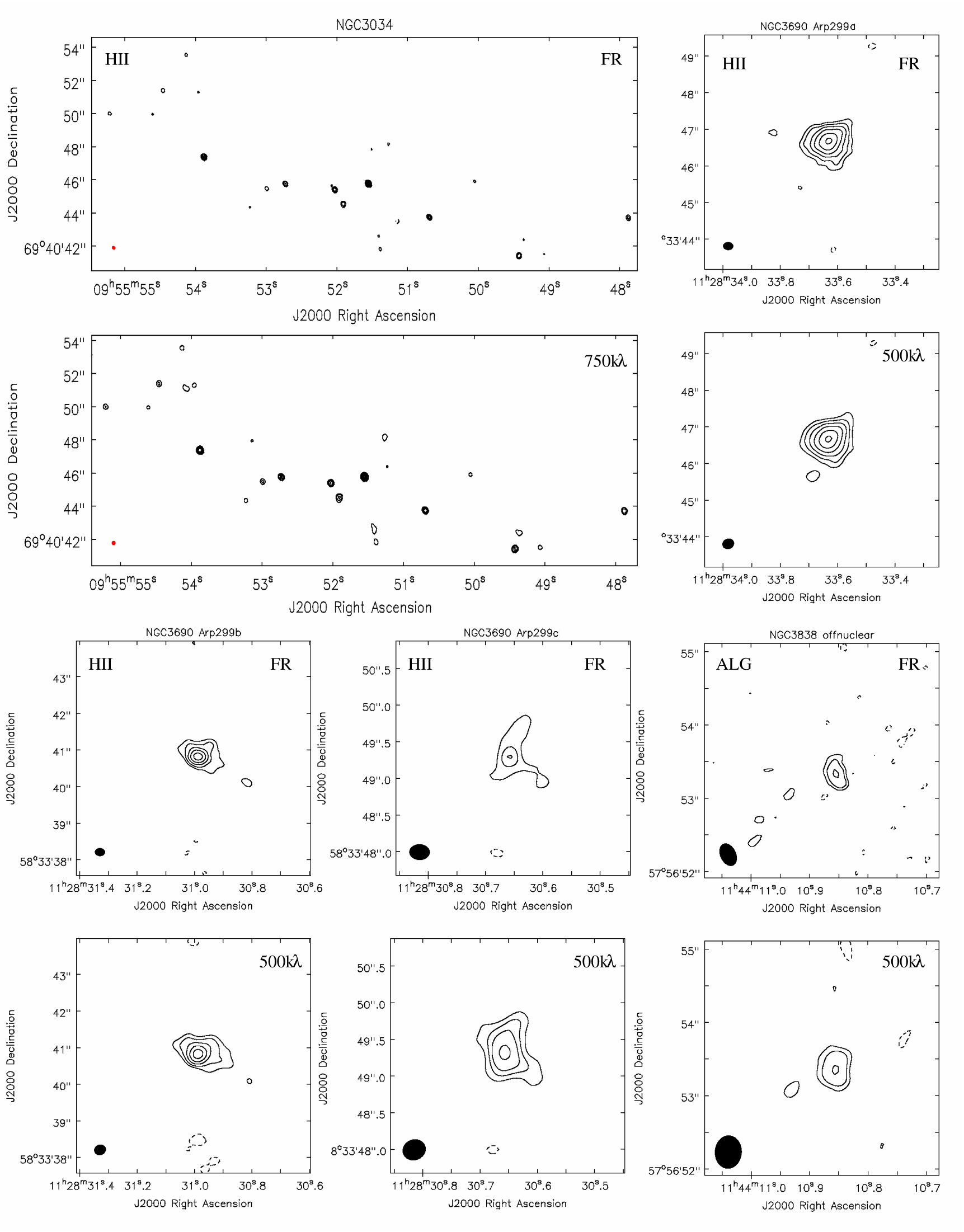}}
\caption{e-MERLIN 1.5-GHz images of the unidentified galaxies. See last page of this figure for detail.}
\end{figure*}

\addtocounter{figure}{-1}
\begin{figure*}
\centering
{\includegraphics[width=0.93\textwidth]{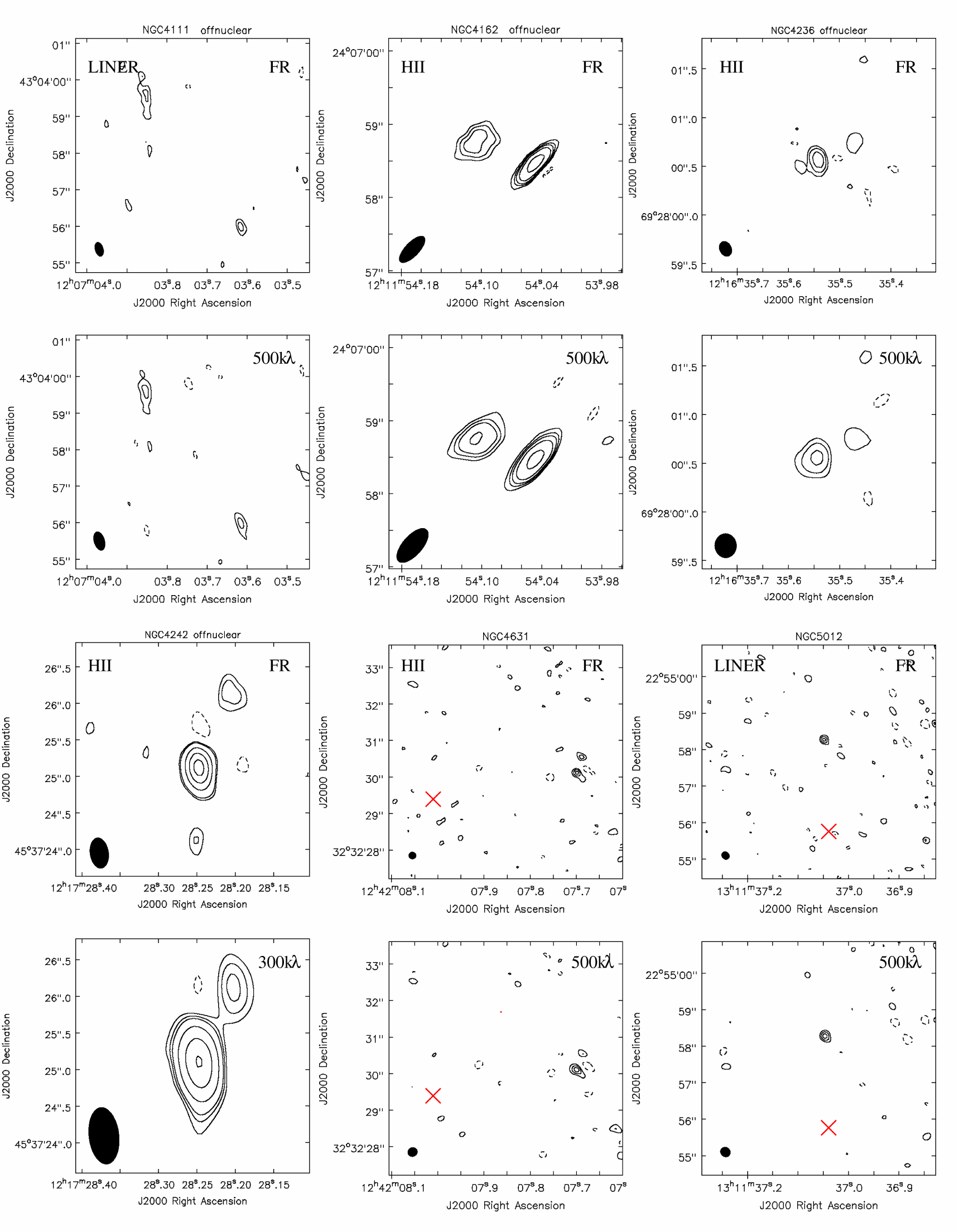}}
\caption{e-MERLIN 1.5-GHz images of the unidentified galaxies. See last page of this figure for detail.}
\end{figure*}

\twocolumn \normalfont\normalsize
\onecolumn
\begin{landscape}

\begin{center}
\begin{longtable}{lccC{0.4cm}C{2.3cm}cC{0.7cm}cC{0.4cm}cccC{0.4cm}C{0.7cm}cc}
\caption[Properties of the  sample.]{Optical and radio properties of the sample.} 
\label{table1} \\

%This is the header for the first page of the table...
\hline 
\hline
Name &  RA & DEC & D & Hubble    & class  & class  &  LEM &  Q  & Beam & PA  & rms & det & morph &  log L$_{\rm core}$  & log L$_{\rm tot}$  \\
 (1) & (2)  & (3)       & (4) & (5)  & (6) & (7) & (8) & (9)  &  (10)          &  (11)     &  (12) &  (13) &  (14)  &  (15) &  (16)           \\
       &      &       & Mpc& type     &    Ho97  &  BPT   &      &           &   arcsec$^{2}$ & degree & $\mu$Jy/b &  &  &  erg s$^{-1}$ & erg s$^{-1}$\\     
\hline 
\endfirsthead

%This is the header for the remaining page(s) of the table...
\multicolumn{3}{c}{{\tablename} \thetable{} -- Continued} \\[0.5ex]
\hline
\hline
Name &  RA & DEC & D & Hubble    & class  & class  &  LEM &  Q  & Beam & PA  & rms & det & morph &  log L$_{\rm core}$  & log L$_{\rm tot}$  \\
 (1) & (2)  & (3)       & (4) & (5)  & (6) & (7) & (8) & (9)  &  (10)          &  (11)     &  (12) &  (13) &  (14)  &  (15) &  (16)           \\
       &      &       & Mpc& type     &    Ho97  &  BPT   &      &           &   arcsec$^{2}$ & degree & $\mu$Jy/b &  &  &  erg s$^{-1}$ & erg s$^{-1}$\\     
\hline 

\endhead

%This is the footer for all pages except the last page of the table...
\hline
  \multicolumn{11}{c}{{Continued on Next Page}} \\
\endfoot

%This is the footer for the last page of the table...
 \\[-1.8ex] 
\endlastfoot

%Now the data..                                                   
NGC~783  &    02 01 06.61 &    +31 52 56.8 &       68.1 & SAc             &             H & H             &  MISC1 &  + & 0.22$\times$0.12 & 34.5   &  53 & U  &$-$    & $<$36.12  & $-$           \\
IC~239       &    02 36 27.82 &    +38 58 09.1 &       16.8 & SAB(rs)cd       &         L2::  & H$^{a}$        &      TEST & ++ & 0.19$\times$0.12 & 40.0   &  63 & U &$-$      & $<$34.98 &  $-$         \\
NGC~1003 &    02 39 17.03 &    +40 52 21.3 &       10.7 & SA(s)cd         &           ALG & H$^{b,c}$      &   TEST    & ++ & 0.19$\times$0.12 & 46.5   &  62 & U &$-$      & $<$34.58 & $-$               \\
NGC~1023 &    02 40 24.01 &    +39 03 47.7 &       10.5 & SB(rs)0-        &           ALG &  ALG$^{b}$     &   TEST    & ++ & 0.19$\times$0.12 & 42.7   &  63 & U &$-$      & $<$34.57 & $-$          \\
NGC~1058 &    02 43 30.01 &    +37 20 28.7 &        9.1 & SA(rs)c         &            S2 & L              & TEST &  ++ & 0.21$\times$0.12 & 41.0  &  57 & U &$-$      & $<$34.41  &  $-$             \\
NGC~1156 &    02 59 42.85 &    +25 14 28.4 &        6.4 & IB(s)m          &             H & H              &    TEST  & ++ & 0.22$\times$0.12 & 33.5  &  64  & U &$-$     & $<$34.15 &  $-$         \\
NGC~1161 &    03 01 14.17 &    +44 53 50.7 &       25.9 & SA0             &         T1.9: &  L             &    TEST   & ++ &0.18$\times$0.12 & 55.8   &  67 & I  & A      & 36.57    & 36.69         \\
NGC~1167 &    03 01 42.34 &    +35 12 19.9 &       65.3 & SA0-            &            S2 &  L             &    TEST   & ++ &0.20$\times$0.20 &  0     & 2700&  I  & A     & 39.65    & 39.76           \\
\hline
\hline
\end{longtable}
\end{center}
Column description: (1) source name; (2)-(3) RA and DEC position
(J2000.0) from NED, using optical or infrared images; (4) distance (Mpc)
from \citet{ho97a}; (5)
 morphological galaxy type given from RC3
\citep{devaucouleurs91}; (6) optical spectroscopic classification
based on \citet{ho97a}: H=H{\sc ii}, S=Seyfert, L=LINER, T=Transition
object, and ALG=Absorption line galaxy. The number attached to the
class letter designates the AGN type (between 1 and 2); quality
ratings are given by ':' and '::' for uncertain and highly uncertain
classification, respectively. Two classifications are given for some
ambiguous cases, where the first entry corresponds to the adopted
choice; (7) optical spectroscopic classification based on BPT diagrams
and from the literature. See the notes for the classification based on
the literature; (8) LeMMINGs observation block;  (9) raw data and calibration quality: `++' = very
good; `+' = good; `m' = moderate; (10) restoring beam size in arcsec in
full resolution map; (11) PA angle (degree) in full resolution map; (12) rms in
full resolution map in $\mu$Jy beam$^{-1}$; (13) radio detection
status of the source: `I' = detected and core identified; `U' =
undetected; `unI' = detected but core unidentified; 'I+unI' =  detected and core identified with additional unknown source(s) in the field; (14) radio
morphological class: A = core/core--jet; B = one-sided jet; C =
triple; D = doubled-lobed ; E = jet+complex (see Sect 4.2); (15)-(16) logarithm of the radio core and total luminosities
(erg s$^{-1}$). To convert the radio luminosities from erg s$^{-1}$ to monochromatic luminosities (W Hz$^{-1}$) at 1.5 GHz, an amount +16.18 should be subtracted to the logarithm of the luminosities presented in the Table.
\\ Notes: letters correspond to following publications: $a$
\citet{keel83}; $b$ \citet{vandebosch15}; $c$ \citet{moustakas06}; $d$
\citet{florido12}; $e$ \citet{buttiglione10}; $f$ \citet{gavazzi13};
$g$ \citet{heckman80}; $h$ \citet{balmaverde13}; $i$ SDSS; $j$
\citet{gavazzi18}; $k$ \citet{cazzoli18}; $l$ \citet{shields07}; $m$
\citet{nyland16}; $n$ \citet{serra18}; $o$ \citet{wegner03}; $p$
\citet{pismis01}; $q$ \citet{kennicutt92}; $r$ \citet{lira07}; $s$
\citet{baldi09}; $t$ \citet{garcia15}.

\end{landscape}

\fontsize{7}{10}\selectfont
\begin{landscape}
\begin{center}
\begin{longtable}{C{1.2cm}|C{0.4cm}ccccccc|C{0.05cm}ccccc|C{1.8cm}}

\caption[Properties of the sample.]{Properties of the detected and core-identified sources.} 
\label{tabdet} \\

%This is the header for the first page of the table...
\hline \hline

%This is the header for the first page of the table...
%\hline %\hline 

     &   \multicolumn{8}{c}{Full resolution}                                                   &  \multicolumn{7}{|c}{Low resolution} \\
name & comp & $\theta_{\rm M}\times\theta_{\rm m}$  & PA$_{\rm d}$ & rms  &  $\alpha$(J2000) & $\delta$(J2000) &  F$_{\rm peak}$ & F$_{\rm tot}$ & &  $u{-}v$     &  $\theta_{\rm M}\times\theta_{\rm m}$ &  PA$_{\rm d}$ & rms &  F$_{\rm peak}$ &morph/size \\
     &        &arcsec$^{2}$ & deg  & mJy/b &                 &                &  mJy/b       &  mJy      & &  k$\lambda$ & arcsec$^{2}$& deg   &  mJy/b &   mJy/b      & \\

\hline	
\endfirsthead

%This is the header for the remaining page(s) of the table...
\multicolumn{3}{c}{{\tablename} \thetable{} -- Continued} \\[0.5ex]
\hline \hline

   &   \multicolumn{8}{c}{Full resolution}                                                   &  \multicolumn{7}{|c}{Low resolution} \\
name & comp & $\theta_{\rm M}\times\theta_{\rm m}$  & PA$_{d}$ & rms  &  $\alpha$(J2000) & $\delta$(J2000) &  F$_{\rm peak}$ & F$_{\rm tot}$ &  & $u{-}v$    &  $\theta_{\rm M}\times\theta_{\rm m}$ &  PA$_{d}$ & rms &  F$_{\rm peak}$  & morph/size \\
     &        & arcsec$^{2}$ & deg  & mJy/b &                 &                &  mJy/b       &  mJy      &  & k$\lambda$ & arcsec$^{2}$ & deg   &  mJy/b &   mJy/b       & \\

\hline
\endhead

%This is the footer for all pages except the last page of the table...
\hline
  \multicolumn{16}{c}{{Continued on Next Page}} \\
\endfoot

%This is the footer for the last page of the table...
  \\[-1.8ex] 
\endlastfoot

%Now the data...&  &      &  0.24$\times$0.17 & 37.2  &        & 
NGC~1161    &  core &    0.06$\times$0.01   &  53.7   &  0.067 &  03 01 14.173  & 44 53 50.66   & 3.09$\pm$0.07  &  3.13   &  & 500  &  $<$0.43$\times<$0.26 & -53.7  &   0.066   &   3.00$\pm$0.07  & core (A) \\
            &       &                       &         &  0.067 & 03 01 14.242 & 44 53 50.77                & $<$0.22        &                   & &  500 &  $<$0.69$\times<$0.26 & 139.8  &   0.066   &   0.61$\pm$0.08  &           \\
\cmidrule(rl){15-15}
            & Tot &                     &          &        &               &              &                &      &     &                      &        &      &    &  4.1$\pm$0.2 &            \\
\hline
NGC~1167    &  core &    0.08$\times$0.07   &  19   &  2.7  &  03 01 42.331  & 35 12 20.30  & 580.1$\pm$10.0  &  695   &  & 500  &  0.20$\times$0.10 & 114  &   3.1   &   618$\pm$11  & \multirow{3}{1.8cm}{\centering core-jet (A) 1$\arcsec\to$310pc} \\
            &       &    0.35$\times$0.14   &  113  &  2.7  &  03 01 42.361  & 35 12 20.10  & 49.0$\pm$11.0   &  121   & &  500 &                    &      &   3.1  &    $<$75       &              \\
            &       &                       &       &  2.7  &  03 01 42.226  & 35 12 21.65  &   $<$30         &  121   & &  500 &   0.18$\times$0.10 & 113  &   3.1  &     20$\pm$5   &              \\
\cmidrule(rl){15-15}
            & Tot &                     &          &        &               &              &                &      &     &                      &        &      &    &  755$\pm$28 &            \\
\hline
NGC~1186    &  core &    $<$0.26$\times<$0.16   &  134.4   & 0.065   &  03 05 30.951  &  42 50 07.95  & 0.36$\pm$0.07  &  0.71   &  & 750 &  $<$0.45$\times<$0.18 & 129.7  &   0.067   &   0.61$\pm$0.07  & \multirow{3}{1.8cm}{\centering complex (E) multi-components 2.5$\arcsec\to$423pc}\\
\hline
\end{longtable}
\end{center}
\vspace*{-1cm} \fontsize{6}{7}\selectfont Column description: (1)
galaxy name; (2) radio component: core, jet, lobe, blob or
unidentified component if not labeled (W or E stand for West or
East); (3) deconvolved FWHM dimensions (major $\times$ minor axes [arcsec$^{2}$],
$\theta_{\rm M}\times\theta_{\rm m}$) of the fitted component, determined from
an elliptical Gaussian fit from the full-resolution radio map; (4) PA
of the deconvolved component, PA$_{\rm d}$, from the full-resolution radio
map (degree); (5) rms of the radio map close to the specific component from the
full-resolution radio map (mJy/b, mJy beam$^{-1}$); (6)-(7) radio position (J2000.0); (8)
peak flux density in mJy beam$^{-1}$, F$_{\rm peak}$, from the full-resolution
radio map: this represents the radio core flux density; (9) integrated flux
density, $F_{tot}$, in mJy, from the full-resolution radio map;
(10) $uv$-taper scale of the low-resolution radio map in k$\lambda$;
(11) deconvolved FWHM dimensions (major $\times$ minor axes [arcsec$^{2}$],
$\theta_{\rm M}\times\theta_{\rm m}$) of the fitted component, determined from
an elliptical Gaussian fit from the low-resolution radio map; (12) PA
of the deconvolved component, PA$_{\rm d}$, from the low-resolution radio
map (degree); (13) rms of the radio map close to the specific component from
the low-resolution radio map (mJy beam$^{-1}$); (14) peak flux density in mJy beam$^{-1}$,
F$_{\rm peak}$, from the low-resolution radio map. For NGC~5194 we give the
total integrated flux densities of the radio lobes instead of the peak
flux densities.  At the bottom of each target the total flux density of the
radio source associated with the galaxy is given in mJy, measured from
the low-resolution map; (15) radio morphology (A, B, C, D, E) and size
in arcsec and pc (see Section~\ref{core-ident}).  
\end{landscape}

\fontsize{7}{10}\selectfont
\begin{landscape}
\begin{center}
\begin{longtable}{C{1.2cm}|C{0.4cm}ccccccc|C{0.05cm}ccccc|C{1.9cm}}

\caption[Properties of the sample.]{Properties of the unidentified sources.} 
\label{tabsfr} \\

%This is the header for the first page of the table...
\hline \hline

%This is the header for the first page of the table...
%\hline \hline 

     &   \multicolumn{8}{c}{Full resolution}                                                   &  \multicolumn{7}{|c}{Low resolution} \\
name & comp & $\theta_{\rm M}\times\theta_{\rm m}$  & PA$_{\rm d}$ & rms  &  $\alpha$(J2000) & $\delta$(J2000) &  F$_{\rm peak}$ & F$_{\rm tot}$ & &  $u{-}v$    &  $\theta_{\rm M}\times\theta_{\rm m}$ &  PA$_{\rm d}$ & rms &  F$_{\rm peak}$ &morph/size \\
     &        & arcsec$^{2}$ & deg  & mJy/b &                 &                &  mJy/b       &  mJy      & &  k$\lambda$ & arcsec$^{2}$ & deg   &  mJy/b &   mJy/b      & \\

\hline	
\endfirsthead

%This is the header for the remaining page(s) of the table...
\multicolumn{3}{c}{{\tablename} \thetable{} -- Continued} \\[0.5ex]
\hline \hline 

   &   \multicolumn{8}{c}{Full resolution}                                                   &  \multicolumn{7}{|c}{Low resolution} \\
name & comp & $\theta_{M}\times\theta_{m}$  & PA$_{d}$ & rms  &  $\alpha$(J2000) & $\delta$(J2000) &  F$_{\rm peak}$ & F$_{\rm tot}$ &  & $u{-}v$    &  $\theta_{M}\times\theta_{m}$ &  PA$_{d}$ & rms &  F$_{\rm peak}$  & morph/size \\
     &        & [arcsec$^{2}$] & deg  & mJy/b &                 &                &  mJy/b       &  mJy      &  & k$\lambda$ & [arcsec$^{2}$] & deg   &  mJy/b &   mJy/b       & \\

\hline
\endhead

%This is the footer for all pages except the last page of the table...
\hline
  \multicolumn{16}{c}{{Continued on Next Page}} \\
\endfoot

%This is the footer for the last page of the table...
  \\[-1.8ex] 
\endlastfoot

%Now the data...&  &      &  0.24$\times$0.17 & 37.2  &        & 
IC~342    &    &   0.13$\times$0.05   &  141   &  0.175 &   03 46 48.464  & 68 05 48.41  & 1.61$\pm$0.18  &  1.79    &  & 500   &  0.21$\times$0.05  &  154   &   0.177   &   1.35$\pm$0.18  &  \multirow{4}{1.8cm}{\centering  multi-components  5.5$\arcsec\to$100pc, $>$2$\arcsec$ offset}\\
          &    &   0.54$\times$0.38   & 163    &  0.175  & 03 46 47.926  & 68 05 47.28 & 1.18$\pm$0.18   &  5.18  &  & 500   &   0.60$\times$0.54  &   57   &   0.177   &   1.69$\pm$0.18  &   \\
       &      &    0.36$\times$0.30  & 142 &  0.175 &   03 46 48.701  & 68 05 43.83  & 0.86$\pm$0.18  &  2.30  &    & 500   &    1.20$\times$0.44  &  165 &   0.177   &   0.60$\pm$0.18  &  \\
       &      &     $<$0.39$\times$ $<$0.20  &  137  &  0.175 &   03 46 47.884  & 68 05 44.14  & 0.94$\pm$0.18  &  1.11  &    & 500   &    $<$0.46$\times$ $<$0.30  &   156 &   0.177   &   0.81$\pm$0.18  &  \\ 
       \cmidrule(rl){15-15}  & Tot  &                     &          &        &               &              &                &      &     &                      &        &      &   &  10.2$\pm$0.9 &           \\     
\hline
NGC~1569  &      &  0.61$\times$0.39 &  99  & 0.115  &   04 30 54.026 & 64 50 42.62 &  0.38$\pm$0.11 & 1.81       & & 300 &  0.84$\times$0.32 & 57 & 0.148  & 0.95$\pm$0.15   &\multirow{4}{1.8cm}{\centering triple source 2$\arcsec\to$14pc, offset $\sim$30$\arcsec$} \\
           & core  &  0.54$\times$0.06 &  98  & 0.115  &   04 30 54.054 & 64 50.43.01 &  0.40$\pm$0.11 & 1.46       & & 300 &                   &      & 0.148  & $<$0.42   & \\
           &      &  0.92$\times$0.29 &  138.0  & 0.115  &   04 30 54.131 & 64 50 43.39 & 0.38$\pm$0.11  & 1.81   & & 300 &  0.88$\times$0.74 & 89 & 0.148  & 0.82$\pm$0.04   & \\
 \cmidrule(rl){15-15}
      & Tot  &                     &          &        &               &              &                &      &  &   &                      &        &        &  3.70$\pm$0.40           &  \\
\hline
IC~467  &      &  0.18$\times$0.15 &  41  & 0.101  &   07 30 19.115 & 79 52 25.99  &  0.44$\pm$0.10 & 0.69   & & 300 &  0.44$\times$0.16 & 136 & 0.102  & 0.63$\pm$0.10   &  offset 5.5$\arcsec$   \\
       \cmidrule(rl){15-15}
           & Tot  &                     &          &        &               &              &                &      &     &                      &        &      &   &  1.0$\pm$0.1 &           \\     
\hline
NGC~2750    & core? & 0.18$\times$0.09      &   167    &  0.072  &  09 05 47.931  &  25 26 15.32   &  0.35$\pm$0.07   &   0.60   &  & 500  &  $<$0.42$\times<$0.22 &  135  & 0.055 &  0.28$\pm$0.07  &  \multirow{5}{1.8cm}{\centering multi-components 2.2$\arcsec\to$425pc, offset $<$2$\arcsec$} \\
            &     &  0.28$\times$0.07      &   127    &  0.072  &  09 05 47.981  &  25 26 15.11   &  0.26$\pm$0.07   &   0.63   &  & 500  &  0.35$\times$0.20     &   125   & 0.055  & 0.35$\pm$0.07  &  \\
            &     &  0.24$\times$0.05      &   10.3   &  0.072  &  09 05 48.014  &  25 26 15.86   &  0.29$\pm$0.07   &   0.46   &  & 500  &  $<$0.31$\times<$0.22  &  123   & 0.055  & 0.19$\pm$0.07  &  \\
            &     &  $<$0.24$\times<$0.15  &   91.4   &  0.072  &  09 05 47.912  &  25 26 14.91   &  0.25$\pm$0.07   &   0.33   &  & 500  &  $<$0.29$\times<$0.20  &  101   & 0.055  & 0.15$\pm$0.08  &  \\
            &     &  0.07$\times$0.03      &   21     &  0.072  &  09 05 47.883  &  25 26 15.45   &  0.23$\pm$0.07   &   0.24   &  & 500  &  $<$0.27$\times<$0.20  &  103   & 0.055  & 0.21$\pm$0.08  &  \\
\cmidrule(rl){15-15}
            &  Tot &                     &          &        &               &              &                &      &     &                      &        &      &    &  1.8$\pm$0.4 &            \\
\hline
NGC~2831   & core & $<$0.19$\times<$0.13      &   44.5    &  0.086  & 09 19 45.483  &  33 44 42.03   &  2.65$\pm$0.09   &   2.75   &  & 500  & 0.18$\times<$0.11 &  56  & 0.103 &  2.39$\pm$0.10  & NGC~2832 companion \\
\cmidrule(rl){15-15}
            & Tot  &                     &          &        &               &              &                &      &     &                      &        &      &    &  3.1$\pm$0.2 &            \\
\hline
NGC~3034  & %[MMW2002] 46.75+67.0 
SNR & $<$0.23$\times$ $<$0.04  &   0.0  &  0.131 & 09 55 55.405  & 69 40 53.15& 1.95$\pm$0.24  & 2.47  &  & 750  & $<$0.28$\times$ $<$0.25  & 0.0  & 0.420  & 2.19$\pm$0.19  &   \multirow{4}{1.8cm}{\centering  multi-components 31.7$\arcsec\to$602pc}\\
       & %[MMW2002] 46.56+73.8 
       SNR &   0.44$\times$0.20  &   40   &  0.131 & 09 55 55.259 &  69 40 59.91  & 2.23$\pm$0.15     &    6.12    &   &750  &   0.56$\times$0.25  &   35   &  0.420  & 2.23$\pm$0.24  &  \\
        & %[MMW2002] 46.52+63.8 
        SNR &   0.27$\times$0.16 &  80 &  0.131 &  09 55 55.214  &  69 40 50.03  &  3.38$\pm$0.27  & 7.08  &  & 750  &   0.27$\times$0.21 &  64 &  0.420  &  3.89$\pm$0.27  & \\
        & %[MMW2002] 45.91+63.8 
        SNR & $<$0.14$\times$ $<$0.08 &  0.0 &  0.131 &   09 55 54.609  &  69 40 49.99  &  3.03$\pm$0.25  & 3.26   &  & 750  &    $<$0.28$\times$ $<$0.25 &  0.0  & 0.420  & 3.17$\pm$0.26  &   \\
       & %[MMW2002] 45.79+65.2 
       SNR &   0.24$\times$0.19 &  132 &  0.131 &  09 55 54.460  &  69 40 51.43  &  3.88$\pm$0.17  & 7.56 &  &  750  & 0.25$\times$0.20    &   133    &   0.420  &  4.52$\pm$0.21  &  \\
        & %[FMB2008] 45.39+60.3 
       &   0.51$\times$0.37 &  168 &  0.131 & 09 55 54.097  &  69 40 46.45  &  1.47$\pm$0.15  & 5.73 &  & 750  &   0.42$\times$0.36 &   177    &  0.420  &  1.92$\pm$0.19  & \\
       & %[MMW2002] 45.44+67.3 
       SNR &   0.13$\times$0.08 &  150 &  0.131 & 09 55 54.133  &  69 40 53.58  &  3.67$\pm$0.16  & 4.56 &   &750  &   0.12$\times$0.04  &   152   &  0.420  & 4.01$\pm$0.18  &  \\
        & %[MMW2002] 45.26+65.3 
        SNR &   0.44$\times$0.24 &  127 &  0.131 &  09 55 53.967  &  69 40 51.31  &  2.06$\pm$0.31  & 8.2 &  & 750  &    0.51$\times$0.30 &  128  & 0.420  & 2.49$\pm$0.34  &   \\
       & %[MMW2002] 45.17+61.2 
       SNR &   0.14$\times$0.13 &  97 &  0.131 & 09 55 53.880  &  69 40 47.40  &  14.38$\pm$0.49  & 20.0 &  &  750  & 0.17$\times$0.15    &   90    &   0.420  &  15.28$\pm$0.48  &  \\
       & %[MMW2002] 44.91+61.1 
       SNR & $<$0.36$\times$ $<$0.07 &  0.0 &  0.131 & 09 55 53.628  &  69 40 47.31  &  1.58$\pm$0.30  & 2.57 &   &750  &   $<$0.37$\times$ $<$0.04 &  0.0  &  0.420  & 1.91$\pm$0.27  &  \\
        & %CXOU J095553.2+694049 
        SNR &   0.26$\times$0.18 &  85 &  0.131 & 09 55 53.145  &  69 40 47.96  &  1.98$\pm$0.31 & 4.09 &  & 750  &    0.31$\times$0.18 &  87  & 0.420  & 2.23$\pm$0.12  &   \\
        & %[MMW2002] 44.52+58.1 
        SNR &   0.19$\times$0.14 &  176 &  0.131 & 09 55 53.229  &  69 40 44.36  &  2.93$\pm$0.17  & 4.39 &  & 750  &   0.19$\times$0.12 &  9 &  0.420  &  3.20$\pm$0.24  & \\
       & %[MMW2002] 44.29+59.3 
       SNR &   0.16$\times$0.15 &  2 &  0.131 & 09 55 52.991  &  69 40 45.49  &  4.26$\pm$0.17  & 6.36 &  &  750  & 0.15$\times$0.13    &   14   &   0.420  &  4.82$\pm$0.23 &  \\
        & %Core Acc.2NED 
        SNR &   0.11$\times$0.03 &  57 &  0.131 & 09 55 52.727  &  69 40 45.77  &  10.63$\pm$0.25  & 11.87 &   &750  &   0.12$\times$0.02  &   57   &  0.420  & 10.83$\pm$0.26  &  \\
        & %*CRATES J0955+6940 
        & $<$0.24$\times$ $<$0.20&  0.0 &  0.131 & 09 55 52.028  & 69 40 45.41  & 22.89$\pm$0.60  & 20.50 &  & 750 & $<$0.28$\times$ $<$0.25 &  0.0 &  0.420  &  22.88$\pm$0.64  & \\
        & %[MMW2002] 43.18+58.3 
        SNR & 0.22$\times$0.17 &  156 &  0.131 & 09 55 51.908  &  69 40 44.54  &  7.87$\pm$0.51  & 14.2 &  & 750  &    0.26$\times$0.19 &  156  & 0.420  & 8.69$\pm$0.48  &   \\
       & {\tiny SN~2008iz} & 0.22$\times$0.17 &  156 & 0.131 & 09 55 51.551  &  69 40 45.78  &  59.00$\pm$0.44  & 56.22 &  &  750  & $<$0.28$\times$ $<$0.25 &  0.0    &   0.420  &  58.61$\pm$0.47  &  \\
       & %[MMW2002] 42.66+56.4 
       SNR & 0.83$\times$0.34 &  18 & 0.131 & 09 55 51.416  & 69 40 42.66  & 2.22$\pm$0.19  & 17.4 &  &  750  & 0.74$\times$0.34   &   20    &   0.420  &  2.95$\pm$0.20  &  \\
       & %[MMW2002] 42.67+55.6 
       SNR & 0.41$\times$0.13 &  22 & 0.131 & 09 55 51.386  & 69 40 41.86  & 2.99$\pm$0.28  & 7.03 &  &  750  & 0.66$\times$0.14    &   22    &   0.420  &  2.94$\pm$0.39  &  \\
       & %[MMW2002] 42.53+61.9 
       SNR & 0.44$\times$0.31 &  174 & 0.131 & 09 55 51.261  & 69 40 48.17  & 2.59$\pm$0.22  & 8.94 &  &  750  & 0.53$\times$0.25    &   172    &   0.420  &  2.95$\pm$0.32  &  \\
       &     & 0.41$\times$0.26 &  57 &  0.131 & 09 55 51.230  & 69 40 46.42  & 1.83$\pm$0.27  & 3.79 &  &  750  & 0.43$\times$0.26    &   64    &   0.420  &  1.81$\pm$0.34  &  \\
       & %*CRATES J0955+6940 NED01
       XS & $<$0.24$\times$ $<$0.20&  0.0 &  0.131 & 09 55 50.688  & 69 40 43.75  & 16.22$\pm$0.36  & 15.33 &  &  750  & $<$0.28$\times$ $<$0.25 &   0.0 & 0.420  &  16.07$\pm$0.41  &  \\
       & %[MMW2002] 41.29+59.7 
       SNR & $<$0.24$\times$ $<$0.20&  0.0 & 0.131 & 09 55 50.049  & 69 40 45.92  & 3.50$\pm$0.24  & 3.18 &  &  750  & $<$0.28$\times$ $<$0.25 &   0.0   &   0.420  &  3.67$\pm$0.29  &  \\
       & %[MMW2002] 40.66+55.2 
       SNR & 0.20$\times$0.12 &  137 & 0.131 & 09 55 49.421  & 69 40 41.43  & 8.93$\pm$0.15  & 13.87 &  &  750  & 10.20$\times$0.13    &   139    &   0.420  &  10.20$\pm$0.13  &  \\
       & %[MMW2002] 40.62+56.0 
       SNR & 0.54$\times$0.34 &  68 &  0.131 & 09 55 49.363 & 69 40 42.42 &  2.27$\pm$0.22  & 11.2  &  &  750  & 0.54$\times$0.43    &   62   &   0.420  &  2.81$\pm$0.32  &  \\
       & %[MMW2002] 40.32+55.1 
       SNR & 0.31$\times$0.20 &  81 &  0.131 & 09 55 49.061 & 69 40 41.52 &  2.51$\pm$0.21  & 5.61 &  &  750  & 0.36$\times$0.20    &   80    &   0.420  &  2.84$\pm$0.21  &  \\
       & %[MMW2002] 39.10+57.4 
       SNR & 0.20$\times$0.15 &  4 & 0.131 & 09 55 47.873 & 69 40 43.72 &  6.14$\pm$0.15  & 10.12 &  &  750  & 0.20$\times$0.15    &   4    &   0.420  &  6.94$\pm$0.18  &  \\
%       &      &      &    &  rms &  RA  &   DEC  &  $<$F PEAK   &     F TOT      &  &  750  & thetaM$\times$thetam    &   PA    &   rms LR  &  F PEAK LR  &  \\
\cmidrule(rl){15-15}
           & Tot  &                     &          &        &               &              &                &      &     &                      &        &      &   & 206$\pm$15 &           \\
\hline    
NGC~3077 &     &     0.28$\times$0.19  & 86 &  0.090 & 10 03 18.808  & 68 43 56.65 & 1.30$\pm$0.08  &  2.92  &    & 500   &   0.26$\times$0.19  &   91 &   0.101 &   1.81$\pm$0.10  &  SNR? 5.6$\arcsec$ offset \\
\cmidrule(rl){15-15}
  &  Tot  &                     &          &        &               &              &                &      &     &             &         &        &        &  2.8$\pm$0.2 & \\
\hline   
NGC~3690 & Arp299C &   0.48$\times$0.28   &  175  & 0.201 & 11 28 30.657 & 58 33 49.28 & 1.33$\pm$0.20 & 7.45 &   & 500  &   0.66$\times$0.40    &  -68.7  & 0.236  & 1.82$\pm$0.24 &   \multirow{3}{1.8cm}{\centering merging system}\\
         & Arp299B &   0.25$\times$0.21   &  59  & 0.190 & 11 28 30.987 & 58 33 40.83 & 14.1$\pm$0.7 & 26.10     &   & 500  &   0.27$\times$0.24   &  79  & 0.231  & 15.21$\pm$0.80 &   \\
         & Arp299A &   0.52$\times$0.41   & 131  & 0.203 & 11 28 33.631 & 58 33 46.68 & 23.14$\pm$0.87 & 108.8 &   & 500  &   0.52$\times$0.42          &  129  & 0.245  & 31.23$\pm$0.25 &    \\
\cmidrule(rl){15-15}
           & Tot  &                     &          &        &               &              &                &      &     &                      &        &      &   &  142$\pm$10 &           \\   
\hline 
NGC~3838    &  core &  $<$0.34$\times<$0.21      &   19.3    &  0.089  &  11 44 10.854   &  57 56 53.34  &  0.99$\pm$0.08   &   1.03  &  & 500  &  $<$0.45$\times<$0.35 &  1.2  & 0.078 &  0.89$\pm$0.08  & offset 23.2$\arcsec$ \\
\cmidrule(rl){15-15}
            &  Tot &                     &          &        &               &              &                &      &     &                      &        &      &    &  1.1$\pm$0.2 &            \\
\hline
NGC~4111 &     &     0.34$\times$0.23  & 19.1 &  0.102 & 12 07 03.614  & 43 03 55.99 & 0.55$\pm$0.10  &  0.78  &    & 500   &  0.57$\times$0.06  &   19.4   &   0.110 &   0.50$\pm$0.11  &  offset 6$\arcsec$ \\
         &     &     0.83$\times$0.32  & 11.3 &  0.102 & 12 07 03.854  & 43 03 59.58 & 0.46$\pm$0.10  &  1.86  &    & 500   &  0.73$\times$0.26  &   1.4   &    0.110 &   0.48$\pm$0.11  &   \\
\cmidrule(rl){15-15}
  & Tot   &                     &          &        &               &              &                &      &     &             &         &        &        &  1.85$\pm$0.13 &           \\
\hline
NGC~4162 &   & 0.36$\times$0.27     &  104  & 0.175 & 12 11 54.104 & 24 06 58.76 & 2.40$\pm$0.24 & 5.35 &  & 500  &   0.48$\times$0.22  &   123   &  0.233  & 3.11$\pm$0.27  &   \multirow{3}{1.8cm}{\centering double source, offset $\gtrsim$30$\arcsec$}\\
         & %SSTSL2 J121154.05+240658.4 
         & $<$0.48$\times <$0.16     &  137  & 0.175 & 12 11 54.046 & 24 06 58.47 & 7.75$\pm$0.28 & 6.70 &  & 500  &   $<$0.60$\times <$0.23  &   139   &  0.233  & 6.96$\pm$0.24  &    \\
\cmidrule(rl){15-15}
            & Tot  &                     &          &        &               &              &                &      &     &                      &        &      &   &  10.2$\pm$1.0 &           \\   
\hline 
NGC~4236   &     & $<$0.19$\times<$0.12      &   12.0    &  0.086  & 12 16 35.542 & 69 28 00.57   &  2.91$\pm$0.20   &   3.23   &  & 500  & 0.13$\times$0.03 &  149  & 0.103 &  2.98$\pm$0.16  & \multirow{2}{1.8cm}{\centering double source, 37.9$\arcsec$ offset} \\
\cmidrule(rl){15-15}
            &  Tot &                     &          &        &               &              &                &      &     &                      &        &      &    &  3.3$\pm$0.2 &            \\
\hline
NGC~4242       &  core    &  0.09$\times $0.06    &  7.5 &  0.279 &  12 17 28.247  &   45 37 25.11  &  11.25$\pm$0.30   & 11.30 &  &  300  & 0.22$\times$0.05    &   7.4  &   0.176  &  10.70$\pm$0.18  &  \multirow{3}{1.8cm}{\centering double source, 26$\arcsec$ offset} \\
               &      &  0.26$\times$0.25    &  168    &  0.279 &  12 17 28.203  &   47 37 26.13  & 1.33$\pm$0.28  &   2.24   &  &  300  & 0.51$\times$0.05    &   16.0    & 0.176 &  1.78$\pm$0.19  &  \\
               &      &  0.42$\times$0.25    &  8.3    &  0.279 &  12 17 28.250  &   47 37 24.12  & 1.18$\pm$0.28  &   1.05    &  &  300  &     &       & 0.176 &  $<$1.2  &  \\
\cmidrule(rl){15-15}
           & Tot  &                     &          &        &               &              &                &      &     &                      &        &      &   &  14.7$\pm$0.5 &           \\   
\hline 
NGC~4631 &      &  0.20$\times$0.21  &  78  & 0.077 &  12 42 07.697 & 32 32 30.08       &   0.34$\pm$0.08  &  0.79   & & 500   &  0.25$\times$0.27  &  -67.7 &  0.076 & 0.40$\times$0.08  &  offset 3.6$\arcsec$ \\
         &      &  0.26$\times$0.02  &  78  & 0.077 &  12 42 07.689 & 32 32 30.57       &   0.31$\pm$0.08  &  0.43   & & 500   &  $<$0.31$\times$0.21  & 57.8 &  0.076 & 0.28$\times$0.08   \\
\cmidrule(rl){15-15}
            & Tot  &                     &          &        &               &              &                &      &     &                      &        &    &      &  1.22$\pm$0.18 &            \\
\hline
NGC~5012 &      & 0.13$\times$0.07  &  45  &   0.079 &  13 11 37.048  &  22 54 58.277      &  0.41$\pm$0.08        &  0.44 & & 500    &  $<$0.30$\times <$0.24 &  31 &  0.068  &  0.41$\times$0.07   & offset 2.6$\arcsec$   \\
\cmidrule(rl){15-15}
            & Tot  &                     &          &        &               &              &                &      &     &                &      &        &          &  0.45$\pm$0.10 &            \\
\hline
\hline
\end{longtable}
\end{center}
\vspace*{-1cm} \fontsize{6}{7}\selectfont Column description: 
(1) galaxy name; 
(2) radio component:  radio component: (core, SNR or X-ray source [XS]) or unidentified component if not labeled; 
(3) deconvolved FWHM dimensions (major $\times$ minor axes [arcsec$^{2}$], $\theta_{\rm M}\times\theta_{\rm m}$) of the fitted component, determined from an elliptical Gaussian fit from the full-resolution radio map; 
(4) PA of the deconvolved component, PA$_{\rm d}$, from the full-resolution radio map (degree); 
(5) rms of the radio map close to the specific component from the full-resolution radio map (mJy/b, mJy beam$^{-1}$); 
(6)-(7) radio position (J2000.0); 
(8) peak flux density in mJy beam$^{-1}$, F$_{\rm peak}$, from the full-resolution radio map; 
(9) integrated flux density, $F_{\rm tot}$, in mJy, from the full-resolution radio map; 
(10) $uv$-taper scale of the low-resolution radio map in k$\lambda$; 
(11) deconvolved FWHM dimensions (major $\times$ minor axes [arcsec$^{2}$], $\theta_{\rm M}\times\theta_{\rm m}$) of the fitted component, determined from an elliptical Gaussian fit from the low-resolution radio map; 
(12) PA of the deconvolved component, PA$_{\rm d}$ from the low-resolution radio map (degree); 
(13) rms of the radio map close to the specific component from the low-resolution radio map (mJy beam$^{-1}$); 
(14) peak flux density in mJy beam$^{-1}$, F$_{\rm peak}$ from the low-resolution radio map. At the bottom of each target the total flux density of the radio source associated with the galaxy is given in mJy, measured from the low-resolution map; 
(15) radio morphology, size in arcsec and pc and offset from the optical centre, if present.  
\end{landscape}

\small
\begin{center}
%\begin{table*}
%\begin{center}
\begin{longtable}{lccp{3cm}|C{0.05cm}cccl}
\caption[]{Radio contour levels} 
\label{contours}\\

%This is the header for the first page of the table...
\hline %\hline 

%This is the header for the first page of the table...
%\hline \hline 

%\begin{tabular}{lccp{3cm}|C{0.05cm}cccl}

name     &  \multicolumn{3}{c|}{full resolution}                      &  \multicolumn{5}{c}{low resolution}  \\
          &   Beam         & PA         & levels       & &  k$\lambda$  & beam & PA &    levels    \\
          
\hline	
\endfirsthead

%This is the header for the remaining page(s) of the table...
\multicolumn{3}{c}{{\tablename} \thetable{} -- Continued} \\[0.5ex]
\hline \hline           
          
name     &  \multicolumn{3}{c|}{full resolution}                      &  \multicolumn{5}{c}{low resolution}  \\
          &   Beam         & PA         & levels       & &  k$\lambda$  & beam & PA &    levels    \\

\hline
\endhead

%This is the footer for all pages except the last page of the table...
\hline
  \multicolumn{9}{c}{{Continued on Next Page}} \\
\endfoot

%This is the footer for the last page of the table...
\hline \hline
\endlastfoot

% \\[-1.8ex] 
%\endlastfoot

NGC~1161 & 0.18$\times$0.12 & 55.8  & 0.20$\times$($-$1,1,2,4,8,13)  &  &  500 &  0.34$\times$0.18 & -39.9 &  0.50$\times$($-$1,1,2,4,5.5) \\
NGC~1167 & 0.20$\times$0.20 &  0    &  15$\times$($-$1,1,2.5,5,10,20,30) &   &  500 & 0.40$\times$0.40 & 0   &  8$\times$(-1, 1, 2, 5, 10, 20, 40, 70)  \\
NGC~1186 &  0.18$\times$0.12 & 52.3 & 0.15$\times$($-$1, 1, 2)      &  &  750 &  0.35$\times$0.19 & -43.4 &  0.15$\times$($-$1,1,2,3) \\
NGC~1275 & 1.00$\times$0.33 & -41.6 &  600$\times$($-$1, 1, 2, 4, 8, 16)  &  &  750 &  1.20$\times$0.39 & -41.0 &   1000$\times$($-$1, 1, 2, 4, 8)  \\
IC~342   & 0.33$\times$0.20 & -57.3 & 0.6$\times$($-$1,1,1.4,2) & & 500 & 0.43$\times$0.35 &  $-$29.5  &  0.5$\times$($-$1,1,1.3,2,3)\\
IC~356   & 0.33$\times$0.20 & -54.5 & 0.25$\times$($-$-1,1,1.5,2) & & 300 & 0.62$\times$0.52 &  5.3  &  0.20$\times$($-$-1,1,1.5)\\
NGC~1569 & 0.33$\times$0.20 & -50.1 & 0.25$\times$($-$1,1,1.5) & & 300 & 0.63$\times$0.52 &  5.0  &  0.30$\times$($-$1,1,2,2.5,3)\\
NGC~1560 & 0.26$\times$0.17 & -53.7 & 0.26$\times$($-$1,1,1.3) & & 300 & 0.52$\times$0.41 &  -174.3  &  0.23$\times$($-$1,1,1.5,2)\\
NGC~1961 & 0.29$\times$0.18 & -35.8 & 0.35$\times$($-$1,1,2,4,7) & & 300 & 0.53$\times$0.41 &  17.6  &  0.28$\times$($-$1,1,2,4,8)\\
NGC~2146 & 0.27$\times$0.18 & -23.7 & 0.7$\times$($-$1,1,2,4) & & 300 & 0.48$\times$0.38 &  31.6  &  0.55$\times$($-$-1,1,2,4,)\\
IC~467   & 0.26$\times$0.19 & -15.7 & 0.26$\times$($-$-1,1,1.5) & & 300 & 0.45$\times$0.39 &  50.6  &  0.30$\times$($-$-1,1,1.5,2)\\
NGC~2683 &  0.26$\times$0.18 &  22.8 &  0.12$\times$($-$1,1,1.5,3,3.9) &  &  500 &  0.38$\times$0.33 &  -57.1 &  0.11$\times$($-$1, 1, 1.5, 2.1, 3.5)  \\
NGC~2750 & 0.19$\times$0.14  &  33.0  &  0.17$\times$($-$1, 1, 1.5, 1.9)     &   & 500  &  0.40$\times$0.24 &  -55.3  &  0.13$\times$($-$1, 1, 1.5, 2.5) \\
NGC~2768 & 0.15$\times$0.15 & 88.6 & 0.8$\times$($-$1,1,2,4,8,13) & & 500 & 0.24$\times$0.21 &  $-$81  &  0.9$\times$($-$1,2,4,8,13)\\
NGC~2782 &  0.24$\times$0.15 & 31.5  &  0.22$\times$($-$1, 1, 2, 4, 8) &  &  750 & 0.27$\times$0.22 & 53.8  &  0.20$\times$($-$1, 1, 2, 4, 8, 12) \\
NGC~2787 & 0.16$\times$0.15 & -53.8 & 0.9$\times$($-$1,1,1.5,3,6) & & 750 & 0.20$\times$0.18 &  $-$47.8  &  0.8$\times$($-$1,1,1.5,3,6)\\
NGC~2832 &  0.18$\times$0.14 &  50.3  &  0.19$\times$($-$1, 1, 1.5, 2)      &   & 500  &  0.42$\times$0.24 &   137.8 &   0.25$\times$($-$1, 1, 1.5, 2)  \\
NGC~2831 &  0.18$\times$0.14 &  50.3  &  0.30$\times$($-$1, 1, 2, 4, 8)    &   & 500  &   0.42$\times$0.24 &   137.8 &   0.30$\times$($-$1, 1, 2, 4, 8) \\
NGC~2964 &  0.17$\times$0.16 &  57.5  &  0.24$\times$($-$1, 1, 2, 4)      &    &  500 &   0.47$\times$0.26 &  -48.5  &   0.32$\times$($-$1, 1, 2, 3, 4) \\
NGC~2985 & 0.16$\times$0.15 & -54.9 & 0.24$\times$($-$1,1,1.5,2,2.5) & & 500 & 0.24$\times$0.22 &  $-$44.5  &  0.26$\times$($-$1,1,1.5,2,2.7)\\
NGC~3031 & 0.25$\times$0.21 & 61.5  & 2.9$\times$($-$1,1,2.5,5,15,30)  & & 500& 0.32$\times$0.30 &  $-$69.5  & 2.9$\times$($-$1,1,2.5,5,15,30)\\    
NGC~3034 & 0.24$\times$0.20 &  60.1  & 0.002$\times$($-$0.5,1,2,4,5, 10,15,20,25,30)  & & 750& 0.28$\times$0.25 &  73.7  &  0.002$\times$($-$0.5,1,2,4,5, 10,15,20,25,30)\\ 
NGC~3077 & 0.24$\times$0.19 & 46.6 & 0.23$\times$($-$1,1,1.5,2,2.7) & & 500 & 0.30$\times$0.28 &  62.3  &  0.24$\times$($-$1,1,2,3)\\
NGC~3077$^{*}$ & 0.24$\times$0.19 & 46.6 & 0.21$\times$($-$1,1,2,3,5,7) & & 500 & 0.30$\times$0.28 &  62.3  &  0.23$\times$($-$1,1,2,4,8)\\
NGC~3079 & 0.30$\times$0.25 & 64.7 & 3$\times$($-$0.8,1,1.5,2,4,8) & & 300 & 0.47$\times$0.38 &  $-$27  &  4$\times$($-$1,1,2,4,8)\\
NGC~3147 & 0.28$\times$0.23 & 67.5 & 0.37$\times$($-$1,1,2,4,8,16) & & 750 & 0.29$\times$0.25 &  65.8  &  0.35$\times$($-$1,1,2,4,8,16)\\
\multirow{2}{*}{NGC~3245}   &\multirow{2}{*}{0.19$\times$0.17} &\multirow{2}{*}{39.1} &\multirow{2}{*}{0.17$\times$($-$1,1,1.5,2,3)} & \rdelim\{{2}{20pt}& 750 & 0.22$\times$0.21 &  86.6  &  0.18$\times$($-$1,1,1.5,2,3)\\
        &                                  &                        &                                                    &  & 300  &  0.49$\times$0.40 &  $-$49.0 & 0.15$\times$($-$1, 1, 1.5, 2, 4, 6) \\
NGC~3301 & 0.21$\times$0.18 & 26.8 & 0.17$\times$($-$1,1,1.5,2,3) & & 750 & 0.23$\times$0.22 &  60.5  &  0.17$\times$($-$1,1,1.5,2,3)\\
NGC~3310 & 0.27$\times$0.23 & 69.4 & 0.3$\times$($-$1,1,1.5,2) & & 750 & 0.29$\times$0.26 &  58.1  &  0.3$\times$($-$1,1,1.5,2)\\
NGC~3348 & 0.23$\times$0.19 & 56.7 & 0.18$\times$($-$1,1,1.5,2,4,7) & & 750 & 0.27$\times$0.24 &  67.0  &  0.19$\times$($-$1,1,1.5,2,4,7)\\      
NGC~3448  &   0.27$\times$0.22 &  71.5 & 0.22$\times$($-$1,1,1.5) &  &  300  &  0.46$\times$0.36 &  -9.9 & 0.25$\times$($-$1,1,1.5,2) \\
NGC~3504 & 0.22$\times$0.19 & -77.4 & 0.22$\times$($-$1, 1, 2, 4, 8, 16, 32, 64) & & 500 & 0.31$\times$0.26 &  $-$81.7  &  0.24$\times$($-$1,1,2,4,8,16,32,64)\\
\multirow{2}{*}{NGC~3516} & \multirow{2}{*}{0.23$\times$0.19} & \multirow{2}{*}{50.9} & \multirow{2}{*}{0.5$\times$($-$1,1,2,3,6)} & \rdelim\{{2}{20pt}& 500 & 0.30$\times$0.28 &  78.7  &  0.5$\times$($-$0.8,1,2,3,4,6)\\
         &                                  &                        &                                                    &  & 300  &  0.38$\times$0.36 &  $-$43.0 & 0.5$\times$($-$1,1,1.5,2,4,6) \\
NGC~3690 Arp299-B & 0.28$\times$0.21 & 88.2 & 0.8$\times$($-$0.8,1,2,4,8,16) & & 500 & 0.32$\times$0.28 &  $-$68.7  &  0.8$\times$($-$0.8,1,2,4,8,16)\\
NGC~3690 Arp299-A & 0.28$\times$0.21 & 88.2 & 0.8$\times$($-$0.9,1,2,4,8,16,28) & & 500 & 0.32$\times$0.28 &  $-$68.7  &  1.1$\times$($-$1,1,2,4,8,16,28)\\
NGC~3690 Arp299-C & 0.28$\times$0.21 & 88.2 & 0.7$\times$($-$0.8,1.2,1.5) & & 500 & 0.32$\times$0.28 &  $-$68.7  &  0.6$\times$($-$1,1,1.5,2,3)\\
NGC~3718 &  0.33$\times$0.20 &  4.1  &  0.25$\times$($-$1, 1, 2, 4, 8, 16, 32) &  &  500  &  0.50$\times$0.35 &  -4.7  & 0.28$\times$($-$1, 1, 2, 4, 8, 16, 32) \\
NGC~3729 &  0.31$\times$0.20 &  2.6  &  0.29$\times$($-$1, 1, 2, 4)            &  &  500  &  0.49$\times$0.35 &  -5.1  & 0.30$\times$($-$1, 1, 2, 4, 7)  \\
NGC~3735 & 0.23$\times$0.19 & $-$59.0 & 0.19$\times$($-$1,1,1.5,2,3) & & 750 & 0.27$\times$0.24 &  68  &  0.2$\times$($-$1,1,1.5,2,3)\\
NGC~3838 &  0.32$\times$0.21 & -155 &  0.22$\times$($-$1, 1, 2, 4)            &  &  500  &  0.45$\times$0.36 &  -3.5  & 0.21$\times$($-$1, 1, 2, 4) \\
NGC~3884 & 0.70$\times$0.34 & -41.5 & 0.45$\times$($-$1,1,2,4,8) & & 750 & 0.47$\times$0.47 &  0  &  0.53$\times$($-$1,1,1.5,3,4.5)\\
NGC~3898 & 0.30$\times$0.19  & -175 &  0.21$\times$($-$1, 1, 2, 3)            &  &  500  &  0.47$\times$0.35 &  -7.2  &   0.30$\times$($-$1, 1, 2, 3)     \\ 
NGC~3941 & 0.20$\times$0.19 & 41.1 & 0.20$\times$($-$1,1,1.5,2) & & 500 & 0.25$\times$0.24 &  37.7  &  0.20$\times$($-$1,1,1.5,2.3)\\
NGC~3945 & 0.17$\times$0.14 & 33.5 & 0.19$\times$($-$1,1,1.5,2,3) & & 300 & 0.44$\times$0.39 &  $-$8.2  &  0.18$\times$($-$1,1,1.5,2,3)\\
NGC~3963 & 0.24$\times$0.15  & -42.8 &  0.20$\times$($-$1, 1, 1.5)            &  &  500  &  0.44$\times$0.35 &  -5.6  &   0.20$\times$($-$1, 1, 1.5)     \\ 
NGC~3982 &  0.18$\times$0.12  & -60.6 &  0.48$\times$($-$1, 1, 1, 2, 4)       &  &  500  &  0.39$\times$0.29 &  -7.7  &   0.34$\times$($-$1, 1, 2, 4, 8) \\
NGC~3998 & 0.32$\times$0.20  &  -4.0  &    5$\times$($-$0.3, 1, 2, 4, 8, 16) &  &  500  &  0.50$\times$0.35  & -9.8  &    1$\times$($-$0.4, 1, 2, 4, 8, 16, 32, 64, 100) \\
\multirow{2}{*}{NGC~4036} & \multirow{2}{*}{0.17$\times$0.14} & \multirow{2}{*}{34.7} & \multirow{2}{*}{0.25$\times$($-$1,1,2,3,7)} & \rdelim\{{2}{20pt}& 500 & 0.28$\times$0.27 &  $-$26  &  0.27$\times$($-$1,1,2,4,8)\\
        &                                  &                        &                                               &  & 300 &  0.43$\times$0.39 & $-$8.4 & 0.20$\times$($-$1,1,2,4,8,14) \\
NGC~4041 & 0.17$\times$0.15 & 30.1 & 0.15$\times$($-$1,1,1.5,2,3) & & 500 & 0.29$\times$0.26 &  $-$27.2  &  0.14$\times$($-$1,1,1.5,3,4)\\  
NGC~4100 & 0.20$\times$0.16 & -55.9 & 0.19$\times$($-$1,1,1.5,2) & & 750 & 0.24$\times$0.19 &  $-$55.1  &  0.20$\times$($-$1,1,1.5,2,2.3)\\
NGC~4102 & 0.20$\times$0.15 & -43.2  & 0.31$\times$($-$1,1,1.5,2,3.5,6)  & & 500 & 0.27$\times$0.22 & $-$39.1  &  0.30$\times$($-$1,1,2,3,4,6,8)\\  
NGC~4111 & 0.40$\times$0.23 & 16.1 & 0.29$\times$($-$1,1,1.4,1.8) & & 500 & 0.53$\times$0.29 &  16.6  &  0.26$\times$($-$1,1,1.4,1.8)\\
NGC~4111$^{*}$ & 0.40$\times$0.23 & 16.1 & 0.30$\times$($-$1,1,1.5) & & 500 & 0.53$\times$0.29 &  16.6  &  0.29$\times$($-$1,1,1.5)\\
NGC~4143 & 0.40$\times$0.22 & 16.0 & 0.50$\times$($-$1,1,1.5,3,4.5) & & 300 & 0.76$\times$0.41 &  9.9  &  0.47$\times$($-$1,1,2,3,4)\\
NGC~4151 & 0.26$\times$0.16 & 38.8 & 1$\times$($-$0.5,1,2,4,9,16, 25,36,49,64) & & 500 & 0.34$\times$0.25 &  47.3  &  1.1$\times$($-$0.5,1,2,4,9,16, 25,36,49,64)\\
NGC~4162 & 0.47$\times$0.19 & -43.8 & 0.9$\times$($-$1,1,1.5,2,3,6) & & 500 & 0.58$\times$0.27 &  $-$41.4  &  0.7$\times$($-$1,1,1.5,2,3,6)\\  
NGC~4203 & 0.22$\times$0.18 & 54.7 & 0.50$\times$($-$1,1,2,4,8,13) & & 500 & 0.28$\times$0.23 &  59.5  &  0.60$\times$($-$1,1,2,4,8,11)\\ 
NGC~4217 & 0.46$\times$0.26 & 8.0 & 0.2$\times$($-$1,1,1.4) & & 300 & 0.73$\times$0.41 &  7.9  &  0.26$\times$($-$1,1,1.4,1.8)\\
NGC~4220 & 0.38$\times$0.21 & 9.8 & 0.21$\times$($-$1,1,1.4) & & 300 & 0.73$\times$0.41 &  $-$174.5  &  0.21$\times$($-$1,1,1.5,2)\\  
NGC~4236 &  0.16$\times$0.12 &  25.0  &  0.40$\times$($-$0.8, 1, 2, 4)      &    &  500 &   0.25$\times$0.22 &  2.2  &   0.60$\times$($-$0.8, 1, 2, 4) \\
NGC~4244 & 0.20$\times$0.20 & 33.7 & 0.20$\times$($-$1,1,1.5) & & 300 & 0.38$\times$0.32 &  6.7  &  0.20$\times$($-$1,1,1.5,2)\\ 
NGC~4242 & 0.42$\times$0.25 & 8.3 & 0.70$\times$($-$1,1,1.4,4,8,13) & & 300 & 0.78$\times$0.41 &  6.7  &  0.80$\times$($-$0.8, 1, 1.5, 2, 4, 8, 13)\\ 
NGC~4258 & 0.42$\times$0.25 & -176.3 & 0.25$\times$($-$1,1,2,4) & & 750 & 0.49$\times$0.28 &  $-$173.9  &  0.25$\times$($-$1,1,2,4)\\ 
NGC~4278 & 0.32$\times$0.32 & 0.0 & 2.5$\times$($-$1,1,2,4,8,16, 32,64) & & 750 & 0.38$\times$0.38 &  0.0  &  2.3$\times$($-$1,1,2,4,8,16,32,6)\\
NGC~4369 &  0.19$\times$0.17 & 78.1  &  0.08$\times$($-$1, 1, 1.5)  &   &  300  &  0.47$\times$0.42 &  -51.6 &  0.09$\times$($-$1, 1, 1.5, 2)  \\
NGC~4395 & 0.22$\times$0.18 & 56.2 & 0.25$\times$($-$1,1,1.5,2.5) & & 500 & 0.27$\times$0.24 &  70.2  &  0.25$\times$($-$1,1,1.5,2,2.5)\\
NGC~4494 & 0.23$\times$0.20 & 49.3 & 0.13$\times$($-$1,1,2) & & 750 & 0.26$\times$0.23 &  57.6  &  0.11$\times$($-$1,1,2,3)\\
NGC~4565 & 0.22$\times$0.20 & 44.4 & 0.26$\times$($-$1,1,2,3,4) & & 500 & 0.28$\times$0.27 &  69.1  &  0.22$\times$($-$0.8,1,2,3,4)\\
NGC~4589 & 0.16$\times$0.12 & 30.9 & 0.65$\times$($-$1,1,2,4,8,16) & & 750 & 0.20$\times$0.16 &  20.9 &  0.70$\times$($-$1,1,2,4,8,16)\\
NGC~4631 & 0.21$\times$0.20 & 77.9 & 0.17$\times$($-$1,1,1.5,2) & & 500 & 0.27$\times$0.25 &  $-$67.7  &  0.18$\times$($-$1,1,2)\\
NGC~4750 & 0.16$\times$0.12 & 27.2 & 0.15$\times$($-$1,1,2,3) & & 500 & 0.25$\times$0.22 &  $-$5.9  &  0.15$\times$($-$0.8,1,2,3,5)\\
NGC~4736 & 0.25$\times$0.16 & 33.1 & 0.30$\times$($-$1, 1, 1.5, 3, 6) & & 500 & 0.34$\times$0.24 &  45.2  &  0.40$\times$($-$1, 1, 1.5, 3, 6)\\
NGC~4826 & 0.22$\times$0.20 & 46.4 & 0.17$\times$($-$1,1,1.5,2,3) & & 300 & 0.39$\times$0.34 &  $-$42.0  &  0.18$\times$($-$1,1,2,3,4)\\
NGC~5012  &    0.23$\times$0.20 &  42.5 & 0.18$\times$($-$1,1,1.5,2) &  & 750  &  0.29$\times$0.26 & 48.5 & 0.17$\times$($-$1,1,1.5,2)  \\      
NGC~5033 &  0.18$\times$0.16 & 52.7  &  0.19$\times$($-$1, 1, 2, 4, 8) &  &  500 & 0.30$\times$0.26 & -50.6  &  0.20$\times$($-$1, 1, 2, 4, 8) \\
\multirow{2}{*}{NGC~5322} & \multirow{2}{*}{0.31$\times$0.22} & \multirow{2}{*}{33.0} & \multirow{2}{*}{0.27$\times$($-$1,1,1.5,2,4,10)} & \rdelim\{{2}{20pt}& 500 & 0.40$\times$0.31 &  17.7  &  0.34$\times$($-$1,1,1.5,2,4,10)\\
        &                                  &                        &                                               &  & 300 &  0.44$\times$0.38 & $-$169 &  0.30$\times$($-$1,,1,1.5,2,4,10) \\
NGC~5354 &  0.19$\times$0.16 & 67.1  &  0.30$\times$($-$1, 1, 2, 4, 8,14) &  &  500 &  0.31$\times$0.26 & -62.4 & 0.30$\times$($-$1, 1, 2, 4, 8,14) \\
NGC~5485 & 0.40$\times$0.20 & 27.1 & 0.44$\times$($-$1, 1, 1.5, 2, 3) &  & 500 & 0.46$\times$0.34 &  13.2  &  0.35$\times$($-$1,1,1.5,2,3)\\
\multirow{2}{*}{NGC~5548}  &    \multirow{2}{*}{0.21$\times$0.14} &  \multirow{2}{*}{-161.6} & \multirow{2}{*}{0.20$\times$($-$1,1,1.5,3,5)} &   \rdelim\{{2}{20pt} & 500   & 0.26$\times$0.22 & 11.2  &  0.20$\times$($-$1,1,2,4)  \\
          &                                  &                        &                                               &  & 300 &  0.35$\times$0.34 & $-$21.2 & 0.25$\times$($-$1,1,1.4,2.2,3.2,5) \\
NGC~6217 & 0.17$\times$0.14 & 72.4 & 0.23$\times$($-$1,1,1.5,2) & & 500 & 0.25$\times$0.23 &  115  &  0.23$\times$($-$1,1,1.5,2)\\
NGC~6340 & 0.17$\times$0.15 & 69.8 & 0.22$\times$($-$1,1,1.5,2) & & 300 & 0.37$\times$0.32 &  $-$21.6  &  0.25$\times$($-$1,1,1.8,2.7)\\
NGC~6482 &  0.19$\times$0.16 &  64.6  &  0.16$\times$($-$1, 1, 1.6, 2.6)  &    &  500 &   0.28$\times$0.21 &  -73.1  &  0.17$\times$($-$1, 1, 2, 2.8)  \\
NGC~7080 & 0.31$\times$0.15 &  4.3  &  0.16$\times$($-$1,1,2.5)     &   &  500  & 0.32$\times$0.21 &  9.6  &  0.16$\times$($-$1,1,1.8) \\\
\end{longtable}
\end{center}
\normalsize
\vspace*{-1cm} \fontsize{6}{7}\selectfont 
Column description: (1) source name; 
(2) FWHM of the elliptical Gaussian restoring beam in arcsec$^{2}$ of the full-resolution maps (Fig.~\ref{ident-maps});
(3) PA of the restoring beam (degree) of the full-resolution maps;
(4) radio contour levels (mJy beam$^{-1}$) of the full-resolution maps; 
(5) uvtaper scale parameter in k$\lambda$ of the low-resolution radio maps (the full sets of figures is in the online supplementary data); 
(6) FWHM of the elliptical Gaussian restoring beam in arcsec$^{2}$ of the low-resolution maps (the full sets of figures is in the online supplementary data); 
(7) PA of the restoring beam (degree) of the low-resolution maps; 
(8) radio contour levels (mJy beam$^{-1}$) of the low-resolution maps. The $*$ tag identifies the secondary radio source detected in the field of the main target.

%%%%%%%%%%%%%%%%%%%%%%%%%%%%%%%%%%%%%%%%%%%%%%%%%%

%\bsp
\label{lastpage}
\end{document}